\documentclass{article}      
\usepackage{graphicx}
\usepackage{epstopdf}

\title{The hypercentral Constituent Quark Model and its application to baryon properties}  
\author{
M.M. Giannini\\
Dipartimento di Fisica dell'Universit\`a di Genova\\
and \\
I.N.F.N., Sezione di Genova\\
E. Santopinto\\
I.N.F.N., Sezione di Genova}
\date{}
\begin{document}             

\maketitle
        
\begin{abstract}
The hypercentral Constituent Quark Model (hCQM) for the baryon structure is reviewed and its applications are systematically discussed. The model is based on a simple form of the quark potential, which contains a Coulomb-like interaction and a confinement, both expressed in terms of a collective space coordinate, the  hyperradius. The model has only three free parameters, determined in order to describe the baryon spectrum. Once the parameters have been fixed, the model, in its non relativistic version, is used to  predict various quantities of physical interest, namely the elastic nucleon form factors,  the  photocouplings and the helicity amplitudes for the electromagnetic excitation of the baryon resonances. In particular, the $Q^2$ dependence of the helicity amplitude is quite well reproduced, thanks to the Coulomb-like interaction. The model is reformulated in a relativistic version by means of the Point Form hamilton dynamics. While the inclusion of relativity does not alter the results for the helicity amplitudes, a good description of the nucleon elastic form factors is obtained.
\end{abstract}

\tableofcontents

\section{Introduction}  

The quark model has been introduced fifty years ago \cite{gm,zw} as a realization of the $SU(3)$ symmetry and it has been used with success for the description of many
important properties of hadrons, as the existence of multiplets, their
quantum numbers  and the magnetic moments \cite{rhd,kok}.
The idea of quarks as effective particles (Constituent Quarks) emerged very early \cite{morp65} and was further developed with the introduction of the colour quantum numbers.

Here we shall concentrate ourselves on Constituent Quark Models (CQM) for baryons.

After the pioneering work of Isgur and Karl (IK) \cite{ik} a series of CQM followed: the relativized Capstick-Isgur model (CI)
\cite{ci}, the algebraic approach (BIL) \cite{bil}, the hypercentral CQM
(hCQM) \cite{pl,nc,es}, the chiral Goldstone Boson Exchange model ($\chi$CQM)
\cite{olof,ple1,ple2,ple3}, the Bonn instanton model (BN) \cite{bn1,bn2,bn3,bn4} and the interacting quark-diquark model \cite{diq}. 

All models reproduce the baryon spectrum, which is the first quantity to be approached when building a model for the baryon structure, but  have been widely used to describe baryon properties. In some cases  the calculations referred to as a CQM one are performed using a simple h.o.\ wave function for the internal quark motion either in a nonrelativistic (HO) or relativistic framework (rHO).

The photocouplings for the excitation of the baryon resonances have been  calculated in various models, among others we quote HO \cite{cko}, IK \cite{ki}, CI  \cite{cap}, BIL  \cite{bil}, hCQM  \cite{aie} (for a comparison among these and other previous 
approaches see e.g.~\cite{aie,cr2}). The calculations   reproduce the overall trend, but the strength is systematically lower than the data. The fact that quite different models lead to similar results  can be ascribed to their  common $SU(6)$ structure.

As for the nucleon elastic form factors there are the calculations performed by BIL \cite{bil,bil2}  with the algebraic method and  by the Rome group \cite{card_95,card_00,demelo_N} within  a light front  approach based on the CI model.
The hCQM has been firstly applied in the nonrelativistic version with Lorentz boosts \cite{mds,rap} and then it has been reformulated relativistically  \cite{ff_07,ff_10}. A quite good description of the elastic form factors is achieved also using the GBE \cite{wagen,boffi} and the BN \cite{mert} models, both being fully relativistic. The same happens for the interacting quark-diquark model \cite{diq}, specially in its relativistic version \cite{diq2}.

A sensible test of both the energy and the short range properties of the quark structure is provided by the  $Q^2$ behaviour of the helicity amplitudes for the electromagnetic excitation to the baryon resonances.

In the HO framework, there are various calculations of the transverse helicity amplitudes, among them we quote refs.~\cite{cko,ki,cl,warns,sw}, while a systematic rHO approach has been used by \cite{ck}. A light cone calculation, using the CI \cite{ci} model, has been performed \cite{cap}  and then successfully applied to the $\Delta$ \cite{card_ND} and Roper excitations \cite{card_Roper}. For more recent light cone approaches, see ref.~\cite{azn-bur} and references therein. The algebraic method has been also used for the calculation of the transverse helicity amplitudes \cite{bil}. The hCQM, in its nonrelativistic version, has produced nice predictions for the transverse excitation of the negative parity resonances \cite{aie2} and, recently, for both the transverse and longitudinal helicity amplitudes of all resonances having a sensible excitation strength \cite{sg}. The calculation of the helicity amplitudes in a relativistic hCQM is in progress and some preliminary results for the $\Delta$ resonance are now available \cite{fb22}.
Helicity amplitudes have been calculated also by the Bonn group, both for the nonstrange \cite{mert,ronn} and strange resonances \cite{caut}.

The models have been applied also to the decays of baryons. The strong decays have been quite soon calculated with the IK model \cite{ki} and in its relativized versions \cite{cr,cr3}. There are also calculations in other models, namely BIL \cite{bil3}, GBE \cite{melde}. As for the hCQM, there are some preliminary calculations \cite{bad}.
There are also calculations of the semileptonic decays of baryons in the BN model \cite{mig}. 

Finally we quote calculations of the  axial nucleon form factors in the GBE  \cite{boffi,gloz} and  BN \cite{mert}.

\section{A review of Constituent Quark Models}

\subsection{Nonrelativistic approach}

The possibility of a nonrelativistic description of the internal quark dynamics was considered very early \cite{morp65} after the introduction of the quark model. In this framework, one can introduce the three-quark wave function $\Psi_{3q} $, factorized according to the various degrees of freedom:
\begin{equation}
\Psi_{3q}~=~\theta_{colour} ~\chi_{spin}~\Phi_{flavour}~\psi_{space}.
\label{3q}
\end{equation}
In agreement with the Pauli principle, the wave function $\Psi_{3q}  $ must be totally antisymmetric for the exchange of any quark pair. Baryons must be colour singlets and the corresponding  wave function $\theta_{colour}$ is by itself antisymmetric, therefore the remaining factors must be completely symmetric. Actually a symmetric quark model has been formulated before the introduction of the colour quantum numbers and the symmetric three-quark states have been classified \cite{ko,hd}.

 Early Lattice QCD calculations \cite{wil} showed that the quark interaction can be split into  a long range part, which is spin and flavour independent and contains confinement, and a short range spin-dependent one \cite{deru}.  This means that one can assume the dominant part to be $SU(6)$ invariant and the wave function of Eq.~(\ref{3q}) becomes
 \begin{equation}
\Psi_{3q}~=~\theta_{colour} ~\Phi_{SU(6)}~\psi_{space}.
\label{3q_2}
\end{equation}
In order to satisfy Pauli principle, the product  
\begin{equation}
~\Phi_{SU(6)}~\psi_{space}
\end{equation}
must be symmetric and then both factors $\Phi_{SU(6)}$ and $\psi_{space}$ must have the same permutation symmetry, that is symmetric (S), antisymmetric (A) or one of the two mixed symmetry types (MS, MA), which are distinguished by the symmetry or the antisymmetry with respect  to a quark pair.

 It should be reminded that each quark belongs to the fundamental $SU(6)$ representation with dimension 6 and that with three quarks one
can obtain
the following $SU(6)$-representations:
\begin{equation}
SU(6):    6 \otimes 6  \otimes 6= 20  \oplus 70  \oplus 70 \oplus 56, 
\end{equation}
the corresponding symmetry type is, respectively, A,  M, M, S.

The spin and
flavour content of each $SU(6)$-representation is well defined, since
the three $SU(6)$ representations can be decomposed according to the
following scheme

\begin{equation}
20~=~^4 \underline{1}~+~^2 \underline{8},
\end{equation}
\begin{equation}
56~=~^2 \underline{8}~+ ~^4 \underline{10} , 
\end{equation}
\begin{equation}
70~=~^2 \underline{1}~+~^2 \underline{8}~+~^4 \underline{8}~+~^2
\underline{10} .
\\
\end{equation}
The suffixes in the r.h.s. denote the multiplicity $2S+1$ of the $3q$ spin
states
and the underlined numbers are the dimensions of the $SU(3)$
representations. This means for instance that the
$56$ representation contains a spin-$1/2$ $SU(3)$ octect and a spin-$3/2$
$SU(3)$ decuplet. 

\begin{figure}[h]

\includegraphics[width=4.5in]{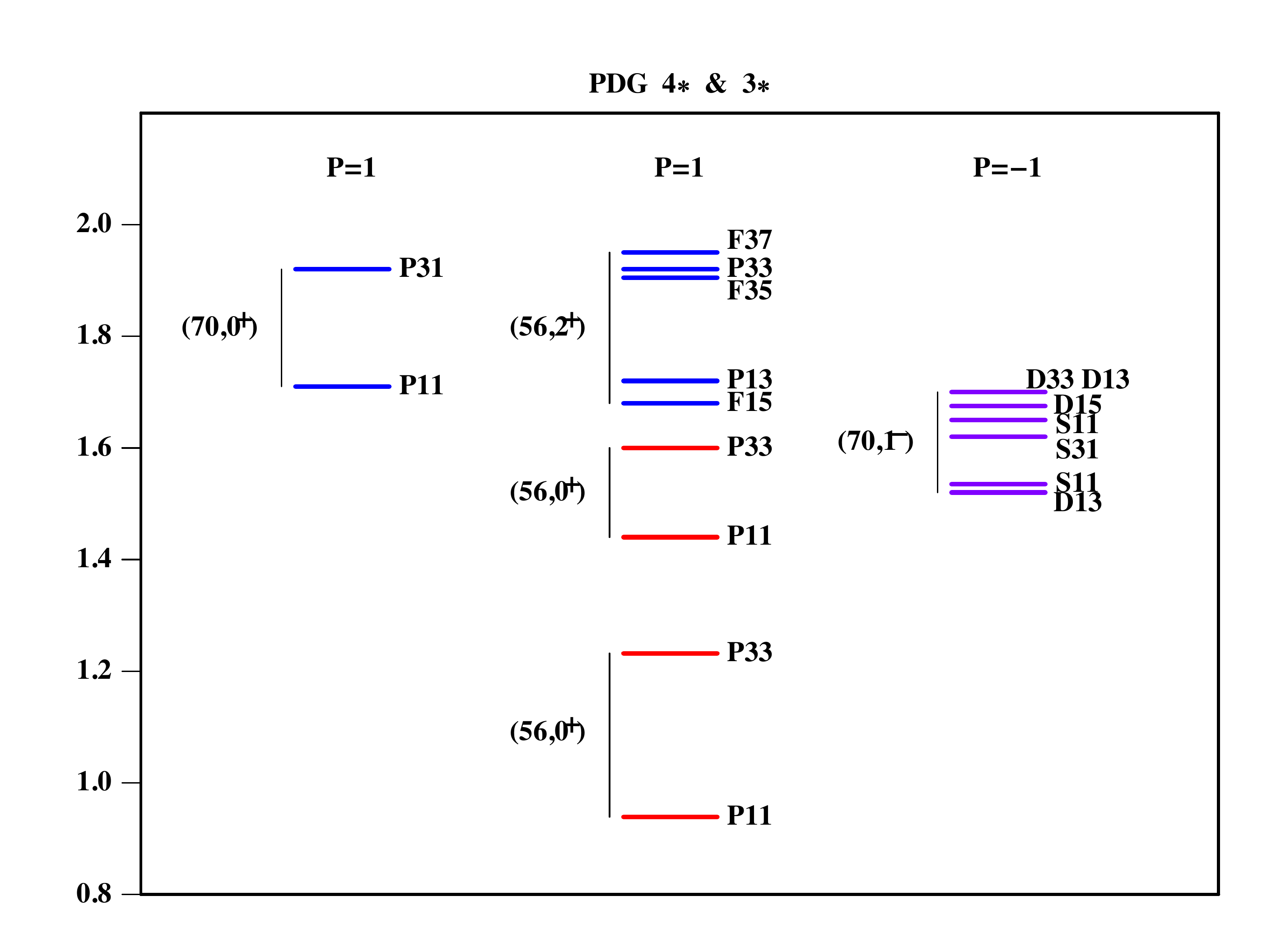}

\caption{ (Color online) The experimental spectrum of the non strange three- and four-star resonances \cite{pdg10}. The states are reported in columns with the same parity P and grouped into $SU(6)$-multiplets.}
\label{baryon}
\end{figure}

The various baryon resonances can be
grouped into
$SU(6)$-multiplets, the energy differences within each multiplet being at
most of the order of $15\%$ as in the case of $N-\Delta$ mass difference 
and of the splittings within the $SU(3)$ multiplets. In Fig.~\ref{baryon} we report the experimental non strange baryon spectrum, including only the three- and four- star states  \cite{pdg10}. The notation for the $SU(6)$-multiplets is $(d, L^P)$, where $d $ is the
dimension of the $SU(6)$-representation, $ L$ is the total orbital
angular momentum of the three-quark state describing the baryon and $P$ 
the corresponding parity.  An alternative but equivalent notation is $ L^P_t$, where t is the symmetry type of the $SU(6)$ representation.

The fact that the $4-$ and $3-$star  non strange resonances can be arranged in 
$SU(6)$ multiplets indicates that the quark dynamics has a
dominant $SU(6)$ invariant part accounting for the average multiplet
energies, while the splittings within the multiplets are obtained by means of
a $SU(6)$ violating interaction, which can be spin and/or isospin
dependent and can be treated as a perturbation.

The various constituent quark models are quite different, but they have a simple
general structure in common, since in any case, analogously to what stated above,  the quark interaction $V_{3q}$  can be split  
into a
spin-flavour independent
part $V_{inv}$, which is $SU(6)$-invariant and contains the confinement
interaction, and a $SU(6)$-dependent part $V_{sf}$, which contains spin
and eventually flavour dependent interactions
\begin{equation} \label{v3q}
V_{3q}~=~V_{inv}~+~V_{sf}.
\end{equation}

\begin{table}
\caption[]{Illustration of the features of various CQMs}
\vspace{15pt}
\begin{tabular}{|c|c|c|c|c|}
\hline
$CQM$ & Kin. Energy & $V_{inv}$ & $V_{sf}$ &   ref. \\
\hline
 Isgur-Karl  & nonrel. & h.o. + shift & OGE &   \cite{ik}\\
 \hline
Capstick-Isgur & rel. & string +coul-like & OGE &  \cite{ci} \\
\hline
$U(7)  B.I.L.$ & $M^2$ & vibr + L &  G\"{u}rsey-Rad &  \cite{bil}\\
\hline
Hypercentral & nonrel./rel. & $O(6)$: lin + hyp.coul & OGE &  \cite{pl} \\
\hline
Glozman-Riska & rel. & h.o. / linear & GBE &  \cite{olof} \\
\hline
Bonn & rel. & linear + $3$ body & instanton & \cite{bn1} \\
\hline
quark-diquark& nonrel./rel. & linear + Coulomb & spin-isospin & \cite{diq} \\
\hline
\end{tabular}
\label{cqm}
\end{table}

In Table \ref{cqm} we report a list of the Constituent Quark Models and their  main features. The order is chronological.

\subsection{The Isgur-Karl model}

The Isgur-Karl \cite{ik} model has some general features that are interesting also for other models, so it is worthwhile to devote to it some attention (more details can be found in \cite{mg}).

The kinetic energy T is assumed to be nonrelativistic
\begin{equation}
T~=~\sum_i (m_i + \frac{\vec{p}_i^{~2}}{2m_i})~=~M_{tot} + \frac{\vec{P}^2}{2 M_{tot}} +T_{intr},
\end{equation}
where $M_{tot}$ is the total mass of the three quarks, $\vec{P}$ is their total momentum and the intrinsic kinetic energy is expressed in terms of the momenta $\vec{p}_\lambda$ and $\vec{p}_\rho$, which are conjugated to the Jacobi coordinates $\vec{\rho}$ and $\vec{\lambda}$,
\begin{equation}
\vec{\rho}~=~ \frac{1}{\sqrt{2}}(\vec{r}_1 - \vec{r}_2) ~,\\
~~~~\vec{\lambda}~=~\frac{1}{\sqrt{6}}(\vec{r}_1 + \vec{r}_2 - 2\vec{r}_3) ~. 
\label{coord}
\end{equation}
In the case of non strange baryons, all quark have the same mass m and then T assumes the form
\begin{equation}
\label{kin}
T~=~3m + \frac{\vec{P}^2}{6m} + \frac{\vec{p}_\rho^{~2}}{2m} + \frac{\vec{p}_\lambda^{~2}}{2m};
\end{equation}
in the case of quarks with different mass, the kinetic energy contains appropriate reduced masses \cite{ik}.

The confining interaction is assumed to be a harmonic oscillator (h.o.)
\begin{equation} 
V_{ho}~=~\sum_{i<j} \frac{1}{2}~K~ (\vec{r}_i - \vec{r}_j)^2,
\end{equation}
which, in terms of the Jacobi coordinates, becomes
\begin{equation} 
V_{ho}~=~\frac{3}{~2}~K~ (\rho^{2} + \lambda^{2}),
\end{equation}
with $\rho^2=\vec{\rho}^{~2}$ and $\lambda^2=\vec{\lambda}^2$. The three-quark interaction is then given by two three-dimensional h.o.\ and the energy levels can be written as $E= (3 +N) \hbar \omega$, with $N=2n+l_\rho +l_\lambda$, where n is a non negative integer number and $l_\rho$ and $l_\lambda$ are the orbital angular momenta associated to the Jacobi coordinates. The h.o.\ parameter $\omega$ is given by $\sqrt{\frac{3K}{m}}$.

\begin{table}[t]
\caption[]{The non strange three-quark states belonging to  the h.o.\ shells up to N=2. The notation for the baryon resonances is $X_{2I 2J}$, where X=S,P,D,F,\ldots denotes the pion wave in the decay channel, I and J are the isospin and the spin of the state, respectively. The numbers within parentheses are the masses of the 4- and 3- star resonances displayed in Fig.~(\ref{baryon}). The asterisk in the first N=2 configuration reminds that the spin-isospin structure is the same as in the ground state but it has a  radial excitation. }
\label{ho}
\vspace{15pt}
\begin{tabular}{|c|c|c|c|c|c|}
\hline
N & $(d,L^P)$ & $^2\underline{8}$ & $^48$ & $^210$ & $^410$ \\
\hline
\hline
 0 & $(56,0^+)$ & $P_{11}(939)$ &  &  &  $P_{33}(1232)$ \\
\hline
\hline
 1 & $(70,1^-)$ & $S_{11}(1535)$ & $S_{11}(1650) $ &$S_{31}(1620)$  &  \\
  &  & $D_{13}(1520)$ & $D_{13}(1700) $ &$D_{33}(1700)$  &  \\
   &  &  & $D_{15}(1675) $ &  &  \\
\hline
\hline
 2 & $(56,0^{+*})$ & $P_{11}(1440)$ &  &  &  $P_{33}(1600)$ \\
\hline
 2 & $(70,0^+)$ & $P_{11}(1710)$ & $P_{13}$ & $P_{31}$ &   \\
\hline
 2 & $(56,2^+)$ & $P_{13}(1720)$ &  &  & $P_{31}(1910)$  \\
  &  & $F_{15}(1680)$ &  &  & $P_{33}(1920)$  \\
  &  &  &  &  & $F_{35}(1905)$  \\
  &  &  &  &  & $F_{37}(1950)$  \\
\hline
 2 & $(70,2^+)$ & $P_{13}$ & $P_{11}$ & $P_{33}$ &   \\
  &  & $F_{15}$ & $P_{13}$ & $F_{35}$ &   \\
  &  &  & $F_{15}$ & &   \\
  &  &  & $F_{17}$ & &   \\
\hline
\hline
 2 & $(20,1^+)$ & $P_{11}$ &  & &   \\
 & & $P_{13}$ &  & &   \\
\hline

\end{tabular}
\end{table}

The general structure of the h.o.\ space wavefunction is 
\begin{equation} 
\psi_{NLt}(\vec{\rho},\vec{\lambda})=C_N P_N(\rho, \lambda) e^{-1/2 \alpha^2  (\rho^2+\lambda^2)} ~Y_{l_\rho}(\Omega_\rho) ~Y_{l_\lambda}(\Omega_\lambda),
\end{equation}
where  $\alpha^2 =\sqrt{3K m} / \hbar$, $C_N$ is a normalization factor, $P_N$ a polynomial of degree N and the spherical harmonics 
have to be combined to a definite total orbital angular momentum L; t (=A, M, S) is 
the symmetry type, the same as the $SU(6)$ states.

In Table \ref{ho} we report the $SU(6)$ states that can be assigned to  the first three shells. All the states reported in Fig.~(\ref{baryon}) fit very well into the scheme. The number of predicted states is however much larger than the observed 4- and 3- star states. The problem of such missing resonances is common to all CQMs and it has been suggested long time ago that some resonances may be observable in electroproduction experiments and not in strong interaction processes \cite{ki}. This statement is supported by the new states reported in the last edition of the PDG review \cite{pdg12}. 

In Table \ref{ho} there is no room for the 3-star $D_{35}(1930)$. In fact, the total spin 5/2 can be obtained combining the L=1 total orbital angular momentum with the total spin 3/2 of the three quark, however the negative parity states belong to the 70-dimensional representation of $SU(6)$, which cannot contain a $\Delta$ state with total spin 3/2. In order to describe this state and  some new 2-star negative parity resonances \cite{pdg12} one should introduce the N=3 shell and the number of missing resonances will be highly increased. Similarly, the shells with N greater than 2 are necessary for the resonances with high spin values,

An important observation regarding the h.o.\ spectrum is the level ordering, which, for any wo body
potential is $0^+, 1^-, 0^+$, while experimentally the $1^-$ states are in average almost degenerate with the first $0^+$ excitation. Moreover, the spacing between two shells is the same over the whole spectrum and the levels are highly degenerate, since the energy depends on the h.o.\ quantum number N only.

In order to avoid the equal spacings and the degeneracy of the levels, in the Isgur-Karl model a shift potential U is added, which simply redefines the energies of the $SU(6)$ states, without any attempt to diagonalize it. In this way the energies of the $SU(6)$ configurations can be written as
\begin{eqnarray} \label{ik_en}
\begin{array} {ccl}
E(0^+_S) & = & E_0 ,\\   
E(0^{+*}_S) & = & E_0+ 2\Omega -\Delta, \\
E(0^-_M) & = & E_0+ 2\Omega -\Delta/2, \\
E(2^+_S) & = & E_0+ 2\Omega -2 \Delta /5, \\
E(2^+_M) & = & E_0+ 2\Omega -\Delta/5 \\
E(1^+_A) & = & E_0+ 2\Omega,
\end{array}
\end{eqnarray}
where 
\begin{equation}
\Omega = \hbar \omega -a_0/2+a_2/3, ~~~~~~~~~~\Delta=- 5/4 a_0+5/3 a_2 -1/3 a_4,
\end{equation}
where the coefficients $a_m$ (m=0,2,4) are determined by the moments of the U potential
\begin{equation}
a_m=3 (\frac{\alpha}{\sqrt{\pi}})^3 \int d^3\rho (\alpha \rho)^m U(\sqrt{2} \rho) e^{-\alpha^2 \rho^2}.
\end{equation}
No explicit form is assumed for the potential U,  but the three coefficients $a_m$ are used as free parameters to be fitted to the experimental spectrum. In this way also the position of the Roper N(1440) resonance is correctly described.

Having assumed the space wave function as given by a h.o.\ three-quark potential, one can build the various $SU(6)$ configurations to be identified, according to Table \ref{ho}, with the observed resonances. The states contained in each multiplet can be denoted as
\begin{equation}
|B~ ^{2S+1}X_J \rangle_t,
\end{equation}
where $B=N, \Delta$ for isospin $1/2,3/2$, respectively, S in the suffix is the total 3q spin, J the spin of the baryon state, X=S,P,D,\ldots according to the total 3q orbital angular momentum and t is the symmetry type. For instance the nucleon is denoted by $|N~ ^{2}S_{1/2}\rangle_S$, the Roper resonance is $|N~ ^{2}S^*_{1/2}\rangle_S$, where the asterisk means that the state is the first radial excitation of the nucleon, the $\Delta$ is $|\Delta~ ^{4}S_{3/2}\rangle _S$ and so on.

Since the quark interaction considered up to now is $SU(6)$ invariant, the energies given by Eqs.~(\ref{ik_en}) are common to all the states in any $SU(6)$ multiplet, at variance with the experimental spectrum (see Fig.~\ref{baryon}). In order to describe the splittings within each multiplet, one has to introduce a $SU(6)$ violating interaction $V_{sf}$, which, in the case of the Isgur-Karl model  \cite{ik} is given by the  hyperfine interaction, in the form proposed in ref.~\cite{deru}
\begin{equation}
H_{hyp} 
= \sum_{i<j} \frac{2 \alpha_S}{3m_i m_j} [\frac{8 \pi}{3}  \vec{S}_i \cdot  \vec{S}_j ~ \delta(\vec{r}_{ij}) +\frac{1}{r_{ij}^3} (\frac{3(\vec{S}_i \cdot \vec{r}_{ij})(\vec{S}_j \cdot \vec{r}_{ij})}{r_{ij}^2} -\vec{S}_i \cdot \vec{S}_j] ,
\label{hyp}
\end{equation}
where $\vec{r}_{ij}=\vec{r}_{i}-\vec{r}_{j}$. Eq.~(\ref{hyp}) is the spin dependent part of the One Gluon Exchange (OGE) interaction between two quarks, the spin independent part being a Coulomb-like term $1/r_{ij}$, which can be considered implicitly taken into account in the shift potential U. The structure of Eq.~(\ref{hyp}) is the same as the Breit-Fermi term in the higher order Coulomb potential for electrons in atoms. The  OGE interaction is in principle valid for short interquark distances, however, it is used just for the determination of the form of the spin-dependent quark interaction and the strong coupling constant $\alpha_S$ is considered as a free parameter to be fitted to the  $N - \Delta$ mass difference. 

The hyperfine interaction is diagonalized in the h.o. \ basis, using as unperturbed  energies the ones given by Eqs.~(\ref{ik_en}). Its matrix elements, in the case of u and d quarks, are given in terms of the quantity (see also Appendix 2 of Ref.~\cite{mg})
\begin{equation}
\delta =\frac{4 \alpha_S \alpha^3}{3 \sqrt{2 \pi} m^2} ,
\end{equation}
which is substantially the $N - \Delta$ mass difference and can be fixed to about 300 MeV.
As for the remaining free parameters, the h.o.\ constant is fitted to the proton r.m.s.\ radius, obtaining $\alpha^2 = 1.23$ fm$^2$, the other parameters are determined by comparison of the theoretical spectrum with the experimental one. The resulting description of the spectrum is quite good, both for non strange and strange resonances \cite{ik}.

An important consequence of the introduction of the hyperfine interaction is that the baryon states are superpositions of $SU(6)$ configurations. For instance, the nucleon is expanded as
\begin{equation}
|N \rangle  = a_S |N ^2S_{1/2} \rangle _S + a'_S |N ^2S^*_{1/2}\rangle_S + a_M |N ^2S_{1/2}\rangle_M + a_D |N ^4D_{1/2}\rangle_M, 
\label{N}
\end{equation}
with $ a_S=0.931, a'_S=-0.274, a_M=-0.233, a_D=-0.067$ \cite{mg}; the asterisk in the second term of Eq.~(\ref{N}) means that the spin-isospin part is the same as the first term, but the space part corresponds to a radially excited wave function. The Roper resonance has a similar expansion, with the dominant component given by $|N ^2S'_{1/2} \rangle_S$: $ a_s=0.281, a'_s=-0.960, a_M=--0.003, a_D=-0.001$. The $\Delta$ resonance is given by
\begin{equation}
|\Delta \rangle = b_S |\Delta  ^4S_{3/2} \rangle_S + b'_S |\Delta ^4S^*_{3/2}\rangle_S + b_D |\Delta ^4S_{3/2}\rangle_S + b'_D |\Delta ^2D_{3/2} \rangle _M ,
\end{equation}
with $ b_S=0.963, b'_S=0.231, b_D=-0.119, b'_D=0.075$. It is well known that with pure $SU(6)$ configurations the E2 electromagnetic $N - \Delta$ transition vanishes \cite{bm}. However, because of the hyperfine interaction, the $\Delta$ state acquires a non-zero D-wave component  and then a small quadrupole strength arises \cite{dg,ikk}. The theoretical estimate of the ratio 
\begin{equation}
R=- \frac{G_{E2}}{G_{M_1}}
\label{E2}
\end{equation}
is about $-0.02$ \cite{ikk,dg}, which compares favourably with the experimental value \cite{pdg12}.

The behaviour of the nucleon electromagnetic form factors in the Isgur-Karl model is dominated by the Gauss factor $e^{-\frac{q^2}{6 \alpha^2}}$, therefore it is too strongly damped for medium-high values of the square momentum of the virtual photon $q^2$. On the contrary, the neutron charge form factor is nicely described  and this is due to the hyperfine interaction \cite{iks}. In fact, if the nucleon state is the symmetric $SU(6)$ configuration $|N ^2S_{1/2} \rangle _S$, the charge form factor is proportional to the total charge of the three-quark system; the hyperfine interaction introduces in the nucleon state a mixed symmetry component $|N ^2S_{1/2}  \rangle _M$, giving rise to a non zero charge form factor for the neutron.

\subsection{The Capstick-Isgur model}

This model \cite{ci} is the extension to the baryon sector of the relativized model for mesons formulated in ref.~\cite{gi}. The three-quark hamiltonian is written as
\begin{equation}
H= T + V_{3q},
\end{equation}
where T is the relativistic kinetic energy
\begin{equation}
T=\sum_{i=1}^3 \sqrt{p_i^2 + m_i^2},
\label{relkin}
\end{equation}
 the three quark potential $V_{3q}$ is separated into two terms, according to Eq.~(\ref{v3q}). In the nonrelativistic limit
\begin{equation}
V_{3q} \rightarrow V_{si} + V_{sd},
\end{equation}
where the spin dependent interaction is 
\begin{equation}
V_{si} = V_{string} + V_{coul}.
\end{equation}
The first term is the three-body adiabatic potential generated by the quantum ground state in
 a $Y-$shaped string configuration and provides confinement;  $V_{string}$ is given by \cite{ci}
\begin{equation}
V_{string} = C_{qqq} + b \sum_{i=1}^3 |\vec{r}_i - \vec{r}_{junction}|,
\end{equation}
where $C_{qqq}$ is an overall constant energy shift and b is the string tension. For practical purposes, $V_{string}$ is split into two-and three-body effective terms
\begin{equation}
V_{string} = C_{qqq} + f b \sum_{i<j}r_{ij} + V_{3b},
\end{equation}
where
\begin{equation}
V_{3b}= b (\sum_{i=1}^2 |\vec{r}_i - \vec{r}_{junction}| -f \sum_{i<j}r_{ij}).
\end{equation} 
The parameter f is chosen to be 0.5493 \cite{dosch} in order to minimize the expectation value of  $V_{3b}$ in the h.o.\ ground state of the baryon. In this way $V_{3b}$ is a small correction and can be treated perturbatively and $V_{string}$ becomes very close to $1/2 b \sum_{i<j}r_{ij}$. 

The potential $V_{coul}$, in the nonrelativistic limit, is given by
\begin{equation}
V_{coul}=  \sum_{i<j} -\frac{2 \alpha_S(r_{ij})}{3 r_{ij}};
\end{equation} 
in ref.~\cite{ci} the momentum (or space) dependence of the strong coupling constant $\alpha_S(r_{ij})$ is properly taken into account.
 
The spin dependent potential $V_{sd}$, again in the nonrelativistic limit, is
\begin{equation}
V_{sd}=  V_{hyp}+ V_{so},
\end{equation} 
where $V_{hyp}$ is the hyperfine interaction of Eq.~(\ref{hyp}) and $V_{so}$ is a spin-orbit interaction containing in particular a Thomas precession term. Please note that the sum of $V_{coul}$ and $V_{hyp}$, together with the corresponding Thomas precession spin-orbit, derive from the nonrelativistic limit of the OGE interaction.

In order to avoid the nonrelativistic approximation one has, according to the discussion reported in ref.~\cite{ci}, to introduce appropriate $\sqrt{\frac{E}{m}}$ factors and to smear  the interactions over a two quark distribution
\begin{equation}
\rho_{ij}(\vec{r}_i-\vec{r}_j)=  \frac{\sigma_{ij}^3}{\pi^{3/2}} e^{-\sigma_{ij}^2 (\vec{r}_i-\vec{r}_j)^2},
\end{equation}  
in particular for the  contact term in the hyperfine interaction.
In this way the factors $m_i$ are substituted with the corresponding energies and the momentum dependence of the interaction is taken into account.
 
The three-body equation, with the relativistic kinetic energy, is solved by means
of a variational approach in a large h.o.\ basis. The result is a good description of the baryon spectrum, including strange, charm and bottom resonances \cite{ci}.

As already mentioned in the introduction, the model by Capstick-Isgur has been successfully applied to the calculation of the electromagnetic amplitudes for the transitions to the $\Delta$ \cite{card_ND} and  Roper resonances \cite{card_Roper}.

\subsection{The U(7) model}

The typical feature of this model \cite{bil} is to describe the state of a three quark system by means of a group theoretic approach. 

In order to describe the space degrees of freedom, the model uses the method of bosonic quantization, similarly to  what has been done in the Interacting Boson model in nuclear  physics \cite{AI} and in molecular physics \cite{iac} as well. The idea is to consider a string-like model with a Y-shaped configuration, in which the vectors $\vec{r}_i (i=1,2,3)$  denote the end points of the string configuration.  To this end, one introduces two vector boson operators defined in terms of the Jacobi coordinates of Eq.~(\ref{coord}) $\vec{\rho}$, $\vec{\lambda}$, together with  their conjugate momenta $\vec{p_\rho}$, $\vec{p_\lambda}$ 
\begin{equation}
b_{\rho, m} = \frac{1}{\sqrt{2}} (\rho_m + i p_{\rho, m}), ~~~~~~~~~~b^\dagger_{\rho, m} = \frac{1}{\sqrt{2}} (\rho_m - i p_{\rho, m}),
\end{equation} 
\begin{equation}
b_{\lambda, m} = \frac{1}{\sqrt{2}} (\lambda_m + i p_{\lambda, m}), ~~~~~~~~~~b^\dagger_{\lambda, m} = \frac{1}{\sqrt{2}} (\lambda_m - i p_{\lambda, m})
\end{equation} 
(with $m=-1,0,+1$) and an auxiliary scalar boson $s$, $s^\dagger$. The bilinear forms $G_{\alpha \alpha'} =c^\dagger_\alpha c_{\alpha'}$, where $c^\dagger_\alpha$ ($\alpha=1, \ldots 7$) is one of the seven creation operators, generate the Lie algebra of U(7). The  choice of U(7) is in agreement with the usual prescription that any problem with $\nu$ space degrees of freedom should be written in terms of the Lie algebra $U(\nu+1)$ \cite{iac93} and all the states are assigned to the totally symmetric representation $[N]$ of $U(\nu+1)$, N being the maximum number of shells. The physical states can be constructed by applying a suitable product of boson operators to the vacuum state
\begin{equation}
\frac{1}{\mathcal{N}} (b^\dagger_\rho)^{n_\rho} (b^\dagger_\lambda)^{n_\lambda} (s^\dagger)^{N-n_\rho-n_\lambda},
\end{equation} 
$\mathcal{N}$ is  a normalization factor. 

The squared mass operator $M^2$ is expressed as the most general combination of $G_{\alpha \alpha'}$, with the condition of being at most  quadratic, preserving angular momentum and parity and transforming as a scalar under the permutation group. The general form of $M^2$ contains several models of baryons structure, including single particle (i.e.\ h.o.) and collective string models. The calculations are performed choosing the latter model,with the consequence that $M^2$ contains also a term of the type $b^\dagger b^\dagger s s + h.c.$, which causes a spread of the wave function over many h.o.\ shells. In order to make more transparent the interpretation of the results, the mass operator is rewritten in terms of vibrational and rotational contributions to the baryon spectrum, using  a procedure already introduced for the Interacting Boson Model \cite{kl}
\begin{equation}
M^2 = M_0^2 + M_{vib}^2 +  M_{rot}^2 + M_{vib-rot}^2;
\end{equation}
in this way the baryon excitation spectrum is determined by vibrations and rotations of the string-like configuration. There seems to be no evidence of excitations due to the term $M_{vib-rot}^2$, therefore it is omitted and the remaining two terms, according to the discussion reported in \cite{bil}, are simplified obtaining the mass formula
\begin{equation}
M^2 = M_0^2 + N[ \kappa_1 n_u + \kappa_2 (n_v + n_w)] + \alpha L,
\label{M_{inv}}
\end{equation}
where $\kappa_1, \kappa_2, \alpha$ are free parameters, L is the total orbital angular momentum and $n_u, n_v, n_w$ are the eigenvalues of  number operators of the type $c^\dagger c$, labeling the vibrational energy levels. 

Eq.~(\ref{M_{inv}}) describes the $SU(6)$ invariant part of the interaction. The splittings within the multiples are introduced with reference to the internal part of the state (see Eq.~(\ref{3q})). The corresponding algebraic 
structure  is
\begin{equation}
G_i = SU_c(3)\otimes SU_s(2) \otimes SU_f(3),
\end{equation} 
describing the colour, spin and flavour degrees of freedom, respectively. Baryons are colour singlets  and then only the spin-flavour degrees of freedom contribute to the energy splittings. The mass squared operator $M^2_{sf}$ is written in a G\"{u}rsey-Radicati form \cite{rad}
\begin{eqnarray} \label{rad}
\begin{array} {ccl}
M^2_{sf}& = &a [C_2(SU(6)_{sf})-45 ] + b [C_2(SU(3)_f)-9 ] +  \\   
 &  & + b' [C_2(SU_I(2))-\frac{3}{4} ] + b'' [C_1(U_Y(1))-1 ]  + \\
  &   & + b''' [C_2(U_Y(1))-1] + c [C_2(SU_S(2))-\frac{3}{4} ].
\end{array}
\end{eqnarray}
The quantities denoted as $C_n(X), n=1,2$ are the Casimir operators of the Group X; the constants in Eq.~(\ref{rad}) are chosen in order that each term vanishes in the nucleon ground state. For non strange baryons the hypercharge Y is equal to 1 and the b and b' terms can be grouped into a single term, therefore one can use the simplified form
\begin{equation} \label{simpl}
M^2_{sf} = a [C_2(SU(6)_{sf})-45 ] + b [C_2(SU(3)_f)-9 ] +    
   c [C_2(SU_S(2))-\frac{3}{4} ].
\end{equation}

The seven parameters in Eqs.~(\ref{M_{inv}}) and (\ref{simpl}) are obtained fitting the non strange baryon spectrum and the results are very good. In the model the number of shell is not limited, therefore one can describe well also the four star resonances with higher values of the spin, such as $G19, H19, G19, I 1 11$ and $H3 11$. The extension of the model to the strange baryons is presented in ref.~\cite{bil_s}.

The model allows to calculate also the elastic nucleon form factors \cite{bil2} and the electromagnetic transition amplitudes for photo-and electroproduction \cite{bil,bil2}, provided that a form for the charge distribution along the string is assumed. In this way both the elastic and inelastic form factors are adequately described.

\subsection{The Goldstone Boson Exchange Model }

 The model is based on the consideration that QCD exhibits an approximate chiral symmetry which is spontaneously broken \cite{olof}. As a consequence of such spontaneous symmetry breaking, quarks acquire an effective mass and Goldstone bosons emerge, which are indentified with the pseudoscalar meson octet. Therefore it is assumed that baryons are considered as a system of three constituent quarks with an effective quark-quark interaction $V_{3q}$, which is split into parts according to Eq.~(\ref{v3q}). While different forms are assumed for $V_{inv}$ in the various versions of the model, the spin-flavour part is always chosen as an exchange of pseudoscalar (Goldstone) bosons between two quarks. The simplest form of this chiral interaction can be written as \cite{olof}
\begin{equation} \label{chir}
H_{\chi} \sim \sum_{i<j} V(\vec{r}_{ij}) \vec{\lambda}^F_i \cdot \vec{\lambda}^F_j  \vec{\sigma}_i \cdot \vec{\sigma}_j ,
\end{equation}
where $\vec{\lambda}^F_i $ are the flavour Gell-Man matrices and $\vec{\sigma}_i $ the quark spin operators. The  interaction has a Yukawa form containing a spin-spin and a tensor part. The spin-spin part is given by
\begin{equation} \label{yuk}
V_P(\vec{r}_{ij}) = \frac{g^2}{4 \pi}  \frac{1}{3} \frac{1}{4 m_i m_j} \vec{\sigma}_i \cdot \vec{\sigma}_j \vec{\lambda}^F_i \cdot \vec{\lambda}^F_j [\mu^2 \frac{e^{-\mu r_{ij}}}{r_{ij}} - 4 \pi \delta(\vec{r}_{ij})],
\end{equation}
where $m_i$ are the quark masses  and P labels  the exchanged boson of mass  $\mu$; the $\delta$ is actually smeared out by the finite size of quarks and mesons. The flavour structure of the quark-quark interaction is then
\begin{equation}
V_{octet}(r_{ij})  = \sum_{a=1}^3 V_\pi (r_{ij}) \vec{\lambda}^a_i \cdot \vec{\lambda}^a_j  + \sum_{b=4}^7 V_K (r_{ij}) \vec{\lambda}^b_i \cdot \vec{\lambda}^b_j  + V_\eta (r_{ij}) \vec{\lambda}^8_i \cdot \vec{\lambda}^8_j  .
\label{oct}
\end{equation}

In ref.~\cite{olof}  $V_{inv}$ is assumed to be  a h.o.\ potential and the boson exchange interaction is considered in the chiral limit, in which case the masses of all the three quarks are equal and $ V_\pi =V_K=V_\eta$. Treating the interaction as a perturbation, the masses of the baryons can be expressed in terms of a limited number of radial integrals, which are used as parameters in order to fit the experimental values. Already in this simplified approach, a reasonable description of the spectrum is obtained, in particular the boson exchange interaction leads to the correct ordering between the excited $0^+$ and the first negative parity levels. As  a further improvement also deviations from the chiral limit and the contribution of the tensor part of the boson exchange interaction are considered.

A substantial improvement of the model has been started in ref.~\cite{ple1}, in the sense that the confinement interaction is assumed to be linear and $V_{3q}$  is inserted into a Faddeev equation for the three quark system to be solved numerically.  The $V_{sf}$ is given by Eq.~(\ref{oct}), to which the exchange of the singlet meson ($\eta '$) is added. In the first application, only nucleon and $\Delta$ states are considered \cite{ple1,ple2}. The interquark potential is then:
\begin{equation}
V(r_{ij})  = V_{octet}(r_{ij}) +V_{singlet}(r_{ij}) +C r_{ij} ;
\label{gbe}
\end{equation}
in $V_{octet}$ only the $\pi$ and $\eta$ potentials contribute and the singlet potential is
\begin{equation}
V_{singlet}(r_{ij}) = \frac{2}{3} \vec{\sigma}_i \cdot \vec{\sigma}_j V_{\eta '} (r_{ij}).
\end{equation}
The quark-eta coupling constant is assumed equal to the quark-pion one, deduced from the pion-nucleon interaction. Keeping  the meson masses equal to their physical values, there remain only four  free parameters, namely the two parameters which determine the smearing of the $\delta$ function in Eq.~(\ref{yuk}), the $\eta '$ coupling constant and the strength of the linear confinement C. The result is a good description of the 14 lowest N and $\Delta$ states, respecting the ordering displayed by the experimental spectra.
 
A unified description of both non strange and strange baryons is finally achieved in \cite{ple3}, where the interaction of Eq.~(\ref{gbe}) is used together with a relativistic kinetic energy as in Eq.~(\ref{relkin}). The three-quark wave equation is solved by means of a variational approach. Again a quite satisfactory description of the low-lying light and strange baryons is achieved, respecting in particular the already mentioned relative ordering of the positive and negative parity states.

\subsection{The Bonn model}

The authors start from the consideration that   the nonrelativistic approach seems to be completely inadequate  for the description of the internal motion of quarks with small constituent  masses and therefore they introduce  a relativistic formulation \cite{bn1}.

The relativistic formulation is performed within quantum field theory and is based on the  six-point Green's function $G_x$ describing three interacting quarks. The infinite series of Feynman diagrams, necessary in order to describe a bound state,
is rearranged in the same way used by Bethe-Salpeter for the two-particle case. The result is that $G_x$ obeys to  an integral equation containing two irreducible kernels $K^{(2)}$ and $K^{(3)}$, describing the two- and three-particle interactions, respectively. Introducing the momentum space representation  $G_P$ of $G_x$, the integral equation can be written in concise form as
\begin{equation}
G_P= G_{0P} - i  G_{0P} K_P G_P,
\end{equation}
where $G_{0P}$ is the three quark propagator and $K_P$ the total kernel, or  equivalently as
\begin{equation}
(G_{0P}^{-1} + i K_P) G_P = I,
\label{G6}
\end{equation}
showing that $G_P$ is the resolvent of the pseudohamiltonian $H_P$
\begin{equation}
H_P = (G_{0P}^{-1} + i K_P).
\label{pH}
\end{equation}

The idea is to extract from  $G_P$ the baryon contributions, meant as real bound states of three quarks with positive energy $\sqrt{\vec{P}^2+M^2}$. To this end, $G_P$ is expanded in a Laurent series, which, near a pole, gives
\begin{equation}
G_P(p_\rho,p_\lambda;p'_\rho,p'_\lambda) = -i \frac{\chi_{\overline{P}}(p_\rho,p_\lambda) \overline{\chi}_{\overline{P}}(p_\rho,p_\lambda)}{P^2-M^2+i\epsilon}  .
\label{laur}
\end{equation}
The quantity $\chi_{\overline{P}}(p_\rho,p_\lambda)$ is the Fourier transform of the Bethe-Salpeter (BS) amplitude $\chi_{\overline{P}}(x_1,x_2,x_3)$ for the bound state $|\overline{P}  \rangle $, defined as transition amplitudes between the state $|\overline{P} \rangle $ and the vacuum $ |0 \rangle $
\begin{equation}
\chi_{\overline{P}}(x_1,x_2,x_3) = \langle 0|T(\Psi_{a_1}(x_1) \Psi_{a_2}(x_2) \Psi_{a_3}(x_3) ) |\overline{P} \rangle ,
\end{equation}
where $\Psi_{a_i}(x_i)$ is the quark field and $a_i$ denotes the  Dirac, flavour and colour indices. Thanks to translational invariance, one gets a BS amplitude $\chi_{\overline{P}}(\rho,\lambda)$ which depends on relative coordinates only and, in momentum space, on the conjugate momenta $p_\rho,p_\lambda$; here $\rho$ 
and $\lambda$ are tetravectors, whose spatial part coincide practically with the Jacobi coordinates defined in Eq.~(\ref{coord}).

The factorization property of the pole residue allows then to extract the BS amplitude $\chi_{\overline{P}}(p_\rho,p_\lambda)$, which satisfies the equation
\begin{equation}
\chi_{\overline{P}} = -i G_{0\overline{P}}~ K_ {\overline{P}}~ \chi_{\overline{P}} ,
\label{BS}
\end{equation}
where $K_{\overline{P}}$ has, as mentioned before, two- and three-body contributions $K^{(2)}$ and  $K^{(3)}$, respectively.

In principle the BS equation (\ref{BS}) allows a covariant description of baryons as bound states of three quarks in the framework of QCD. However it cannot be used practically because the single quark propagators and the kernels $K^{(2)}$,  $K^{(3)}$ are only formally defined in perturbation theory as an infinite sum of Feynman diagrams and are not deducible from QCD. Moreover, the dependence on the relative energy (or relative time) leads to a complicate analytical structure. Therefore an adequate parametrization is necessary, to be introduced after having having obtained a six-dimensional reduction of the full eight dimensional BS equation, the so called Salpeter equation, trying to preserve the covariance of the theory and  keep it as close as possible to the quite successful nonrelativistic quark model. 

This can be achieved following the lines of what has already 
performed in the covariant quark model for meson case using an instantaneous $q\overline{q}$ BS equation \cite{bn-sal}. To this end, the quark propagators are assumed to be given by their free forms with effective constituent masses. Moreover the kernels $K^{(2)}$ and  $K^{(3)}$ are approximated by effective interactions which are instantaneous in the baryon rest frame, that is
\begin{equation}
K^{(3)}_P(p_\rho,p_\lambda; p'_\rho,p'_\lambda) ~ = ~V^{(3)}_P(\vec{p_\rho},\vec{p_\lambda}; \vec{p'_\rho},\vec{p'_\lambda}),
\label{k3}
\end{equation}
\begin{equation}
K^{(2)}_{\frac{2}{3}P+p_\lambda}p_\rho, p'_\rho) ~ = ~V^{(2)}_P(\vec{p_\rho},;\vec{p'_\rho}),
\label{k2}
\end{equation}
where $P=(M,0)$. This instantaneous approximation can be formulated in a covariant way following the method proposed in ref.~\cite{wm}. 
The reduction to the six dimensional Salpeter equation can be performed more easily if only the three-body kernel is present; the introduction of the two body kernel is possible provided that it is substituted by a suitable effective interaction $V_{eff}$. In any case the reduction to the Salpeter amplitude $\Phi_M$ is achieved by means of an integration over the energy variables, a procedure which, thanks to Eqs.~(\ref{k3}) and (\ref{k2}) affects the BS amplitude only
\begin{equation}
\Phi_M(\vec{p}_\rho,\vec{p}_\lambda)=\int \frac{dp_\rho^0}{2 \pi}\frac{dp_\lambda^0}{2 \pi} \chi_M(p_\rho^0,p_\lambda^0,\vec{p}_\rho,\vec{p}_\lambda),
\end{equation}
where $\chi_M$ is the BS amplitude in the rest frame.

Finally the Salpeter equation is written in a Hamiltonian formulation
\begin{equation}
\mathcal{H}_M \Phi_M^\Lambda= M \Phi_M^\Lambda,
\end{equation}
where $\Lambda$ is a projector operator over positive energy states.

In order to perform explicit calculations of the baryon spectrum one has to assume some specific form of the hamilton operator $\mathcal{H}_M$. In agreement with the previous discussion, $\mathcal{H}_M$ contains two-and three-body potentials \cite{bn2}.

The confining  three-body potential is chosen within a string-like picture, where the quarks are connected by gluonic strings (flux tubes) and the potential increases linearly with a collective radius $r_{3q}$
\begin{equation}
V^{(3)}_{conf}(\vec{r}_1,\vec{r}_2,\vec{r}_3)\sim r_{3q}(\vec{r}_1,\vec{r}_2,\vec{r}_3).
\end{equation}
There are three different ways to define $r_{3q}$ \cite{bn2}. The first one is the $Y$-type \cite{carl}
\begin{equation}
r_{3q}=r_Y=min \sum_{i=1}^3 |\vec{r}_i-\vec{r}_0|,
\end{equation}
$\vec{r}_0$ is the position where the flux tubes can merge and is chosen in order to minimize $r_{3q}$. A second possibility is given by the so called $\Delta$-type
\begin{equation}
r_{3q}=r_\Delta=\sum_{i<j}^3 |\vec{r}_i-\vec{r}_j|;
\end{equation}
rescaling $r_\Delta$ by  a factor f one gets a good approximation of $r_Y$ (\cite{bn2} and references quoted therein), provided that $1/2 < f < 1/\sqrt{3}$. The third choice is provided by the hypercentral one \cite{pl}
\begin{equation}
r_{3q}=r_H=\sqrt{\vec{\rho}^2+\vec{\lambda}^2}.
\end{equation}
Having tested that the structure of  spectra depends only slightly on the choice of  $r_{3q}$, the authors of ref.~\cite{bn2} use the three-body $\Delta$-shape string potential rising linearly with $r_\Delta$, which provides the $SU(6)$ invariant part of the three-quark interaction.

The two-body potential is taken from the instanton interaction introduced by \'{}t Hooft \cite{tH} in the $SU(2)$ case (extended to $SU(3)$ in ref.~\cite{svz}), in which the $\delta$ term is smeared with an effective range $\lambda$. Such an interaction acts only on flavour antisymmetric states and therefore it does non act on $\Delta$ states, thereby leading to a $N-\Delta$ mass splitting. 

The model hamiltonian depends on seven free parameters, which are used to describe both the non-strange \cite{bn2} and strange \cite{bn3} baryon sector. The results are quite satisfactory.

The model has been successfully applied to the description of the elastic form factors \cite{mert,ronn}, the helicity amplitudes for both the nonstrange \cite{ronn} and strange resonances \cite{caut}, the semileptonic decays of baryons  \cite{mig} and the axial form factors \cite{mert}.

\subsection{The interacting quark-diquark model}

The Interacting quark quark model \cite{diq} and  its relativistic version \cite{rel_diq}
give a good  reproduction of the spectrum, moreover they have much
less missing resonances than a normal three quark model. 
 In particular, we report here the
rest frame mass operator of the  Relativistic quark-diquark model :
\begin{eqnarray}
 M=E_0 +\sqrt{q^2+m^2_1}+\sqrt{q^2+m^2_1}+ M_{\rm dir}(r)+ M_{\rm cont} +M_{\rm ex}(r),
\end{eqnarray}
where $E_0$ is a constant, $M_{\rm dir} (r)$ and $M_{\rm ex} (r)$, respectively, are
the direct and the exchange diquark-quark interaction, $m_1$ and
$m_2$ stand for the diquark and quark masses, where $m_1$ is either
$m_S$ or $m_{AV}$ according if the mass operator acts on a scalar or
an axial vector diquark, and $M_{\rm cont}(r)$ is a contact interaction.
The direct term is a Coulomb-like interaction with a cutoff
plus a linear confinement term
\begin{eqnarray}
 M_{\rm dir}= -\frac{\tau}{r}\left(1-e^{-\mu r}\right) + \beta r.
\end{eqnarray}
A simple mechanism that
generates a Coulomb-like interaction is the one-gluon exchange.
One needs also an exchange interaction. This is indeed the crucial ingredient of a
quark-diquark description of baryons and  has the form
\begin{eqnarray}
 M_{\rm ex}(r)=  (-1)^{l+1}2Ae^{-\sigma r}\left[A_S(\vec{s}_1\cdot\vec{s}_2)+A_I(\vec{t}_1\cdot\vec{t}_2)+
 A_{SI}(\vec{s}_1\cdot\vec{s}_2)
 (\vec{t}_1\cdot\vec{t}_2)\right],
\end{eqnarray}
where $\vec{s}$ and $\vec{t}$ are the spin and the isospin operators.
Moreover, we consider a contact interaction 
\begin{eqnarray}
 M_{\rm cont}(r) = \left(\frac{m_1m_2}{E_1E_2}\right)^{1/2+\epsilon}\frac{\eta^3D}{\pi^{3/2}}e^{-\eta^2 r^2}
 \delta_{L,0}\delta_{s_1,1}
 \left(\frac{m_1m_2}{E_1E_2}\right)^{1/2+\epsilon},
\end{eqnarray}
where $E_i=\sqrt{q^2+m^2_i}(i=1,2), \epsilon, \eta$ and $D$ are parameters of the model.
The relativistic Interacting quark-diquark model is a relativistic
version of the Interacting model  of
Ref.~\cite{diq}. The Interacting quark-diquark model
hamiltonian  is 
\begin{eqnarray} 
 H&=&E_0+\frac{q^2}{2\mu}-\frac{\tau}{r}+\beta r+ [B+C \delta_{L,0}]\delta_{s_1,1}\nonumber\\
 &+& (-1)^{l+1}2Ae^{-\alpha r}\left[(\vec{s}_1\cdot\vec{s}_2)+(\vec{t}_1\cdot\vec{t}_2)+(\vec{s}_1\cdot\vec{s}_2)
 (\vec{t}_1\cdot\vec{t}_2)\right],
\end{eqnarray}
where $\vec{s}_1$ and $\vec{s}_2 $  are the spin of the quark and of
the diquark respectively, while  $\vec{t}_1$ and $\vec{t}_2 $  the the
same for  the isospin. The contact interaction $ C \delta_{L,0}$ acts only on the spatial
ground state, while the $\delta_{s_1,1}$ on the axial diquark.

\begin{figure}[h]

\includegraphics[width=4.5in]{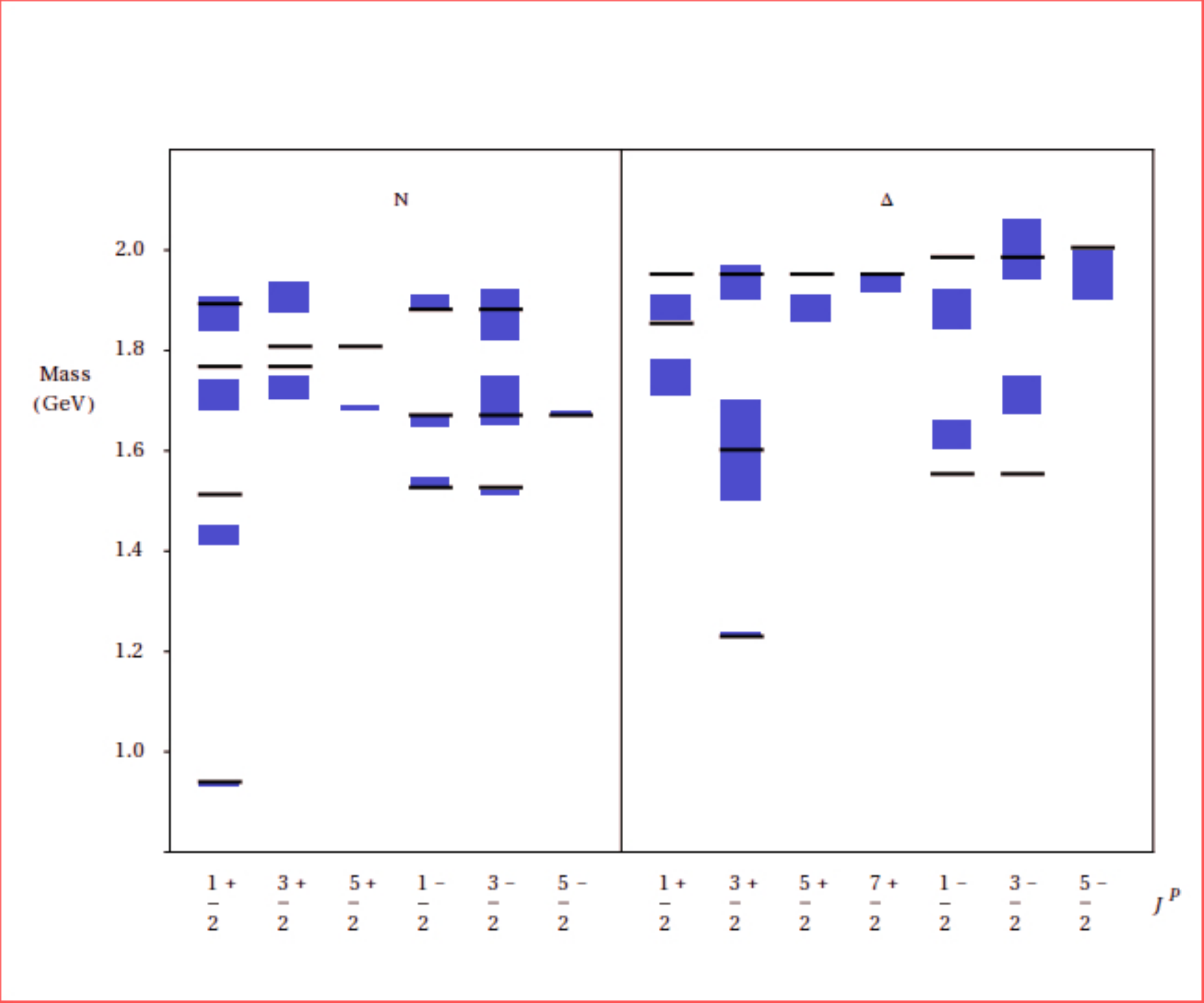}

\caption{ (Color online) The experimental spectrum of the non strange three- and four-star resonances \cite{pdg10} in comparison with the results of the interacting quark-diquark model \cite{diq}. }
\label{q-diq}
\end{figure}

\section{The hypercentral Constituent Quark Model}

\subsection{The hyperspherical coordinates}

The starting point of the hypercentral Constituent Quark Model (hCQM) is the introduction of the hyperspherical coordinates \cite{morp52,sim,baf}, which are given by the  angles ${\Omega}_{\rho}=({\theta}_{\rho},{\phi}_{\rho})$ 
and  ${\Omega}_{\lambda}=({\theta}_{\lambda},{\phi}_{\lambda})$ 
together with the hyperradius, $x$, and the hyperangle, $\xi$, defined 
 in terms of the absolute values $\rho$ and $\lambda$ of the Jacobi coordinates of Eq.~(\ref{coord})
\begin{equation}
x = \sqrt{\vec{\rho}^2 + \vec{\lambda}^2}, ~ ~ ~ ~ ~ ~ ~ ~ ~ ~ \xi = \arctan {\frac{\rho}{\lambda}}.
\end{equation}
The hyperradius x is a collective variable, which gives a measure of the dimension of the three-quark system, while the hyperangle $\xi$ reflects its  deformation.

Using these variables, the nonrelativistic kinetic energy operator of Eq.~(\ref{kin}), after having separated the c.m.\ motion,  can be written as
\begin{equation}
- \frac{\hbar^2}{2m} (\Delta_\rho + \Delta_\lambda) = - \frac{\hbar^2}{2m} ( \frac{\partial ^2}{\partial x^2}+\frac{5}{x} \frac{\partial}{\partial x}  + \frac{L^2(\Omega)}{x^2}).
 \end{equation}
The grand angular operator $L^2(\Omega)=L^2(\Omega_\rho, \Omega_\lambda,\xi)$ is the six-dimensional generalization
of the squared angular momentum operator and is a representation of the quadratic Casimir operator of the rotation group in six dimensions O(6).  Its eigenfunctions are the so called hyperspherical harmonics (h.h.) \cite{baf} $Y_{[\gamma]l_{\rho}l_{\lambda}}({\Omega}_{\rho},{\Omega}_{\lambda},\xi)$
\begin{equation}
L^2(\Omega_\rho, \Omega_\lambda,\xi)~Y_{[{\gamma}]l_{\rho}l_{\lambda}}({\Omega}_{\rho},{\Omega}_{\lambda},\xi)~=~-\gamma(\gamma+4) Y_{[{\gamma}]l_{\rho}l_{\lambda}}({\Omega}_{\rho},{\Omega}_{\lambda},\xi);
\label{grand}
 \end{equation}
the grand angular quantum number $\gamma$ is given by $\gamma = 2 n + l_\rho + l_\lambda$, where n is a nonnegative integer  and  $ l_\rho$, $l_\lambda$ are the angular momenta corresponding to the Jacobi coordinates of Eq.~(\ref{coord}).

The h.h.\ describe the angular and hyperangular part of the three-quark wave function and are written as 
\cite{baf}
\begin{equation}
{Y}_{[{\gamma}]l_{\rho}l_{\lambda}}
({\Omega}_{\rho},{\Omega}_{\lambda},\xi)~=~
{Y}_{l_{\rho}m_{\rho}}({\Omega}_{\rho})~{Y}_{l_{\lambda}
m_{\lambda}}({\Omega}_{\lambda})~~ 
^{(2)}{P_{\gamma}^{l_{\rho}l_{\lambda}}}~(\xi).\label{yg}
\end{equation} 
\noindent
where the hyperangular 
functions $^{(2)}{P_{\gamma}^{l_{\rho}l_{\lambda}}}(\xi)$ are given in 
terms of trigonometric functions and Jacobi polynomials \cite{baf}.

The h.h.\ form a complete orthogonal basis in the space of the functions of $\Omega_\rho, \Omega_\lambda,\xi $ and  then any three-quark wave function can be expanded as a series of h.h.

\begin{equation}
\Psi(\vec{\rho},\vec{\lambda})~~=~~\sum_{\gamma,l_{\rho},l_{\lambda}}~
c_{[\gamma]l_{\rho}l_{\lambda}}~
{\psi}_{\gamma}(x)~{Y}_{[{\gamma}]l_{\rho}l_{\lambda}}(\Omega),
\label{span}
\end{equation} 
the hyperradial wave function ${\psi}_{\gamma}(x)$, depending on the hyperradius x only,  is completely symmetric for the exchange of the quark coordinates.

\subsection{The hypercentral approximation}

An expansion similar to Eq.~(\ref{span}) is valid for any quark interaction, the first term depending on the hyperradius $x$ only:
\begin{equation}
\Sigma_{i<j} ~V(r_{ij})~=~V(x) + \cdots .
\end{equation}
 
Retaining the first term only one gets the so called hypercentral approximation. Such approximation has been applied with success to the description of few-nucleon systems \cite{hca,ffs}, while in the baryon case it has been shown that the matrix elements of the currently used two-body qq potentials in the 3q space exhibit an almost perfectly hypercentral behaviour \cite{has}.

\begin{figure}[t]

\includegraphics[width=4in]{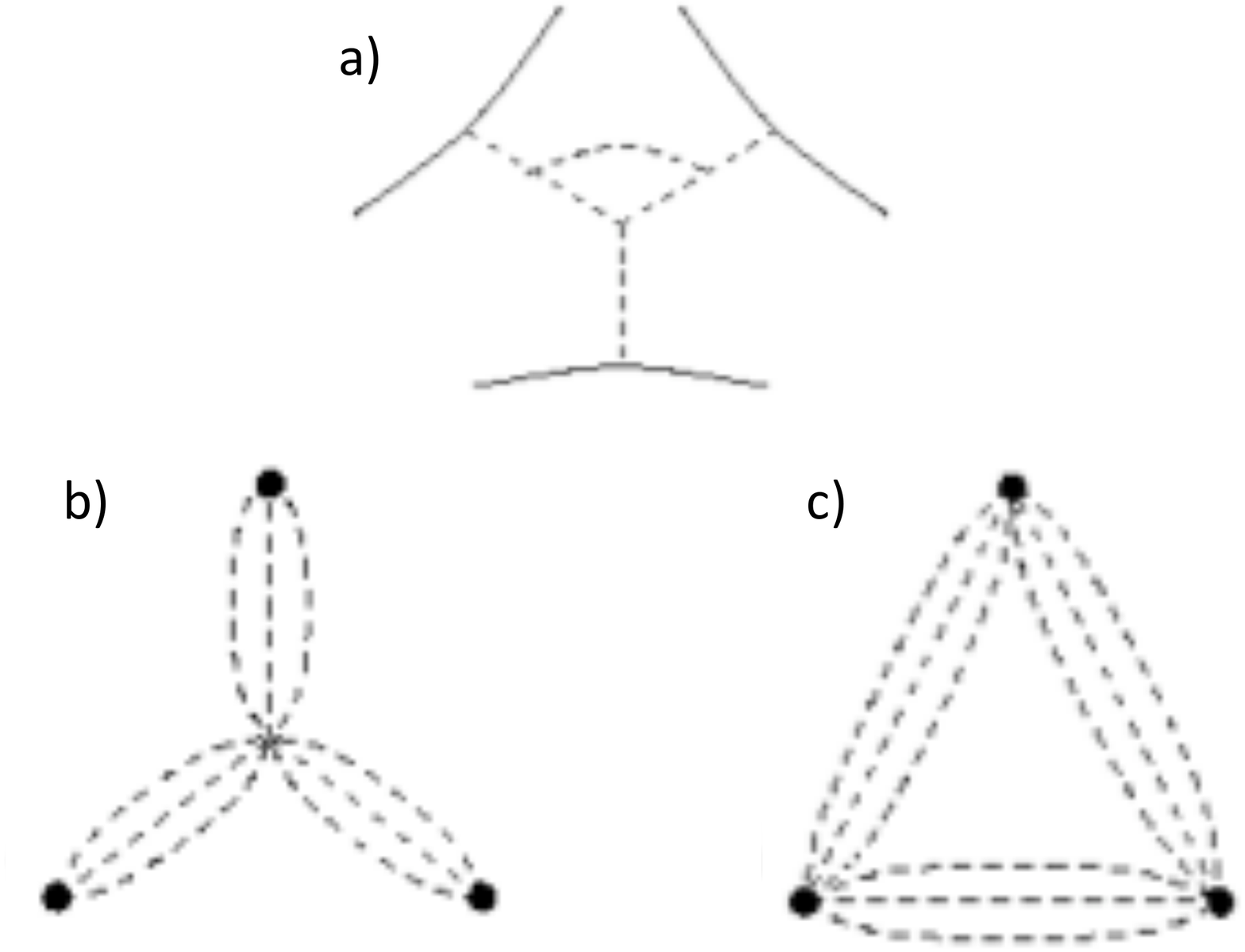}

\caption{  a) The lowest order diagram leading to a non zero three-body interaction of quarks in a baryon. b) The -Y-shaped string configuration. c) The $\Delta$-shaped string configuration.}
\label{string}
\end{figure}

In the hypercentral approximation, the three-quark potential depends on the hyperradius x only and therefore it has a three-body character, since the dependence on the single pair coordinates cannot be disentangled from the third one. The possibility of three-body forces is strictly related to the existence of a direct gluon-gluon interaction, which is one of the fundamental features of QCD. The diagram shown in Fig.~\ref{string} a is the lowest order one leading to a non vanishing three-body interaction among quarks in a baryon, but of course many others can be considered. A three-quark mechanism
 is considered also in flux tube models, which have been proposed as a QCD-based description of quark interactions \cite{ip},  leading to the Y-shaped three-quark configuration of Fig.~\ref{string}b \cite{carl},  besides the standard $\Delta$-like  
two-body one of Fig.~\ref{string}c. Furthermore, a Born-Oppenheimer treatment of the confinement potential in a QCD motivated bag model leads quite naturally to three-body forces \cite{has,hel} which increases linearly with some 'collective' radius.

In the hCQM the three-quark interaction is assumed to be hypercentral
\begin{equation}
V_{3q}(\vec{\rho},\vec{\lambda})~=~V(x) ,
\end{equation}
as a consequence, the three-quark wave function is factorized
\begin{equation}
\psi_{3q}(\vec{\rho},\vec{\lambda})~=
\psi_{\gamma \nu}(x)~~
{Y}_{[{\gamma}]l_{\rho}l_{\lambda}}({\Omega}_{\rho},{\Omega}_{\lambda},\xi);
\label{psi}
\end{equation}
the hyperradial wave function $\psi_{\gamma \nu}(x)$  is labeled by the grand angular quantum
number $\gamma$ defined above and  by the number of nodes $\nu$. The angular-hyperangular part of the 3q-state is completely described by the h.h.\ and is the same for any hypercentral potential. The dynamics is contained in the hyperradial  wave function $\psi_{\gamma \nu}(x)$, which, because of  the factorization Eq.~(\ref{psi}),   is obtained as a solution of the hyperradial equation
\begin{equation}
[~\frac{{d}^2}{dx^2}+\frac{5}{x}~\frac{d}{dx}-\frac{\gamma(\gamma+4)}{x^2}]
~~\psi_{\gamma \nu}(x)
~~=~~-2m~[E-V_{3q}(x)]~~\psi_{\gamma \nu}(x).
\label{hyrad}
\end{equation}

The Eq.~(\ref{hyrad}) can be solved analytically in two cases. The
first one is the six-dimensional harmonic oscillator (h.o.)
\begin{equation}
\sum_{i<j}~\frac{1}{2}~k~(\vec{r_i} - \vec{r_j})^2~=~\frac{3}{2}~k~x^2~=
~V_{h.o}(x),
\label{eq:ho}
\end{equation}
which is exactly hypercentral. The eigenvalues are given, as already mentioned, by $E= (3 +N) \hbar \omega$, where N can be written as $N=2 \nu + \gamma$ and the hyperradial wave functions are reported in Appendix A.

The second analytical case is given by the hyperCoulomb (hC) potential \cite{nc,hyc,breg,sig,sig2}
\begin{equation}
V_{hyc}(x)= -\frac{\tau}{x}. 
\label{hyc}
\end{equation}

The eigenvalues of the hyperCoulomb problem can be obtained by 
generalizing to six dimensions the calculations performed 
in three dimensions, obtaining
\begin{equation}
E_{n,\gamma}~=~-\frac{{{\tau}^2}m}{2n^2},
\label{en}
\end{equation}
\noindent where $n~=~N+\frac{5}{2}$ is the principal quantum number and $N=\gamma+\nu$, where
$\nu~= 0,1,2,~\ldots$ is the radial quantum number that counts the number of 
nodes of the wave function. 

The fundamental reason why the three-body 
problem with h.o.\ or hC interaction is 
exactly solvable is that they have a dynamic symmetry, $U(6)$ and $O(7)$ 
respectively. 

The dynamic symmetry $O(7)$ of the hC problem 
can be used to obtain the eigenvalues using purely algebraic 
methods.  The 
hyperCoulomb Hamiltonian can be rewritten as \cite{breg}

\begin{equation}
H~=~-\frac{{{\tau}^2}m}{2~[C_{2}(O(7))+\frac{25}{4}]},
\end{equation}

\noindent where $C_{2}(O(7))$ is the quadratic Casimir operator of $O(7)$. It can be shown \cite{sig2} that the eigenvalues
of $C_{2}(O(7))$ are given by $(\nu+\gamma)(\nu+\gamma+5)$, obtaining
\begin{equation}
E~=~-\frac{{{\tau}^2}m}{2~[(\nu+\gamma)(\nu+\gamma+5)+25/4]},
\end{equation}
which coincides with Eq.~(\ref{en}).

The eigenfunctions of Eq.~(\ref{hyrad}) with the hyperCoulomb potential
can be obtained analytically and are \cite{sig2}
\begin{equation}
\psi_{\gamma \nu}(x)~~=~~
\left[\frac{\nu!~(2g)^6}{(2 \gamma + 2 \nu+5)(\nu+ 2 \gamma+4)!^3}
\right]^
{\frac{1}{2}}~(2gx)^{\gamma}~e^{-gx}~~
L^{2\gamma+4}_{\nu}(2gx)
,\label{eigom}
\end{equation}
where for the associated Laguerre polynomials 
the notation  of Ref.~\cite{mf}  is used and 
$g~=~\frac{\tau m}{\gamma+\nu+\frac{5}{2}}$. The explicit expression of the hyperradial wave functions are obtained in ref.~\cite{sig2} and reported in Appendix A.

A complete solution of the hyperCoulomb problem, using the SO(7,2) dynamical group, has been worked out in ref.~\cite{bis}, where the states, the elastic and inelastic form factors have been also developed.

\begin{figure}[h]

\includegraphics[width=4.5in]{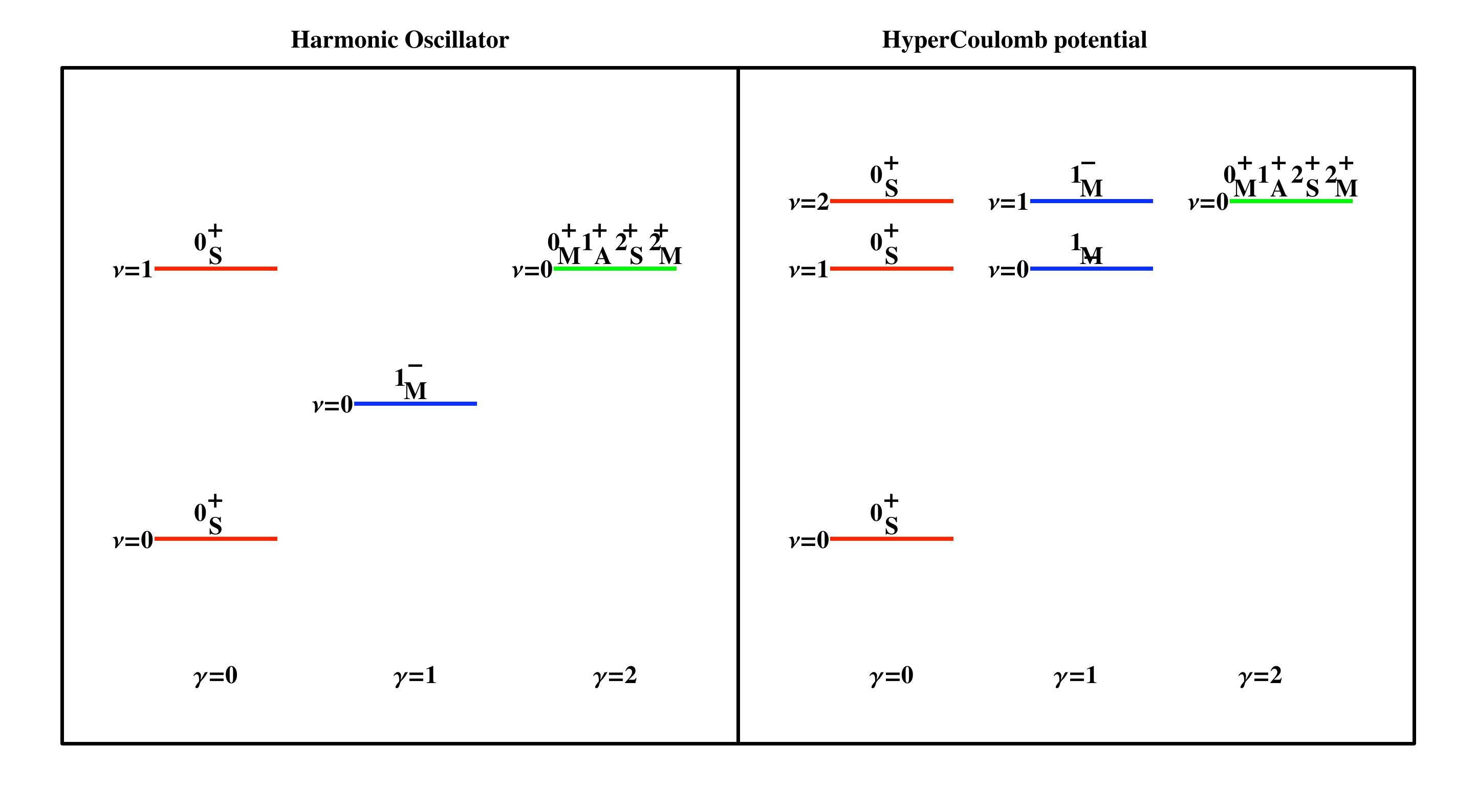}

\caption{ (Color online) Qualitative structure of  theoretical spectra for the h.o.\ (left) and for the hC potentials (right), up to the first three shells. The energy units are arbitrary and different for the two potentials.}
\label{ho-hc}
\end{figure}

The hC potential has important features \cite{pl,sig,sig2}. The energy eigenvalues depend on $\nu + \gamma$ and then  the negative parity states are exactly degenerate with the positive parity excitations, as is shown in Fig.~(\ref{ho-hc}). The observed Roper resonance is somewhat lower with respect to the negative parity baryon resonance, at variance with
the prediction of any $SU(6)-$invariant two-body potential, therefore the hC
 potential provides a better starting point for the description of the spectrum. The spectrum of Fig.~(\ref{ho-hc}) shows that within the first three shell the hC potential exhibits two more states with respect to the h.o.  The first extra level has positive parity and contains a further N and $\Delta$ state, thus enhancing slightly the number of theoretical states, but the second one has negative parity and allows to insert the recently observed states, already mentioned in Sec.~2.2.
 
 Another interesting property of the hC potential is that the form factors calculated with its wave functions have a power-law behaviour \cite{nc,sig,sig2}, leading to an improvement with respect to the widely used harmonic oscillator, for which the form factors are  too strong damped for increasing momentum transfer.

\begin{figure}[h]

\includegraphics[width=4.in]{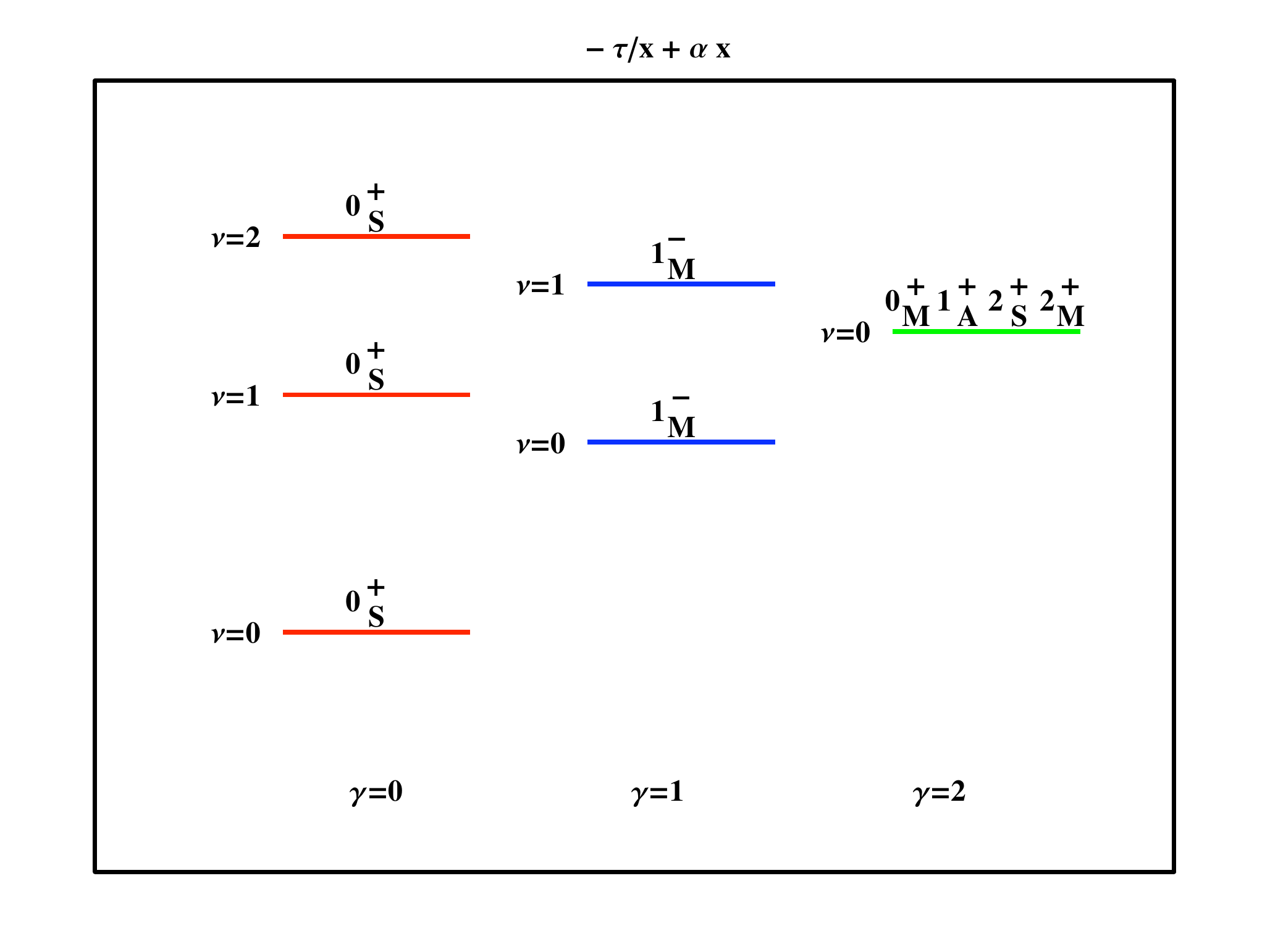}

\caption{ (Color online) Qualitative structure of  theoretical spectra for the potential of Eq.~(\ref{conf}), up to the first three shells. The energy units are arbitrary.}
\label{conf}
\end{figure}

\subsection{The hypercentral Constituent Quark Model}

The hC potential has interesting features, however it is not confining. In the hCQM, the
$SU(6)$ conserving part of the potential is then assumed to be the sum of the hC interaction and a linear confinement term \cite{pl,es}
\begin{equation}
V_{inv}~=~-\frac{\tau}{x} + \alpha x;
\label{h_pot}
\end{equation}
the hyperradial equation (\ref{hyrad}) must be solved numerically and the presence of the confinement removes the degeneracies typical of the hC potential, as shown in Fig.~(\ref{conf}), however the general structure is only slightly modified with respect to the hC potential.

\begin{figure}[h]

\includegraphics[width=4.5in]{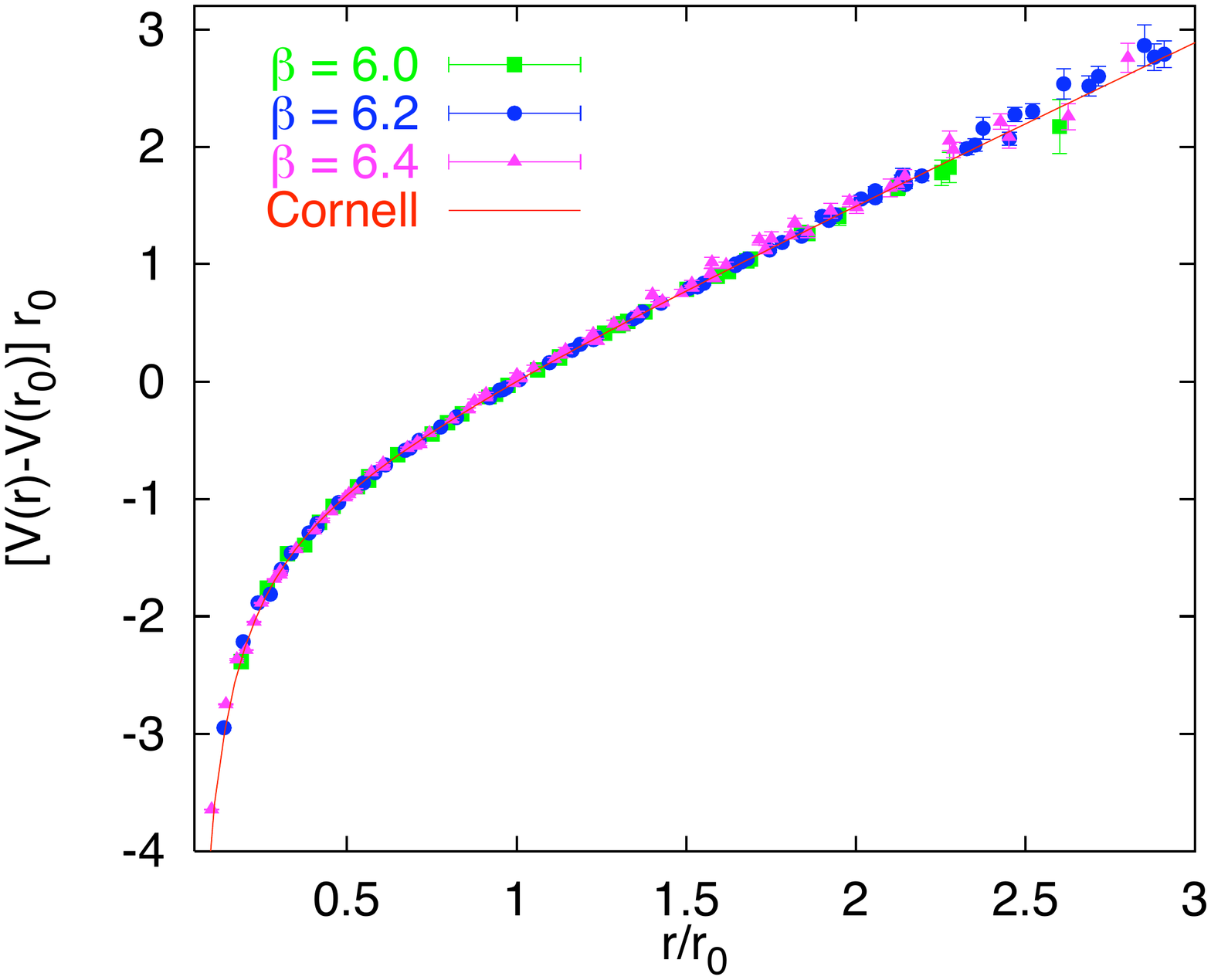}

\caption{ (Color online) The quark-antiquark potential calculated in LQCD \cite{LQCD} for static quarks in the $SU(3)$ limit. The constant $\beta$ is the inverse QCD coupling and $r_0\sim 0.5$ fm.}
\label{bali}
\end{figure}

The structure of the potential of Eq.~(\ref{h_pot}) is formally similar to the quark-antiquark Cornell potential \cite{corn} widely used for the description of mesons. It is noteworthy that  Lattice QCD calculations \cite{LQCD} are able to reproduce  the Cornell potential, as it is seen in Fig.~(\ref{bali}), where the LQCD results for static quark-antiquark pairs in the $SU(3)$ limit are reported. The model potential of Eq.~(\ref{h_pot}) can then be considered as the hypercentral approximation of the Cornell potential.

Having chosen the form for the hypercentral potential, the solutions of the hypercentral equation (\ref{hyrad}) produce a series of wave functions $\psi_{\gamma \nu}(x)$ and then one can build up the model $SU(6)$ states.  Taking advantage of the fact that also the h.o.\ potential is hypercentral, one can start from the states of the Isgur-Karl model \cite{ik}, express them in terms of the hyperspherical coordinates and substitute the h.o.\ hyperradial wave functions with those determined by the potential of Eq.~(\ref{h_pot}). The complete $SU(6)$ configurations for non strange baryons are reported in Appendix B.

\begin{figure}[t]

\includegraphics[width=4.5in]{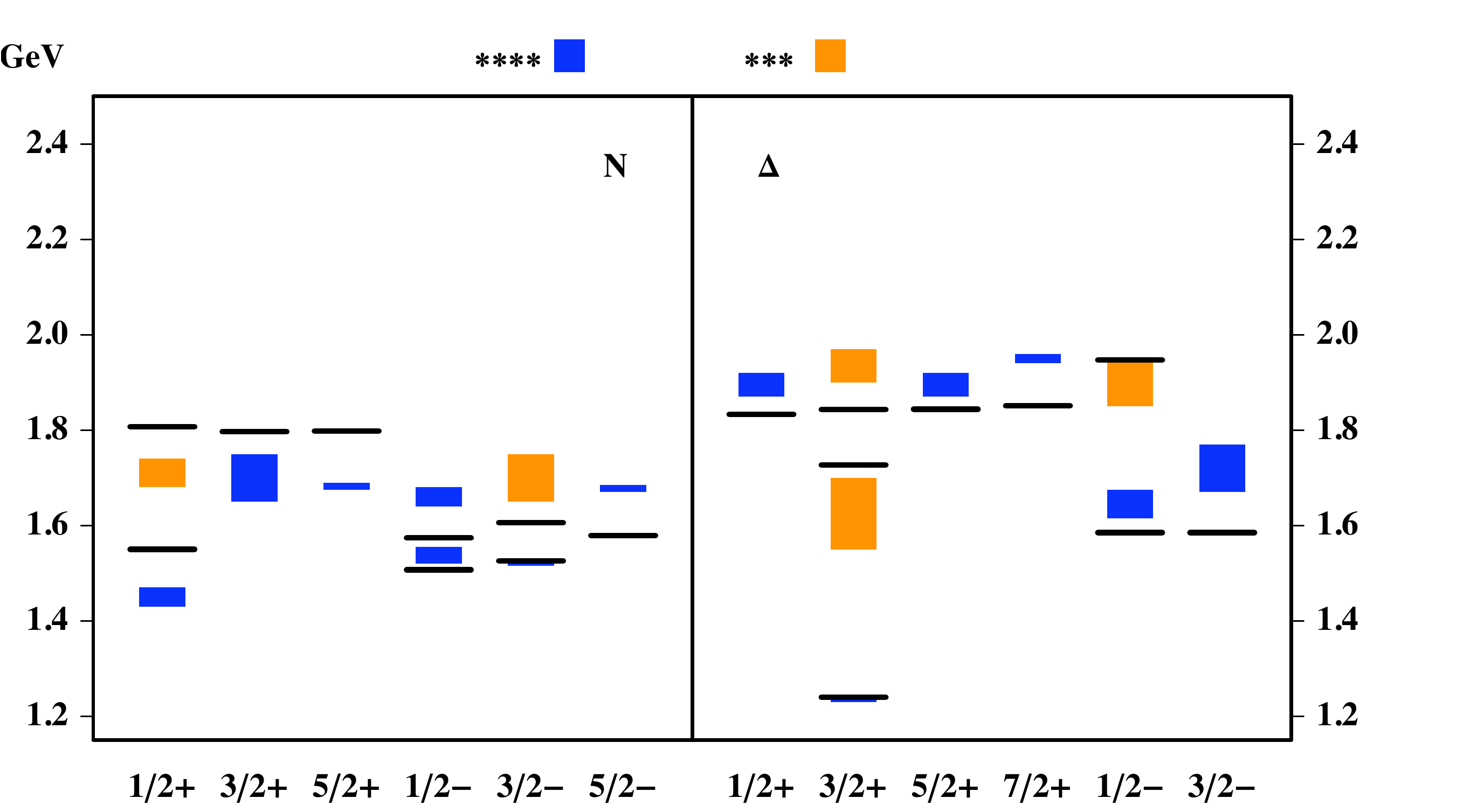}

\caption{ (Color online) The spectrum obtained using the hCQM hamiltonian of Eq.~(\ref{H_hCQM}). The free parameters are fitted to the experimental values of the 4* and 3* resonances reported in the PDG \cite{pdg10}.}
\label{hCQM}
\end{figure}

In order to complete the model hamiltonian one has to add a term violating the $SU(6)$ symmetry. In hCQM such term is chosen to be the standard hyperfine interaction \cite{deru,ik} of Eq.~(\ref{hyp}). The hamiltonian for the three quark system in the hCQM is then
\begin{equation}
H_{hCQM}~=~3m + \frac{\vec{p}_\rho^{~2}}{2m} + \frac{\vec{p}_\lambda^{~2}}{2m}-\frac{\tau}{x} + \alpha x + H_{hyp}.
\label{H_hCQM}
\end{equation}
In this way the baryon states are superpositions of the $SU(6)$ configurations reported in Appendix B.

Using the notation introduced with the Isgur-Karl model, the nucleon state can be written as
\begin{eqnarray}
  \label{nucl}
  |N \rangle  = &a_S |N ^2S_{1/2} \rangle _S + a'_S |N ^2S^*_{1/2} \rangle _S + a''_S |N ^2S^{**}_{1/2} \rangle _S+
 \nonumber \\ 
 &+ a_M |N ^2S_{1/2} \rangle _M + a_D |N ^4D_{1/2} \rangle _M ,
\end{eqnarray}
with $ a_S=0.976, a'_S=-0.196, a''_S=-0.043, a_M=-0.051, a_D=-0.070$ \cite{pl,es}; the asterisks in the second and third term of Eq.~(\ref{nucl}) mean that the spin-isospin part are the same as in the first term, but the space part corresponds to  hyperradially excited wave functions. One should not forget that in the hCQM there are two hyperradial excitations of the nucleon within the first three shells (see Fig.~(\ref{ho-hc}). The Roper resonance has a similar expansion, with the dominant component given by $|N ^2S'_{1/2} \rangle _S$: $ a_s=0.183, a'_s=0.967,  a''_s=0.185, a_M=0.021, a_D=-0.176$. The $\Delta$ resonance is given by
\begin{equation}
|\Delta \rangle  = b_S |\Delta  ^4S_{3/2} \rangle _S + b'_S |\Delta ^4S'_{3/2} \rangle _S + b_D |\Delta ^4S_{3/2} \rangle _S + b'_D |\Delta ^2D_{3/2} \rangle _M ,
\end{equation}
with $ b_S=0.976, b'_S=0.166, b''_S=-0.042, b_D=0.026, b'_D=-0.132$.

The quark mass is taken to be 1/3 of the proton mass, as prescribed in order to reproduce the proton magnetic moment (see e.g.~\cite{mg}). In this way in the model there are only three free parameters: $\tau$, $\alpha$ and the strength of the hyperfine interaction, the latter being mainly determined form the $N-\Delta$ mass difference. The parameters are found by fitting the energies of the 4$*$ and 3$*$ non strange baryons relative to the nucleon and are given by
\begin{equation}\label{par}
\alpha= 1.16fm^{-2},~~~~\tau=4.59~.
\end{equation}
The resulting spectrum is reported in Fig.~(\ref{hCQM}).

Having fixed the parameters of the three-quark hamiltonian from the spectrum, the baryon states are completely determined and can be used for the calculations of various properties. In the rest of the paper, the results obtained by the present nonrelativistic hCQM are given by parameter free calculations, that is they are  {\bf predictions}.

\subsection{An analytical model}

The hyperradial equation with the hypercentral potential of Eq.~(\ref{h_pot}) cannot be solved analytically unless the linear confinement term is treated as a perturbation \cite{sig,sig2}. This situation is reasonably valid for the lower states since they are confined in the low x region where the hC term is dominant. With this assumption, the perturbative contributions to the energies of the states are determined by the integral
\begin{equation}
\int_{0}^{\infty}~{dx}~x^{6}~|\psi_{\gamma \nu}(x)|^{2},
\end{equation}
which is given by \cite{sig,sig2}
\begin{equation}
\frac{1}{2g}
\frac{\nu!}{(\nu+2\gamma+5){(\nu+2\gamma+4)!}^3}
\left[2~\Gamma(\nu+2\gamma+5)\right]^{2}
\sum_{\sigma}
\frac{\Gamma(7+2\gamma+\sigma)}{\sigma!(\nu-\sigma)!^2
(\sigma-\nu+2)!^{2}}, \label{int}
\end{equation}
where 
\begin{equation}
\nu-2~\leq~\sigma~\leq~\nu.
\end{equation}

\begin{figure}[t]

\includegraphics[width=4.5in]{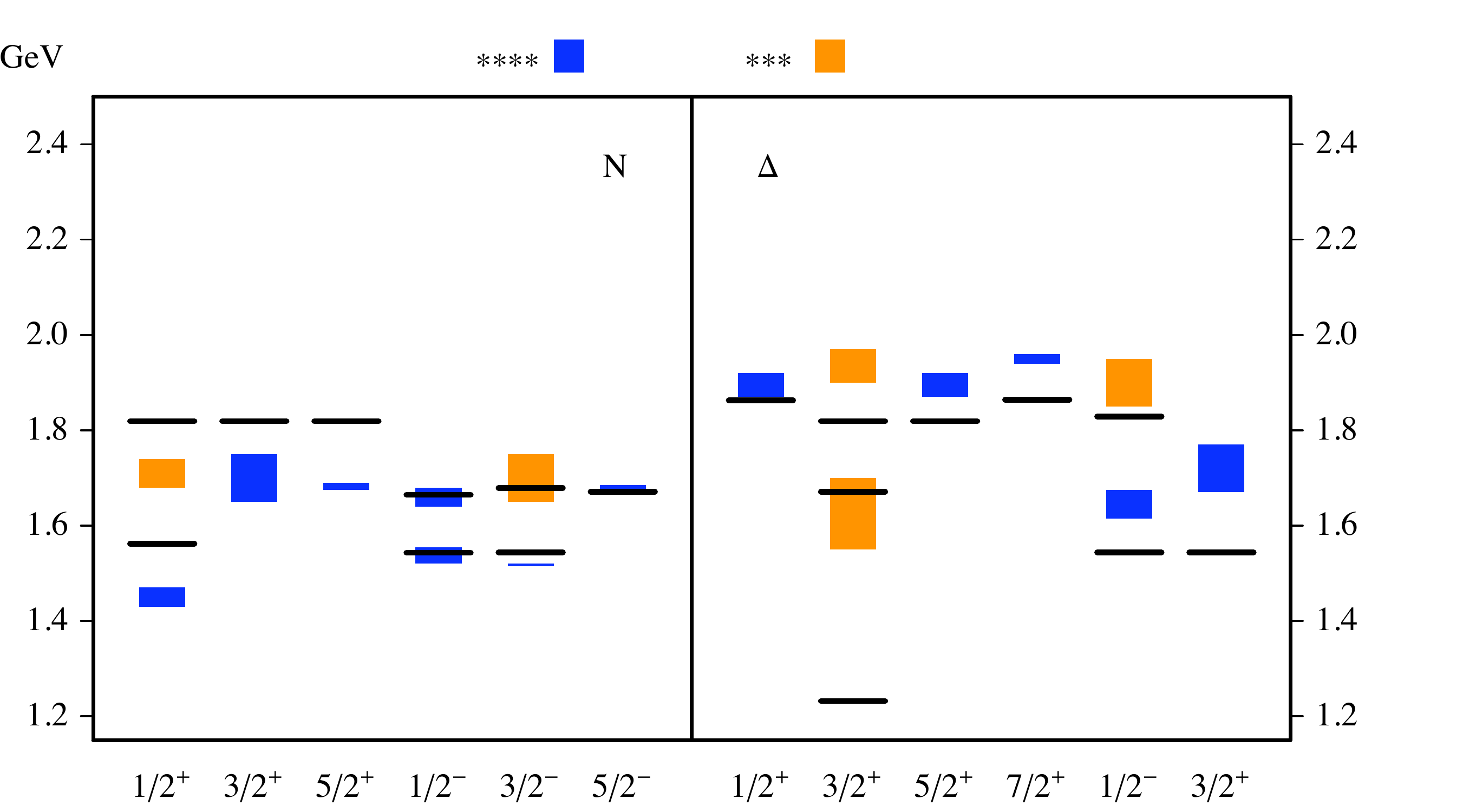}

\caption{ (Color online) The spectrum obtained using the analytical model. The free parameters are fitted to the experimentl values of the 4* and 3* resonances reported in the PDG \cite{pdg10}.}
\label{solv}
\end{figure}

The energy eigenvalues are then
\begin{equation}
E_{n,\gamma}~=~~-\frac{\tau^{2}m}{2n^{2}}+\frac{\alpha}{2m\tau}
[3n^2-\gamma(\gamma+4)-\frac{15}{4}],
\label{ener}
\end{equation}
where, according to the definition introduced in Sec.~3.2, n is given by $\nu+\gamma+5/2$. This formula, to which a constant $E_0$ should be added, is used to reproduce the energies averaged over the states with the same quantum numbers $\nu,\gamma$. In the fitting procedure, the Roper resonance is not taken into account because the presence of the confinement pushes it upwards with respect to the negative parity resonances. Assuming a quark mass about 1/3 of the proton mass, the result of the fit leads to \cite{sig,sig2}
\begin{equation}
E_0=2.152 \mbox{\ GeV}, ~~~~\tau=6.39,~~~~\alpha=0.148 \mbox{\ fm}^{-2},
\end{equation}
the latter two values are not much different from the ones of the non perturbative analysis reported in Eq.~(\ref{par}).

In order to describe the splittings within the $SU(6)$ multiplets, one has to add a $SU(6)$ violating interaction to the potential of Eq.~(\ref{h_pot}). For simplicity, such interaction is assumed \cite{sig,sig2} to contain a spin-spin term
\begin{equation}
V^{S}(x)~=~A~e^{-\beta x}~
\sum_{i<j}~\vec{\sigma_{i}}\cdot\vec{\sigma_{j}}~
=~A~e^{-\beta x}~[2~S^2-\frac{9}{4}]
,\\
\label{spin}
\end{equation}
where S is the total spin of the 3-quark system, and a tensor interaction
\begin{equation}
V^{T}(x)~=~B~\frac{1}{x^3}~\sum_{i<j}~\left[\frac{\left(\vec{\sigma_{i}} \cdot 
(\vec{r_{i}}-\vec{r_{j}})\right)
~\left(\vec{\sigma_{j}}\cdot(\vec{r_{i}}-\vec{r_{j}})\right)}
{|\vec{r_{i}}-\vec{r_{j}}|^{2}}
-\frac{1}{3}(\vec{\sigma_{i}}\cdot\vec{\sigma_{j}})\right].
\label{tens} \end{equation}

The parameters of the spin-spin interaction Eq.~(\ref{spin}) are determined from the $N(939)~-~\Delta (1232)$ and the $N(1535)~-~N(1650)$ splittings, obtaining $A~=~140.7~$MeV and $\beta~=~1.53 ~$fm$^{-1}$. We observe that, at variance with the OGE hyperfine interaction, the spin-spin term has a non-zero range, a feature that seems to be necessary since the hC wave functions are not so concentrated near the origin.

The effect of the tensor interaction Eq.~(\ref{tens}) is very small in the case of the hC wave functions, so there is no way to determine the parameter B directly from the spectrum. However, assuming for B a value of about 1/10 A, the description of the spectrum is slightly improved. The final results for the spectrum are reported in Fig.~(\ref{solv}).

The description of the spectrum is particularly good for the negative parity resonances. The analytical model has been also used for the calculation, without free parameters, of the electromagnetic transition amplitudes to some negative parity baryon resonances \cite{sig,sig2}. 

The success of the 1/x model in describing the spectrum indicates that, at least for the inner states, the confinement is provided by the hC potential \cite{sig2}, as further supported by ref.~\cite{adam}.

\section{The baryon spectrum}

\subsection{Results from the hCQM}

As discussed in the previous section, the free parameters of the hCQM are determined by fitting the masses of the 4* and 3* resonances \cite{pdg10} reported in Fig.~(\ref{baryon}). Of course the model can predict the masses of all other resonances belonging to the first three energy shells and the number of theoretical states exceeds the observed one, leading to the problem of the missing resonances, a problem in common with other CQM, in particular the h.o.\ one. In this respect, it is interesting to compare the number of states predicted by the two potentials,  the h.o.\ (see Table \ref{ho}) and the hCQM, which, as shown  in Table  \ref{comp} are 30 and 39 respectively. These two values are certainly larger than the number of observed 4* and 3* states, however the situation becomes different if one considers separately the positive and negative parity states and if also the new results from PDG2012 \cite{pdg12} are taken into account.

\begin{table}[h]
\caption[]{The number of states predicted by the h.o.\ and hCQM models, reporting separately  the positive and negative parity N and $\Delta$ states.  In the last four columns, the number of states listed in the 2010 \cite{pdg10} and 2012 \cite{pdg12} PDG editions  are reported.}
\label{comp}
\vspace{0.4cm}
\begin{center}
\begin{tabular}{|c|c|c||c|c||c|c|}
\hline
& & & & & &  \\
 & h.o. & hCQM & PDG10  &PDG10  &PDG12   & PDG12   \\
 & & & 4* + 3* & 4*+3*+2* & 4* + 3*  & 4*+3*+2* \\
& & & & &  & \\
\hline
& & & & & &  \\
$N^+$ & 14 & 15 &  5 & 8 &6 & 11  \\
& & & & & &  \\
 $N^-$ &5 & 10 & 5 & 7 & 6 & 9  \\
& & & & & &  \\
$ \Delta^+$& 9& 10 & 6 & 7 & 6 & 7 \\
& & & & & &  \\
$ \Delta^-$& 2 & 4 & 2 & 3 & 2 & 4 \\
& & & & &  & \\
Total& 30 & 39 & 18 & 25 & 20 & 31  \\
& & & & &  & \\
\hline
\end{tabular}
\end{center}
\end{table}

The positive parity states allowed by the h.o.\ model are abundant, but the negative ones are just what is necessary for the description of  the observed 4* and 3* states of PDG2010 \cite{pdg10}. In the hCQM there are two positive parity states more and the
 negative parity ones are doubled, in agreement with the fact that in the hCQM spectrum (see Fig.~\ref{conf}) there are two extra levels, one $0^+_S$ with a P11 and a P33 state, and a $1^-_M$ level, where a further series of negative parity states can be settled.

In the last edition of the PDG \cite{pdg12}, some new states are reported. This achievement has been possible also thanks to the availability of very precise cross section and polarization data from photoproduction experiments at CLAS \cite{clas}
and by the most recent coupled- 
channel analysis of the Bonn-Gatchina group \cite{anis} (for a discussion see \cite{vb_new}). In particular there is a resonance $D_{13}(1875)$ with a 3* status. The negative parity states allowed by the h.o.\ model are all already occupied and only the hCQM, with its further negative parity level, can describe such a  state. Moreover, if one considers also the new 2* states \cite{pdg12}, there are 13 negative parity resonances, 9 of the N and 4 of the $\Delta$ type, to be compared with the allowed values of the hCQM, that is 10 and 4, respectively. Finally, the total number of observed 4*, 3* and 2* resonances is 31, greater than the number allowed by the h.o.\ and not so far from 39, the hCQM value. 

\begin{figure}[h]

\includegraphics[width=4.5in]{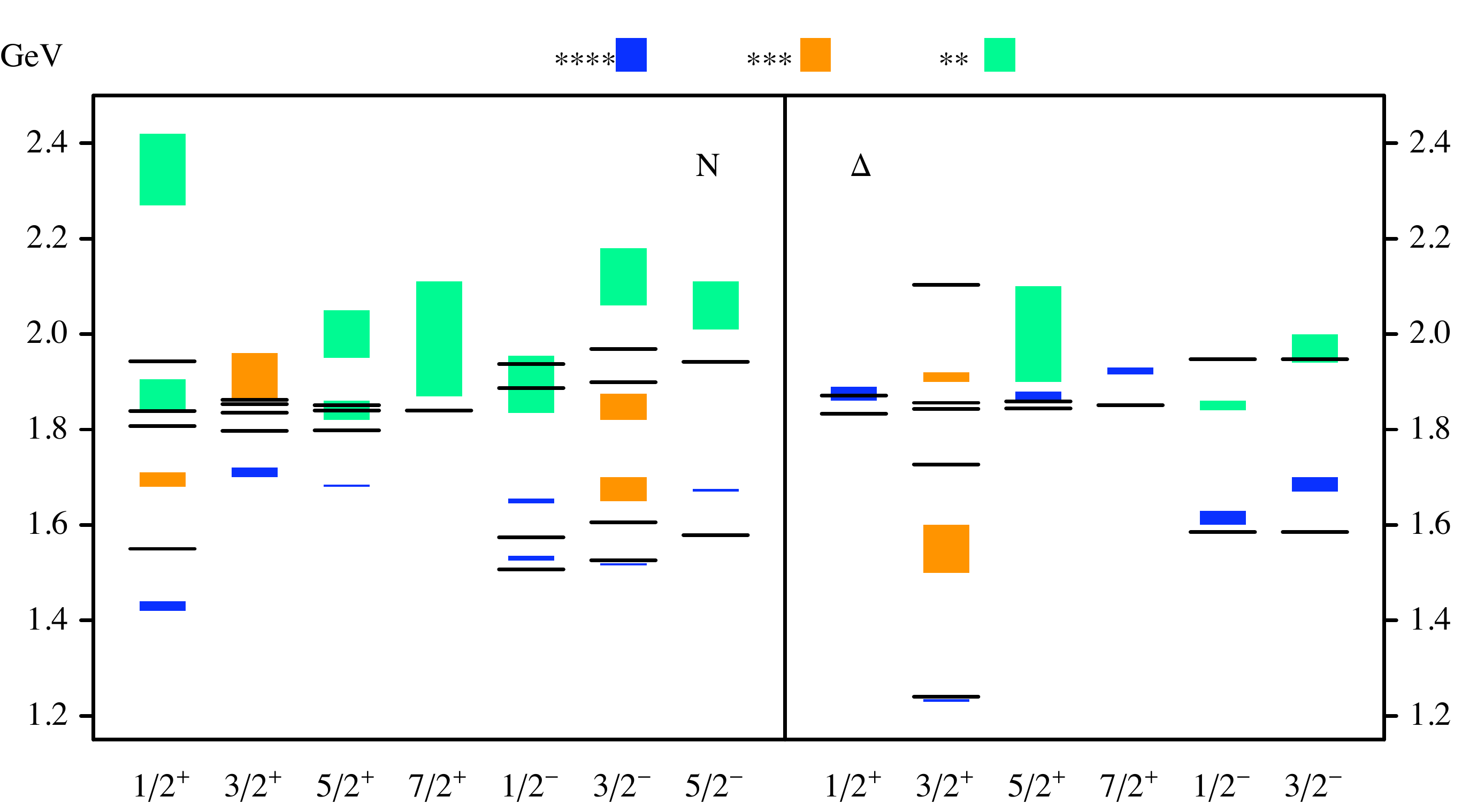}

\caption{ (Color online) The spectrum obtained with the hCQM described in Sec.~3 in comparison with all the 4*, 3* and 2* resonances reported in the 2012 edition of the PDG \cite{pdg12}.}
\label{hCQM_12}
\end{figure}

The comparison of  the theoretical spectrum with all the 4*, 3* and 2* listed in PDG \cite{pdg12} is shown in Fig.~(\ref{hCQM_12}) \cite{gs15}. In this Figure two theoretical levels are not shown, that is one $N1/2$ and one $N3/2$ state belonging to the $(20, 1^+)$ $SU(6)$ multiplet; they are mixed by the hyperfine interaction only with the states with $\gamma+\nu$ greater than 2 and do not contribute to the structure of the nucleon.

The overall description of the spectrum is quite good, considering that the model has only three free parameters. As for the Roper resonance, it is practically degenerate with the negative parity states, thanks to the behaviour of the hC potential, which is only slighlty modified by the confinement term. However, the theoretical Roper mass is still to high with respect to the experimental data. This problem is common to various CQM and it has some time ago suggested the idea that the Roper is not simply a ``breathing'' mode of the nucleon, but it is a hybrid state qqqG \cite{lb}, that is three quark plus a gluon component. However this model for the Roper has been ruled out by the recent results on the $\gamma* p \rightarrow N(1440)$ data \cite{ab_rop}, which showed that the longitudinal electroexcitation is significantly non zero while the hybrid model predicts that it should be vanishing.

The theoretical states in the higher part of the spectrum are somewhat compressed as an effect of the hC interaction. However, the masses of the resonances in this region are determined with large uncertainties and, because of their strong decays, the states are expected to have large widths and to be partially overlapped. When comparing the theoretical baryon spectra with the experimental data one should not forget that up to now CQM models predict states with zero width, since no coupling to the continuum has been consistently introduced. Some time ago the Isgur-Karl model has been implemented with quark-meson couplings \cite{blask}, allowing to calculate both the strong decay widths and the effects of the continuum on the resonance energies. This approach considers   quark and meson degrees of freedom on the same footing, it would be desirable on the contrary to have an approach in which quark-antiquark pair mechanisms are consistently taken into account. In this respect an important improvement has been achieved by a recent work \cite{sb1,bs,sb2}, in which an unquenched constituent quark model for baryons has been formulated and the quark-antiquark pair contributions are taken into account consistently.

\subsection{The hCQM with isospin}

In the hCQM hamiltonian, the $SU(6)$ violating term is provided by the hyperfine interaction of Eq.~(\ref{hyp}), similarly to what happens with the Isgur-Karl model \cite{ik} and its semirelativistic extension \cite{ci}. 
However the $SU(6)$ violation can be given also by a flavour dependent term, which is more or less explicitly included
in other approaches. In fact, in the algebraic model  (BIL \cite{bil})  the quark energy is written in terms of Casimir operators of symmetry groups which are relevant for 
the three-quark dynamics (see Sec.~2.4). In particular, for the internal degrees of freedom the 
G\"{u}rsey-Radicati mass formula \cite{rad} is used, leading to  an isospin dependent term which turns out to be important for the description of the spectrum. In the $\chi$CQM \cite{olof}, the quark-quark interaction is provided by one meson exchange and therefore the corresponding potential is spin-flavour dependent and is crucial for the description of baryons up to 1.7 GeV. As for the BN \cite{bn2}) model, the $SU(6)$ violation arises from the instanton interaction which does not act on $\Delta$ states. Moreover, it has been pointed out  that an isospin dependence of
the quark potential can be obtained by means of quark exchange \cite{dm}.

Therefore there are many motivations for the introduction of a flavour dependent 
term in the three-quark interaction and for this reason also in the  case of the hCQM 
an isospin dependent term has been included in the quark interaction \cite{vass,iso}.

The $SU(6)$ violation coming from the hyperfine interaction is still present, with one important modification, namely in the spin-spin interaction the $\delta$-like term is substituted with a smearing factor given by a gaussian function of the quark pair relative distance \cite{vass}:
\begin{equation}\label{srho}
H_{S}=~A_{S}~\sum_{i<j}~\frac{1}{(\sqrt{\pi}\sigma_S)^{3}}~
e^{-\frac{r_{ij}^2}{\sigma_S^2}}~({\vec{s}}_{i}\cdot {\vec{s}}_{j}).
\end{equation}

\begin{figure}[h]
\includegraphics[width=4.5in]{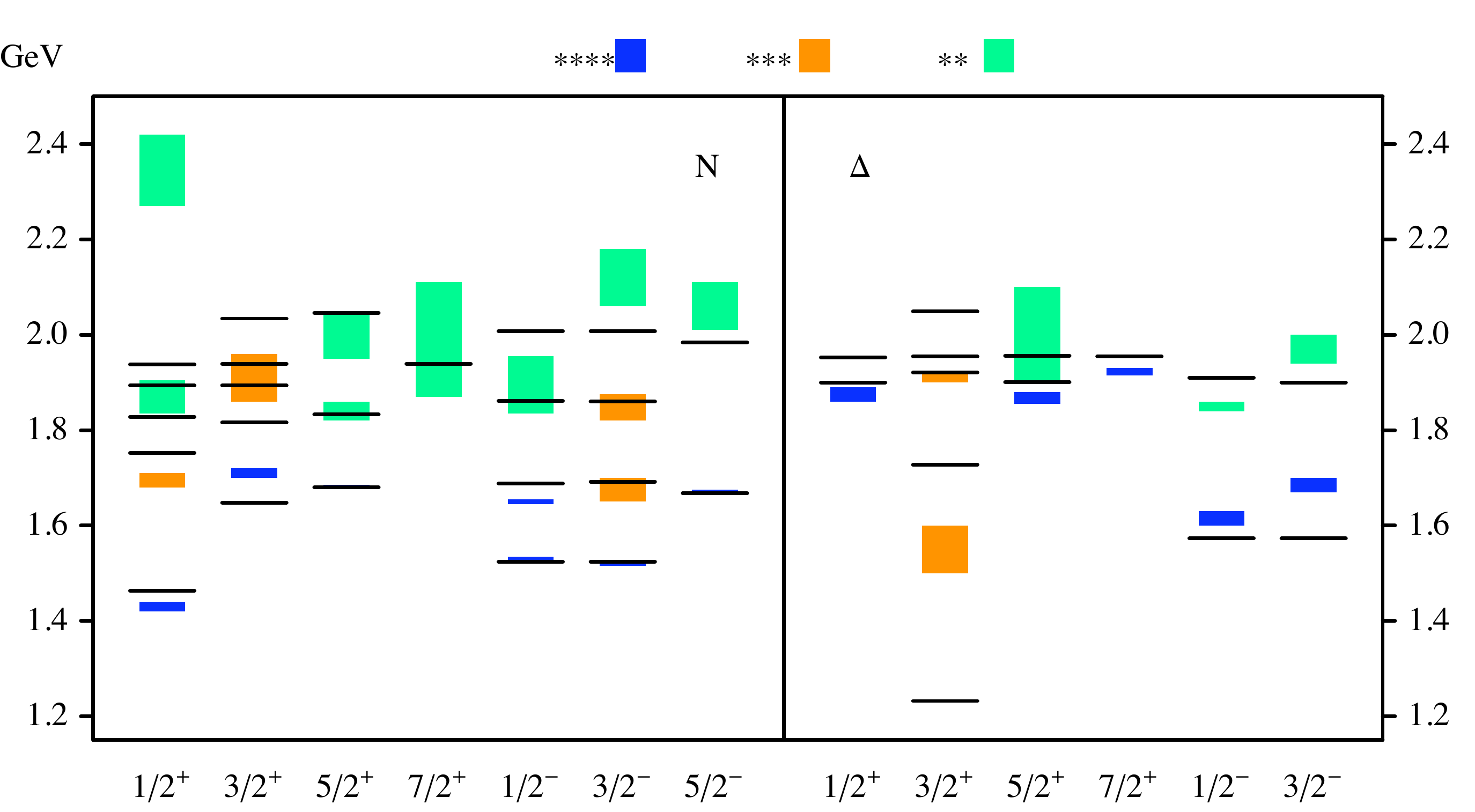}

\caption{ (Color online) The spectrum obtained with the hCQM with the spin and isospin dependent interactions of Eq.~(\ref{tot}) in comparison with all the 4*, 3*  and 2* non strange resonances reported in the 2012 edition of the PDG \cite{pdg12}.}
\label{hCQM_iso_12}
\end{figure}

The remaining $SU(6)$ violation comes from two terms. The first one is isospin dependent
\begin{equation}\label{taurho}
{H}_{I}=
~A_{I} \sum_{i<j}\frac{1}{(\sqrt{\pi}\sigma_I)^3}~
e^{-\frac{{\bf r}^2_{ij}}{\sigma_I^2}}({\vec{t}}_i \cdot {\vec{t}}_j),
\end{equation}\label{st}
where $\vec{t}_i$ is the isospin operator of the i-th quark and 
 $r_{ij}$ is the relative quark pair coordinate.
 The second one is a spin-isospin interaction, given by
\begin{equation}\label{si}
{H}_{SI}=
A_{SI}~\sum_{i<j}\frac{1}{(\sqrt{\pi}\sigma_{SI})^3}~e^{-\frac{r^2_{ij}}
{\sigma^2_{SI}}}({\vec{s}}_i \cdot {\vec{s}}_j)({\vec{t}}_i \cdot {\vec{t}}_j),
\end{equation}
\noindent where $\vec{s}_i$ and $\vec{t}_i$ are respectively the spin and isospin
operators of the i-th quark and $r_{ij}$ is the relative quark pair coordinate. 
The complete interaction is then given by
\begin{equation}\label{tot}                                         
H_{int}~=~V(x) +H_{S} +
H_{I} +H_{SI},
\end{equation}          
where V(x) is the hypercentral potential of Eq.~(\ref{h_pot}). 
The resulting spectrum for the 3*- and 4*- resonances is shown in Fig.~(\ref{hCQM_iso_12}).

The $N-\Delta$ mass difference is no more due only to the 
hyperfine interaction. 
In fact, in this model its contribution is only about $35\%$, 
the remaining splitting coming from the
spin-isospin term $(50\%)$ and from the isospin one $(15\%)$. 

It should also be noted that the negative parity resonances are again well described. In this model however there is the correct inversion between  the Roper and the 
negative parity resonances and this is almost entirely due to the spin-isospin 
interaction, as stated in Ref.~\cite{olof}. In general, the position of the Roper resonance is reproduced in all models containing an isospin dependent interaction \cite{bil,olof,bn2}.

Also the higher states are  fairly described and slightly less compressed than in the standard hCQM.

The tensor part of the hyperfine interaction, which is omitted for simplicity in Eq.~(\ref{tot}), is taken into account in the calculation, however its contribution to the spectrum is negligible.

\subsection{An extension to  strange baryons}

The hypercentral interaction of Eq.~(\ref{h_pot}) describes the average energies of the $SU(6)$ multiplets, while the splittings within each multiplets are generated by the hyperfine interaction Eq.~(\ref{hyp}) or by the spin-isospin interaction of Eq.~(\ref{tot}). In the latter case the flavour dependence is due only to the isospin operators, provided that the interest is limited to the non strange baryons.  In order to describe the spectrum of strange baryons as well, it is necessary to introduce a flavour dependence which involves both isospin and strangeness. This can be achieved in the hCQM in a similar manner to the algebraic model \cite{bil} quoted in Sec.~2.4, that is describing the $SU(6)$ violation by means  of a G\"{u}rsey-Radicati (GR) mass formula \cite{gr}. 

The original  GR mass formula \cite{rad} can be rewritten in terms of Casimir operators \cite{gr}
\begin{equation}
  \label{grorigcasimir}
  M=M_0+C~C_2[SU_S(2)]+D~C_1[U_Y(1)]+E~\left [C_2[SU_I(2)]-
\frac{1}{4}(C_1[U_Y(1)])^2 \right],
\end{equation}
where $C_2[SU_S(2)]$ and $C_2[SU_I(2)]$ are the $SU(2)$ (quadratic) Casimir operators for
spin and isospin, respectively, $C_1[U_Y(1)]$ is the Casimir for the $U(1)$ subgroup generated by
the hypercharge $Y$. 

However, in the framework of the CQM, the underlying symmetry is provided by $SU(6)$ and Eq.~(\ref{grorigcasimir}) is not the most general formula that can be written on the basis
of a broken $SU(6)$ symmetry. It can then be generalized as follows \cite{gr}
\begin{eqnarray}
  \label{grfull}
  M=M_0&+A~C_2[SU_{SF}(6)]+B~C_2[SU_F(3)]+C~C_2[SU_S(2)]+
 \nonumber \\ 
 &+D~C_1[U_Y(1)]+E~\left(C_2[SU_I(2)]-\frac{1}{4}(C_1[U_Y(1)])^2\right),
\end{eqnarray}
where $M_0$ is the $SU_{SF}(6)$ invariant mass.

The idea is then to consider the energy levels provided by the hypercentral potential of Eq.~(\ref{h_pot})  as the values of the central masses of the $SU_{SF}(6)$ multiplets and to use  the generalized mass formula in order to describe the spin-flavour splittings within the multiplets \cite{gr}.  The hamiltonian is assumed to be
\begin{equation}
H = H_0 + H_{GR},
\label{HGR}
\end{equation}
where $H_0$ is the hCQM hamiltonian without the hyperfine interaction
\begin{equation}
H_0 = 3m +  \frac{\vec{p}_\rho^{~2}}{2m} + \frac{\vec{p}_\lambda^{~2}}{2m} -\frac{\tau}{x} + \alpha x
\label{h0}
\end{equation}
and $H_{GR}$ is given by the spin-flavour dependent part of Eq.~(\ref{grfull})
\begin{eqnarray}
 \label{H_GR}
H_{GR}  = & A~C_2[SU_{SF}(6)]+B~C_2[SU_F(3)]+C~C_2[SU_S(2)]+
 \nonumber \\ 
 &+D~C_1[U_Y(1)]+E~\left(C_2[SU_I(2)]-\frac{1}{4}(C_1[U_Y(1)])^2\right).
\end{eqnarray}

\begin{table}[h]
\caption{The eigenvalues of the quadratic  Casimir operators for the groups $SU_{SF}(6)$ (left) and $SU_F(3)$ (right). } 
\vspace{0.4cm}
\begin{center}
\begin{tabular}{|c c||cc|}
\hline
$SU_{SF}(6)$ & $C_2$   & $SU_F(3)$ & $C_2$ 	\\
\hline
 & & & \\
$[56]$ & $\frac{45}{4}$ & $[8]$ & 3  \\
& & & \\
$[70]$ & $\frac{33}{4}$ & $[10]$ & 6  \\
& & & \\
$[20]$ & $\frac{21}{4}$ & $[1]$ & 0  \\
& & & \\
\hline
\end{tabular}
\end{center}
\label{casim}
\end{table}

In order to apply the generalized GR mass formula to the baryon spectrum it is necessary to assume that the coefficients A,B, \ldots in Eq.~(\ref{H_GR}) be the same in the various $SU_{SF}(6)$ multiplets. This actually seems   to be the case, as shown by the algebraic approach to the baryon spectrum \cite{bil}, where a formula similar to Eq.~(\ref{H_GR}) has been applied.

The  matrix elements of $H_{GR}$ are completely determined by the values of the various Casimir operators \cite{gr}: for the $SU_{SF}(6)$ and $SU_F(3)$ groups the values of the Casimir operator $C_2$ are reported in Table \ref{casim}, while for the $SU(2)$ and $U_Y(1)$ groups one has
\begin{equation}
\langle C_2[SU_I(2)] \rangle  =  I(I+1), ~~~\langle C_1[U_Y(1)] \rangle  =  Y, ~~~\langle C_2[SU_S(2)] \rangle  =  S(S+1).
\end{equation}

The mass of each baryon state is then
\begin{equation}
  \label{masses}
  \langle B \vert H \vert B \rangle =E_{\gamma \nu}+ \langle B \vert H_{GR} 
\vert B \rangle ,
\end{equation}
where $E_{\gamma \nu}$ are the eigenvalues of the hypercentral potential of Eq.~(\ref{h0}).

The parameters $\alpha$ and $\tau$ of the hypercentral potential have been fitted in Sec.~3.3 in presence of the hyperfine interaction. Here the $SU(6)$ violation is provided by a different mechanism and then these parameters must fitted to the spectrum together with those introduced in Eq.~(\ref{H_GR}).

Such fit can be performed in two ways. The first one is  an analytical procedure which consists in choosing a limited number of well known resonances and expressing their mass differences in terms of the Casimir operator values. In this way a part of the unknown coefficients is evaluated directly, while the remaining ones is fitted to the experimental spectrum. A possible choice of resonance pairs is given by the following ones
\begin{eqnarray}
\nonumber
\langle N(1650) S11 - N(1535) S11 \rangle & = & 3C,  \\ 
\nonumber
\langle \Delta(1232) P11 - N(938) P11 \rangle & = & 9B + 3C + 3E,\\
\nonumber
\langle N(1535) S11 - N(1440) P11 \rangle & = &  E_{10} - E_{01}  +12 A, \\
\nonumber
\langle \Sigma(1193) P11 - N(938) P11 \rangle & = & 3/2 E - D, \\
\langle \Lambda(1116) P01 - N(938) P11 \rangle & = &  - D - 1/2 E .
\label{analyt}
\end{eqnarray}

\begin{table}[h]
\caption{The fitted values of the parameters of the Hamiltonian (\ref{HGR}) \cite{gr}. Column (I) reports the values given by the analytical procedure of Eq.~(\ref{analyt}),  while Column (II) contains the results of the direct fit to the experimental spectrum}. 
\vspace{0.4cm}
\begin{center}
\begin{tabular}{|c|cc|}
\hline
Parameter & (I)   & (II) 	\\
\hline
 & &  \\
$\alpha$ & $1.4 fm^{-2}$ & $2.1 fm^{-2}$   \\
$\tau$ & 4.8 & 3.9   \\
$A$ & -13.8 MeV & -11.9 MeV \\
$B$ & 7.1 MeV MeV & 11.7 MeV  \\
$C$ & 38.3 MeV & 30.8 MeV  \\
$D$ & -197.3 MeV & -197.3 MeV  \\
$E$ & 38.5 MeV & 38.5 MeV  \\
& &  \\
\hline
\end{tabular}
\end{center}
\label{param}
\end{table}

 \begin{figure}[]
\includegraphics[width=4.5in]{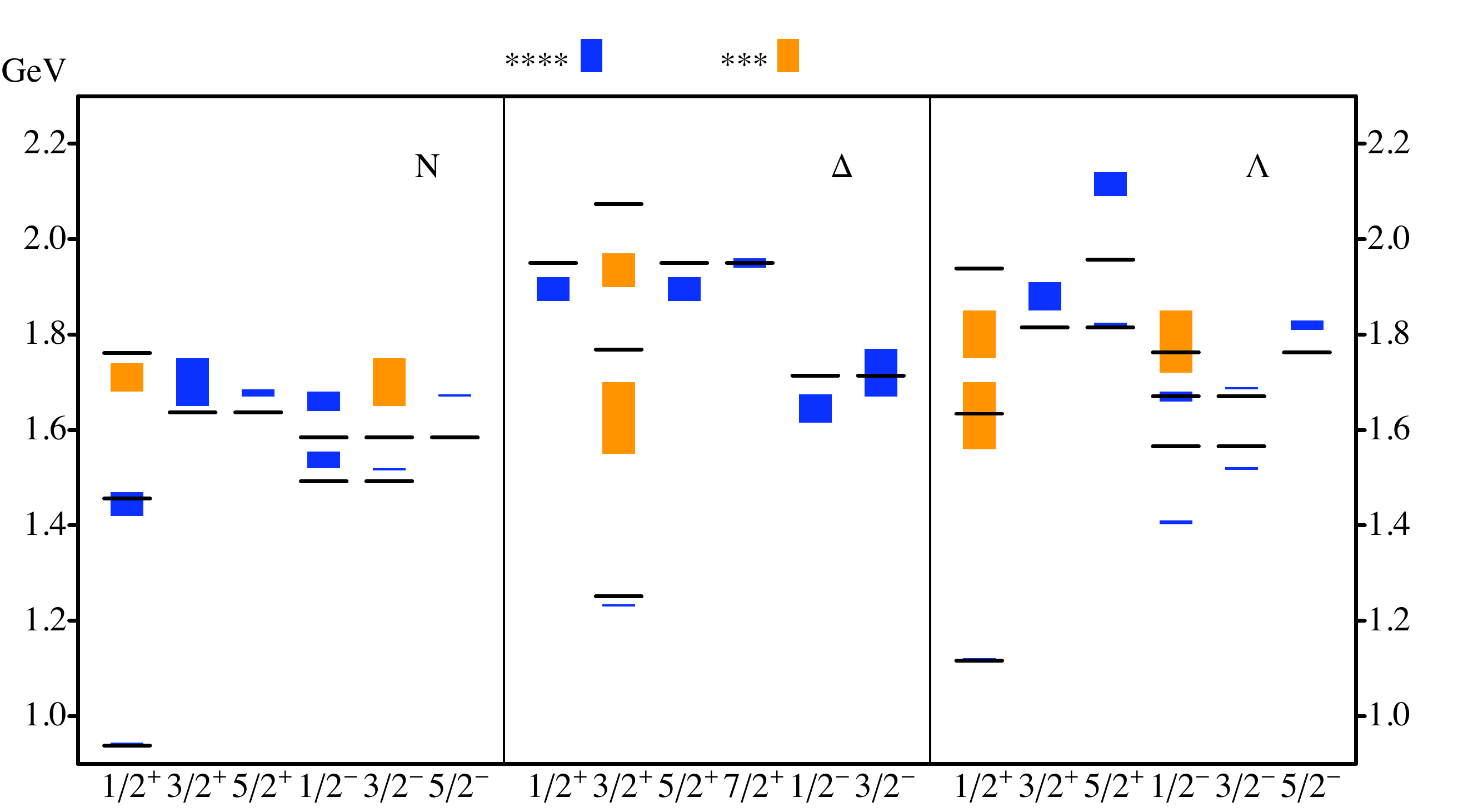}

\caption{ (Color online) The spectrum obtained with the hCQM and the generalized G\"ursey-Radicati formula of Eq.~(\ref{grfull}) in comparison with all the 4* and 3* N, $\Delta$ and $\Lambda$ resonances reported in the 2000 edition of the PDG \cite{pdg00}. The parameters given by the direct fit are used (see Table \ref{param}, column (II)).}
\label{hCQM_GR}
\end{figure}

\begin{figure}[]
\includegraphics[width=3.5in]{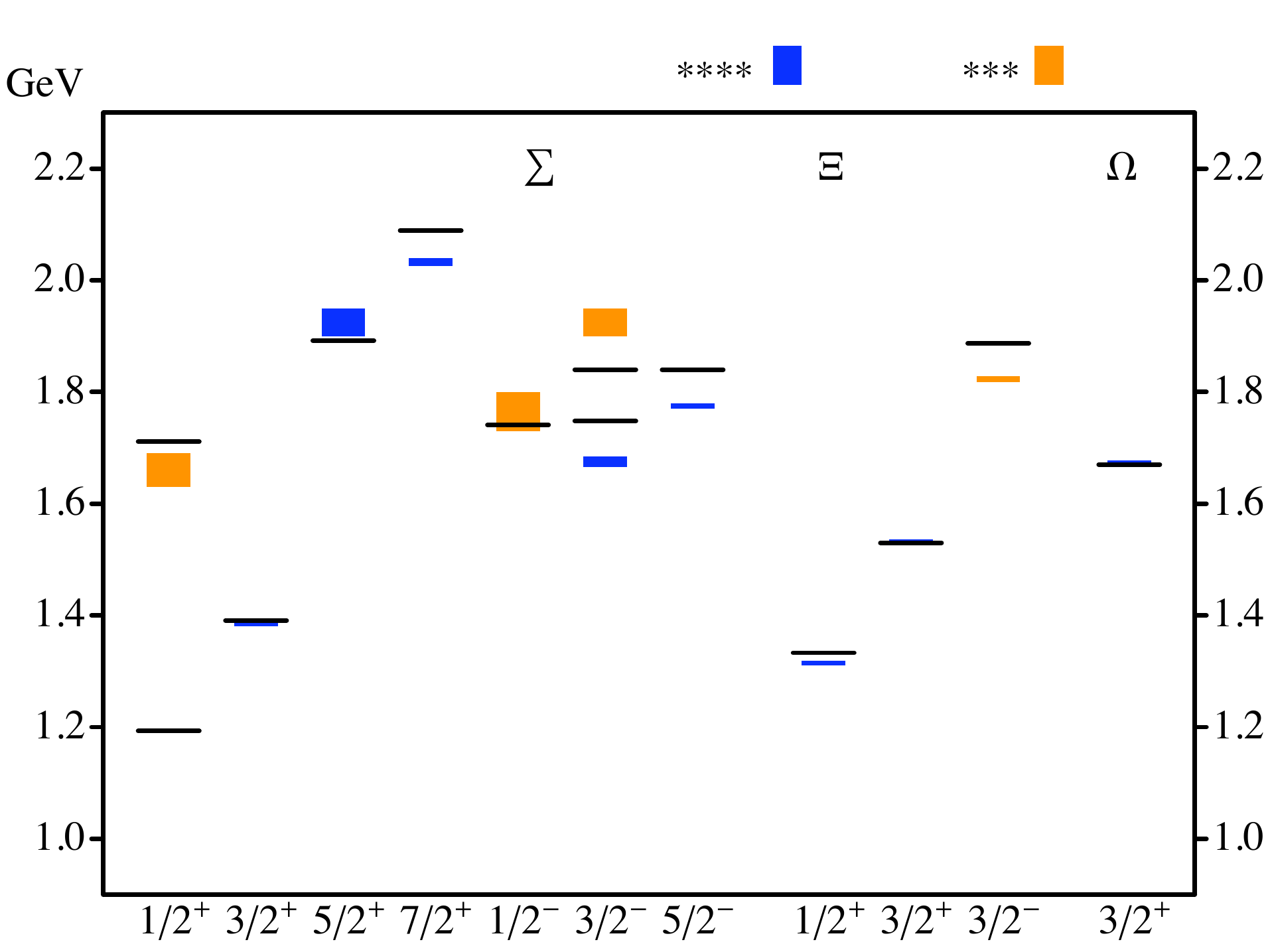}

\caption{ (Color online) The same as in Fig.~\ref{hCQM_GR} for the $\Sigma$, $\Xi$ and $\Omega$ resonances.}
\label{hCQM_GR2}
\end{figure}

In the second procedure all the parameters are fitted in order to reproduce the baryon spectrum. The resulting values of the parameters are reported in Table \ref{param} and the corresponding spectrum is shown in Fig.~\ref{hCQM_GR}. The overall description of the baryon spectrum of the  4* and 3* resonances is quite good, specially considering the simplicity of the model.

Using the analytical procedure the overall agreement with the spectrum is slightly worsened, but the strange sector is better described.

In both procedure, there is the need  of a non zero value of the parameter A in order to reproduce the spectrum. An attempt to fit the data fixing A=0 has been tried, however the resulting parameters $\alpha$ and $\tau$ are quite different form those reported in Table \ref{param}. Furthermore, the correct ordering of the Roper resonance and the negative parity resonances is lost. The presence of the Casimir $C_2[SU_{SF}(6)]$ is essential in order to shift down the energy of the first excited $0^+$ state   with respect to the $1^-$, an effect which is similar to the one produced by the U-potential in the Isgur-Karl model and by the presence of a flavour dependent interaction in the previously quoted CQMs.

\section{The electromagnetic excitation of baryon resonances}

\subsection{The transition amplitudes}

The electromagnetic excitation of the baryon resonances is an important source of information concerning the nucleon structure. The absorption of  real photons is a direct measure of the  excitation strength while the inelastic electron scattering is a probe of the excited nucleon structure at short distances. There are presently many experimental data taken at various laboratories (Jlab, Mainz, Bonn, \ldots) but a systematic study of the electromagnetic excitation of the nucleon at high $Q^2$ is expected to be performed by the upgraded 12 GEV beam at Jlab \cite{wp,wp2}.

There is an intense theoretical and phenomenological activity  which aims at extracting the transition amplitudes from the experimental data on photo- and electro-production of mesons off nucleons using mainly the Partial Wave Analysis \cite{brag07}. Various groups have devoted much effort in this sense, using different techniques. Among them  we quote the analyses  made by  the following groups: George Washington University (SAID) \cite{gwu}, Mainz University (MAID) \cite{uim,dmt,maid07},  Dubna-Mainz-Taipei (DMT) \cite{dmt2,dmt} Bonn-Gatchina  (BnGa) \cite{bnga}, EBAC at Jefferson Lab \cite{ebac},  J\"{u}lich \cite{jul}, Giessen \cite{gies}, Zagreb-Tuzla \cite{zt}.

From the theoretical point of view, the photo- and electro-excitations of the nucleon to the various baryon resonances are described by the helicity amplitudes, defined as
 the matrix elements of
the electromagnetic interaction, $A_{\mu} J^{\mu}$, between the nucleon, 
$N$, and the resonance, $B$, states:
\begin{equation} \label{hel}
\begin{array}{rcl}
\mathcal{A}_{1/2}&=&   \sqrt{\frac{2 \pi \alpha}{k_0}}   \langle B, J', J'_{z}=\frac{1}{2}\ | J_{+}| N, J~=~
\frac{1}{2}, J_{z}= -\frac{1}{2}\
\rangle,\\
& & \\
\mathcal{A}_{3/2}&=& \sqrt{\frac{2 \pi \alpha}{k_0}}  \langle B, J', J'_{z}=\frac{3}{2}\ |J_{+} | N, J~=~
\frac{1}{2}, J_{z}= \frac{1}{2}\
\rangle,\\
& & \\
\mathcal{S}_{1/2}&=&   \sqrt{\frac{2 \pi \alpha}{k_0}}   \langle B, J', J'_{z}=\frac{1}{2}\ | J_{0}| N, J~=~
\frac{1}{2}, J_{z}= \frac{1}{2}\
\rangle,\\
\end{array}
\end{equation}
where $k_0$ is the photon energy and, for the transverse excitation, the photon has been assumed,  without loss of generality, as left-handed.

The hCQM has been completely specified in Sec.~3.3. We recall that the Hamiltonian is given by Eq.~(\ref{H_hCQM})
\begin{equation}
H_{hCQM}~=~3m + \frac{\vec{p}_\rho^{~2}}{2m} + \frac{\vec{p}_\lambda^{~2}}{2m}-\frac{\tau}{x} + \alpha x + H_{hyp},
\end{equation}
where $\alpha = 1.16$ fm$^{-2}$, $\tau=4.59$ and the strength of the hyperfine interaction is fixed by the $N-\Delta$ mass difference.

The states of the various resonances have been explicitly built up and therefore they can be used for the calculation of any quantity of physical interest. In order to proceed to the calculation of the helicity amplitudes, one has to specify the current in the electromagnetic interaction. In the framework of the  hCQM, the current $J^{\mu}$ is simply given by the sum of the quark currents $ j_\mu(i)$
\begin{equation}
J^{\mu} = \sum_{i=1}^3 j_\mu(i)
\label{qcurr}
\end{equation}
and will be used in its nonrelativistic  form \cite{cko, ki}
\begin{equation}
\rho(\vec{k})= \sum_{i=1}^3 e_i e^{\vec{k} \cdot \vec{r}_i}, ~~~~\vec{j}(\vec{k}) = \frac{1}{2m} \sum_{i=1}^3 (\vec{p}_i' + \vec{p}_i + i \vec{\sigma}^q(i) \times \vec{k}) e^{\vec{k} \cdot \vec{r}_i},
\label{q_curr}
\end{equation}
where $e_i$ is the charge of the i-th quark
\begin{equation}
e_i=\frac{1}{2} [\frac{1}{3} + \tau^q_3 (i) ],
\end{equation}
$\vec{r}_i$ is the quark coordinate and $\vec{\sigma}^q(i)$, $\vec{\tau}^q(i)$   are, respectively, the quark spin and isospin operators. 

In order to compare the theoretical results with the experimental data, the calculation should be performed in the rest frame of the resonance (see e.g.~\cite{azn-bur-12}). The nucleon and resonance wave functions are actually calculated in their respective rest frames and, before evaluating the matrix elements given in Eqs.~(\ref{hel}), one should boost the nucleon to the resonance c.m.s.. However, in order to minimize the discrepancy between the nonrelativistic and the relativistic boosts in comparing with the experimental data,  we can  consider the Breit frame, as in refs.~(\cite{aie2,bil,sg}). In this frame $\vec{p}_N = -\vec{p}_R=-\vec{k}/2$, where $\vec{p}_N$, $\vec{p}_R$ and $\vec{k}$  are, respectively, the nucleon, resonance and photon trimomenta. The relation of the latter with the momentum transfer squared $Q^2$ is given by:
\begin{equation}
{\vec{k}}^2 = Q^2 + \frac{(W^2 - M^2)^2}{2(M^2 + W^2) + Q^2},
\label{breit}
\end{equation}
where $M$ is the nucleon mass, $W$ is the mass of the resonance and $Q^2~=
~{\vec{k}}^2- k_0^2$, $k_0$ being the photon energy.

\begin{figure}[h]

\includegraphics[width=3.5in]{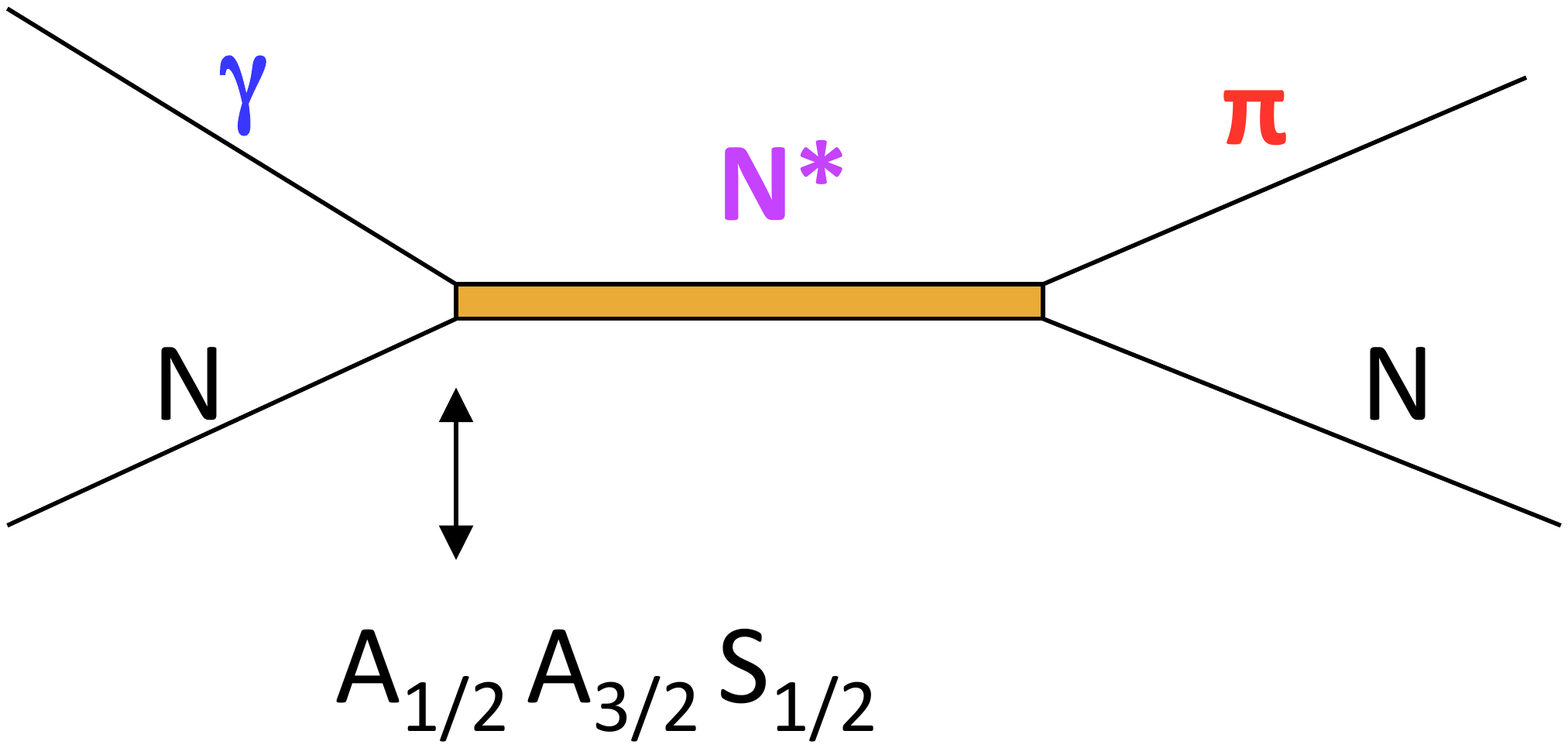}
\centering
\caption{ (Color online) The electromagnetic excitation of nucleon resonances. The decay can occur in more than one pion. }
\label{helamp}
\end{figure}

Furthermore, one has to consider that the helicity amplitudes extracted from the photoproduction contain also the sign of the $\pi N N^*$ vertex (see Fig.~\ref{helamp}). The theoretical helicity amplitudes are then  defined up to a common phase factor $\zeta$
\begin{equation}
A_{1/2,3/2}~=~\zeta~ \mathcal{A}_{1/2,3/2} ~~~~~~~~~~~S_{1/2}~=~\zeta~ \mathcal{S}_{1/2} .
\end{equation}
The factor $\zeta$ can be taken \cite{sg} in agreement with the choice of ref.~\cite{ki}, with the exception of the Roper resonance, in which case the sign is in agreement with the analysis performed in \cite{azn07}.

In the following Subsections the results for the photo- and electro-excitation of the baryon resonances calculated with the hCQM will be presented. 
For consistency reasons, in the calculations  the values of $W$ given by the model  are used instead of  the phenomenological ones.

The calculations regarding the photocouplings have already been published in \cite{aie}, those for the transverse excitation of the  negative parity resonances in \cite{aie2}, while a systematic study of all the electromagnetic excitations has been reported in \cite{sg}.

 It should be stressed that the results are obtained with a  parameter free calculation, that is they are {\bf predictions of the model}. 

\begin{table}
\centering
\caption[]{Photocouplings (in units $10^{-3}$ GeV$^{-1/2}$) predicted by the hCMQ \cite{sg} in comparison with PDG data \cite{pdg12} for proton excitation to N*-like resonances. The proton transitions to the S11(1650), D15(1675)  and D13(1700) resonances vanish in the $SU(6)$ limit.}
\vspace{15pt}
\label{photo}
\begin{tabular}{|c|rr|rr|}
\hline
& & & &   \\
 &$A_{1/2}^p $ & $A_{1/2}^p $ & $A_{3/2}^p$ & $A_{3/2}^p$  \\
 & & & &   \\
$Resonance$ & hCQM & PDG & hCQM & PDG \\
& & &  &  \\
\hline
& & &  & \\
P11(1440)&$88$ & $-65 \pm 4$  & &  \\
D13(1520)&$-66$ & $-24 \pm 9$ & $67$ & $ 166 \pm 5$  \\
S11(1535)&$109$ & $90 \pm 30$ &  & \\
S11(1650)&$69$ & $53 \pm 16$ &  & \\
D15(1675)&$1$ & $19 \pm 8$ & $2$ &$15 \pm 9$   \\
F15(1680)&$-35$ & $ -15 \pm 6$ & $24$ & $ 133 \pm 12$  \\
D13(1700)&$8$ & $ -18 \pm 13$ & $-11$ &$ -2  \pm 24$  \\
P11(1710)&$43$ &$ 9 \pm 22$ & & \\
P13(1720)&$94$ & $ 18 \pm 30 $ &$-17$ & $ -19 \pm 20$ \\
& & & & \\
\hline
\end{tabular}
\end{table}

\begin{table}
\centering
\caption[]{The same as in Table \ref{photo} but for the $N-\Delta$ excitation.}
\vspace{15pt}
\label{delta}
\begin{tabular}{|c|rr|rr|}
\hline
& & & &   \\
 &$A_{1/2}^p $ & $A_{1/2}^p $ & $A_{3/2}^p$ & $A_{3/2}^p$  \\
 & & & &   \\
$Resonance$ & hCQM & PDG & hCQM & PDG \\
& & &  &  \\
\hline
& & &  & \\
P33(1232)&$-97$ & $-135 \pm 6$ & $-169$ &$-250 \pm 8$  \\
S31(1620)&$30$ &$27 \pm 11$ &  & \\
D33(1700)&$81$ & $104 \pm 5$ &$70$ & $85 \pm 2$ \\
F35(1905)&$-17$ & $26 \pm 11$ &$-51$ & $-45 \pm 20$ \\
F37(1950)&$-28$ & $-76 \pm 12$ & $-35$ & $-97 \pm 10$ \\
& & & &  \\
\hline
\end{tabular}
\end{table}

\subsection{The photocouplings in the hCQM}

The resonances which have been considered are those which, according to the PDG classification \cite{pdg12}, have an electromagnetic decay with a three- or four- star status. This happens for twelve resonances, namely the positive parity $I=\frac{1}{2}$ states 
\begin{equation}
P11(1440), F15(1680), P11(1710),
\end{equation}
the negative parity $I=\frac{1}{2}$ states
\begin{equation}
D13(1520), S11(1535), S11(1650), D15(1675)
\end{equation}
 and the $I=\frac{3}{2}$ ones 
 \begin{equation}
 P33(1232), S31(1620), D33(1700), F35(1905), F37(1950).
 \end{equation}
 Besides these states, we  have considered also the two resonances D13(1700) and P13(1720), which are excited in an energy range particularly interesting for the phenomenological analysis.

\begin{table}[t]
\centering
\caption[]{ Photocouplings (in units $10^{-3}$ GeV$^{-1/2}$) predicted by the hCMQ \cite{sg} in comparison with PDG data \cite{pdg12}  and the recent Bonn-Gatchina analysis \cite{bn_n} for neutron excitation to N*-like resonances. }
\vspace{15pt}
\label{photo_n}
\begin{tabular}{|c|rrr|rrr|}
\hline
& & & & & &    \\
 &$A_{1/2}^n $ & $A_{1/2}^n $ & $A_{1/2}^n $ & $A_{3/2}^n$ & $A_{3/2}^n$ & $A_{3/2}^n$   \\
 & & & & & &    \\
$Resonance$ & hCQM & PDG & BnGa & hCQM  & PDG & BnGa\\
& & & & & &    \\
\hline
& & & & & &    \\
P11(1440)&$58$ & $40 \pm 10$  & $43 \pm 12  $& & & \\
D13(1520) &$-1$ & $-59 \pm 9$ & $-49 \pm 8$ & $-61$ & $ -139 \pm 11$ & $-113 \pm 12$ \\
S11(1535)&$-82$ & $-46 \pm 27$ & $-93 \pm 11$ & & & \\
S11(1650)&$-21$ & $-15 \pm 21$ & $-25 \pm 20$ & & &  \\
D15(1675)&$-37$ & $-43 \pm 12$ & $-60 \pm 7$ & $-51$ &$-58 \pm 13$ & $ -88 \pm 10$   \\
F15(1680)&$38$ & $ 29 \pm 10$ & $34 \pm 6$ & $15$ & $ -33 \pm 9$ & $-44 \pm 9$  \\
D13(1700)&$12$ & $ 0 \pm 50$ &  & $70$ &$ -3  \pm 44$  & \\
P11(1710)&$-22$ &$ -2 \pm 14$  & $-40 \pm 20 $ & & & \\
P13(1720)&$-48$ & $ 1 \pm 15 $ & $-80 \pm 50 $ &$4$ & $ -29 \pm 61$ & $-44 \pm 9$ \\
& & & & & &    \\
\hline
\end{tabular}
\end{table}

The proton and neutron photocouplings predicted by the hCQM \cite{aie} are reported in Tables \ref{photo}, \ref{delta} and \ref{photo_n} in comparison with the PDG data \cite{pdg12} and the analysis by the Bonn-Gatchina group \cite{bn_n}. The overall behaviour is fairly well reproduced, but in general there is a lack of strength. The proton transitions to the S11(1650), D15(1675) and D13(1700) resonances vanish exactly in absence of hyperfine mixing and are therefore entirely due to the $SU(6)$ violation.

 As already noted in Sec.~2.2,  the hyperfine interaction is responsible for a deformation of the $\Delta$ resonance and therefore the ratio of Eq.~(\ref{E2}) is different from zero. This ratio can also be expressed in terms of the helicity amplitudes
\begin{equation} \label{REM}
R_{EM}~=~- ~\frac{G_{E}}{G_{M}}~=~ ~\frac{\sqrt{3} ~ A_{1/2}~-~A_{3/2}}{\sqrt{3}~A_{1/2}~+~3 ~A_{3/2}};
\end{equation}
with the theoretical values reported in Table \ref{photo}, $R_{EM}$ turns out to be smaller than the experimental one. The point is that the E2 transition strength predicted by the hCQM is too low and a possible explanation of this result will be discussed later.

The results obtained with other calculations are qualitatively not much different \cite{aie,cr2} and this is because the various CQM models have the same $SU(6)$ structure in common.

 It should be reminded that in previous nonrelativistic calculations with h.o.\ wave functions \cite{cko}, it was necessary to assume a proton radius of the order of 0.5 fm in order to  ensure a vanishing $A_{1/2}^p$ for the resonances D13(1520) and F15(1680), whose peaks are absent in the forward photoproduction \cite{cko}. The proton radius calculated with the hCQM is actually 0.48 fm and this explains why the predictions of the hCQM do not differ too much from the other calculations. A too low proton radius is of course a problem if one wants to calculate the elastic form factors of the nucleon, but  for the description of the helicity amplitudes it is beneficial and, as we shall see later, it plays an important role in the discussion concerning the mechanisms which are missing in any CQM.

 There are now  many new analyses concerning the neutron helicity amplitudes (\cite{bn_n} and references quoted therein). In Table \ref{photo_n} we report also   the results of the  Bn-Ga analysis  \cite{bn_n}. The hCQM predictions are in fair agreement with these data, perhaps better than with the PDG ones.

\begin{figure}[h]

\includegraphics[width=2.3in]{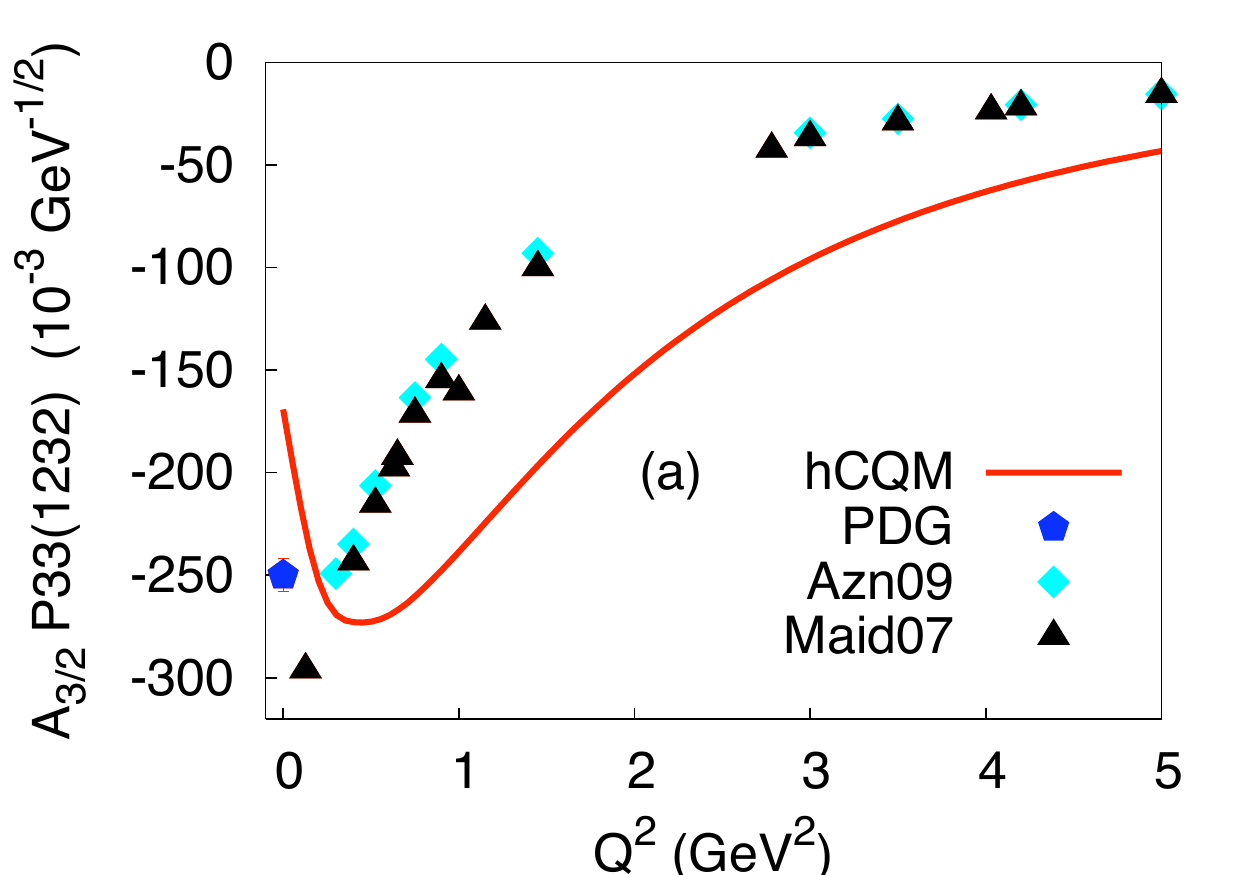}
\includegraphics[width=2.3in]{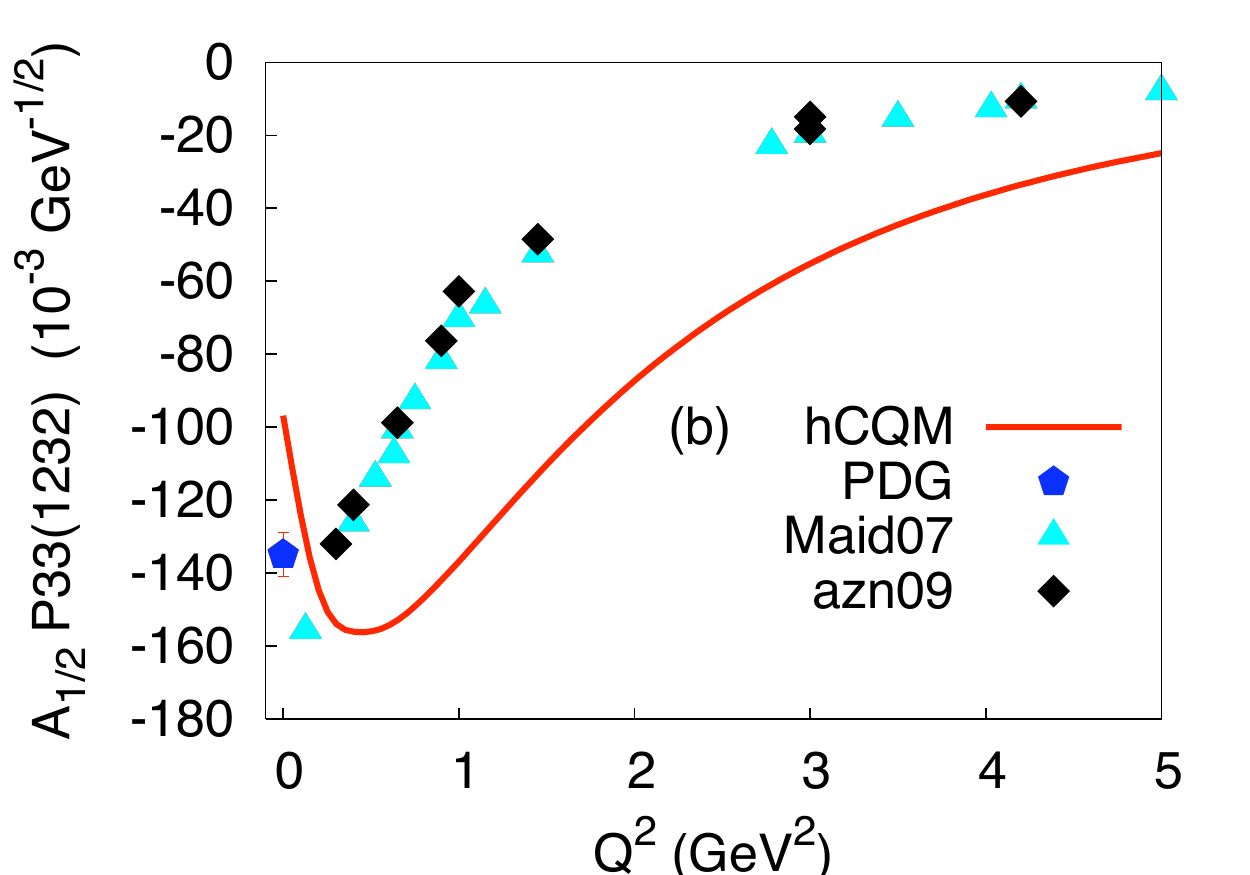}
\includegraphics[width=2.3in]{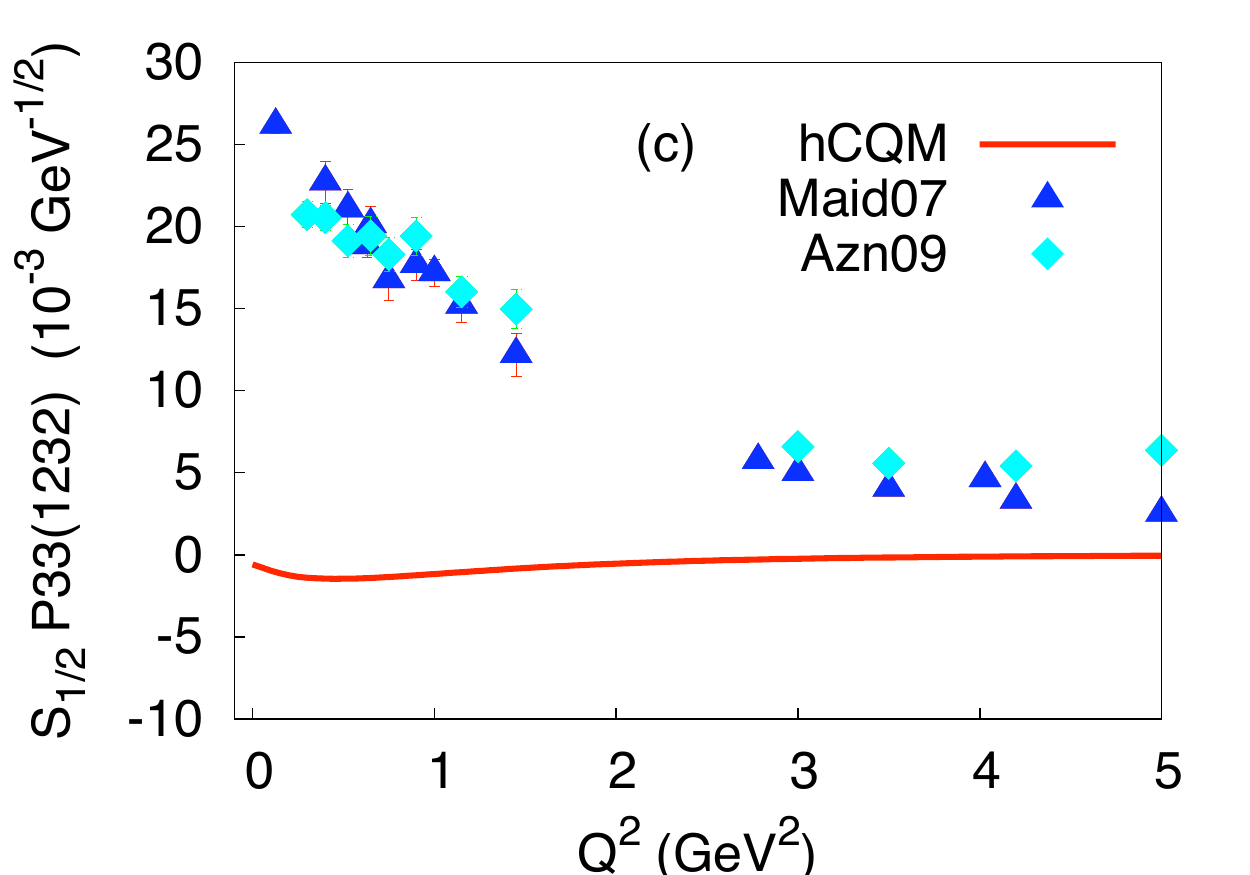}

\caption{(Color on line) The P33(1232) helicity amplitudes predicted by the hCQM (full curves)  $A_{3/2}$ (a), $A_{1/2}$ (b) and $S_{1/2}$ (c), in comparison with the data of ref.~\cite{azn09} and with  the Maid2007 analysis \cite{maid07} of the data by refs.~\cite{joo02} and \cite{lav04}. The PDG points \cite{pdg12} are also shown. The figure is taken from ref.~\cite{sg} (Copyright (2012) by the American Physical Society)}
\label{p33}
\end{figure}

\subsection{The helicity amplitudes in the hCQM}

As already mentioned, a systematic review of the hCQM predictions for the transverse and longitudinal helicity amplitudes and their  comparison with the experimental data is reported in ref.~\cite{sg}. Here we limit ourselves to some of the most important excitations.

\begin{figure}[h]

\includegraphics[width=2.3in]{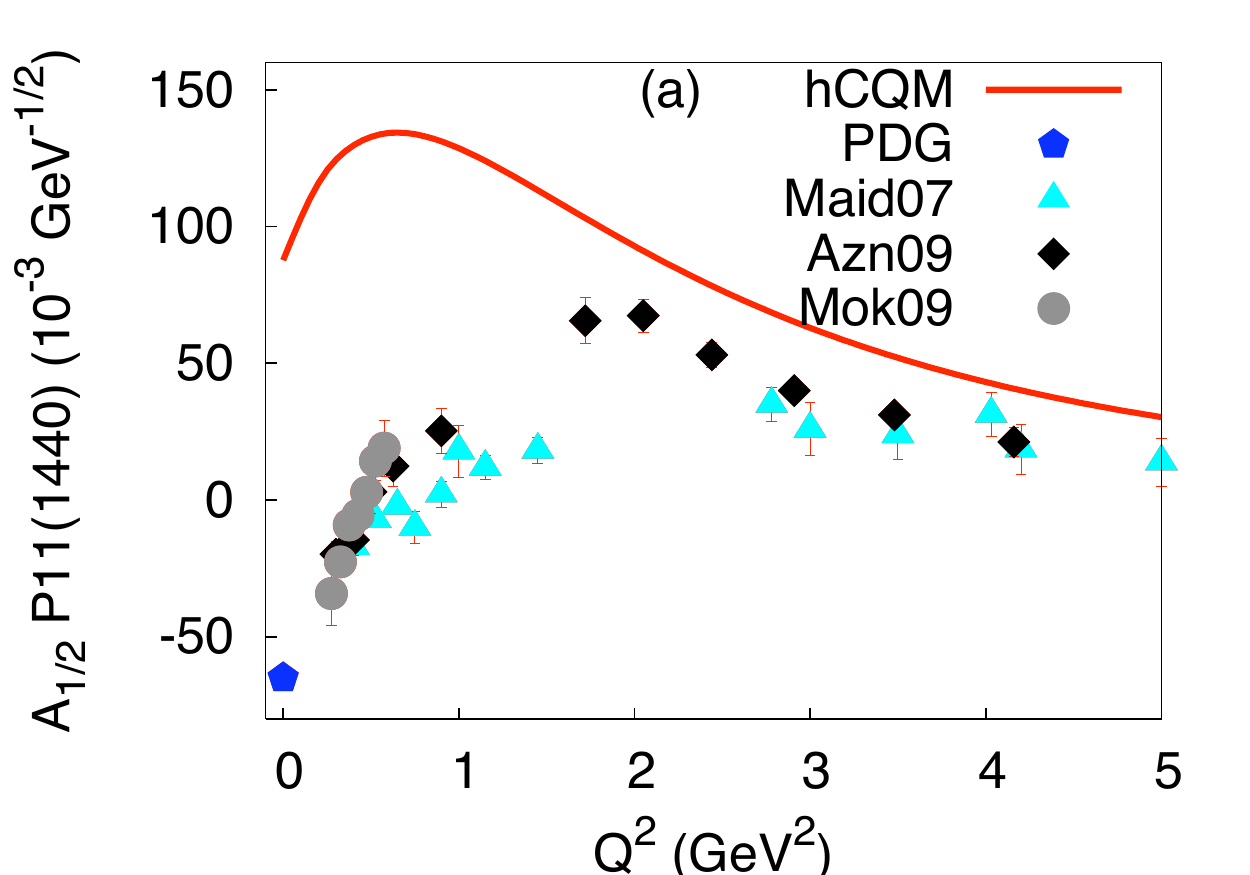}
\includegraphics[width=2.3in]{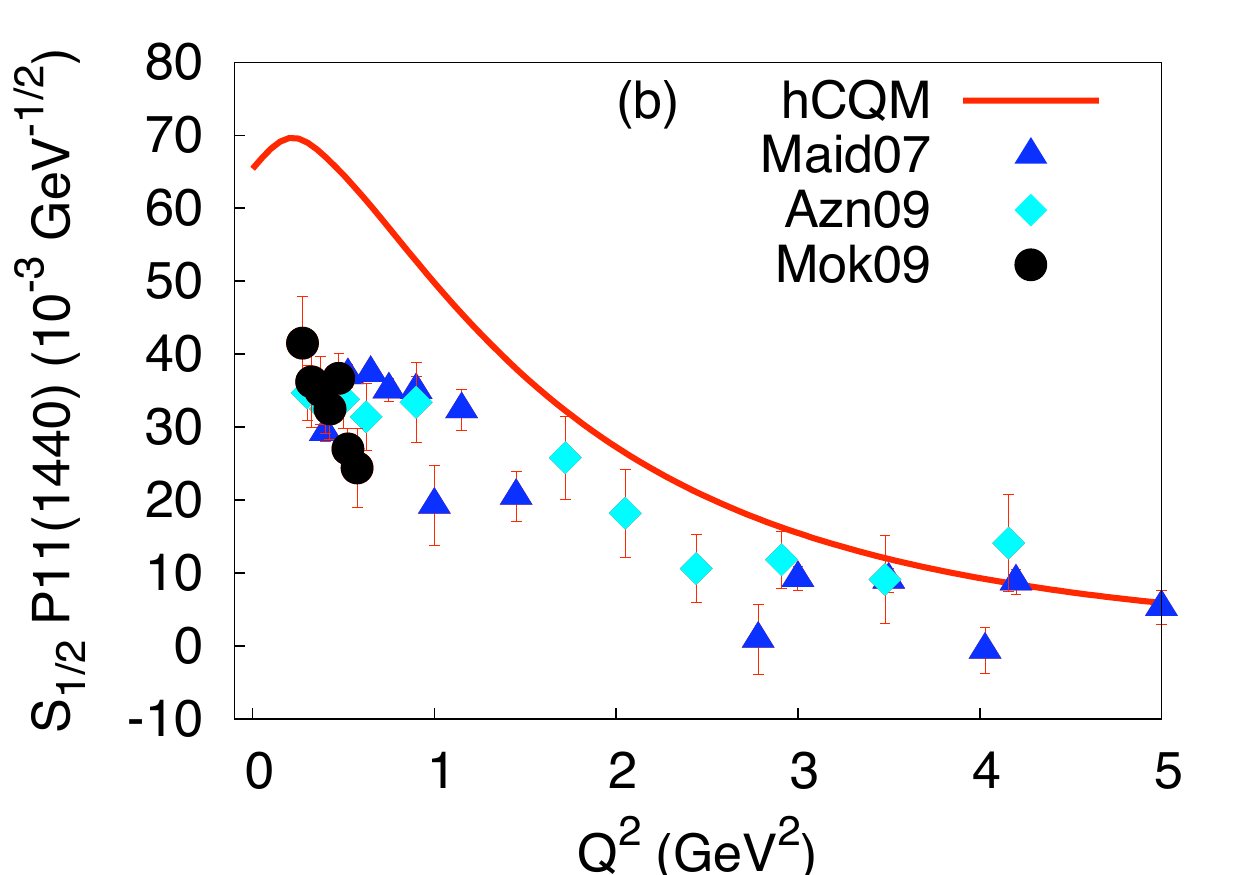}

\caption{(Color on line) The P11(1440) proton transverse (a) and longitudinal (b) helicity amplitudes predicted by the hCQM (full curves), in comparison with the data of refs.~\cite{vm09}, \cite{azn09} and the Maid2007 analysis \cite{maid07}  of the data by refs.~\cite{fro99},\cite{joo02},  \cite{lav04} and \cite{ung06}. The PDG point \cite{pdg12} is also shown. The figure is taken from ref.~\cite{sg} (Copyright (2012) by the American Physical Society).}
\label{p11}
\end{figure}

The transition amplitudes for the excitation of the P33(1232) resonance are given in Fig.~\ref{p33}.

The transverse  excitation to the $\Delta$ resonance has a lack of strength at low $Q^2$, a feature in common with all CQM calculations. The medium-high  $Q^2$ behavior is decreasing too slowly with respect to data, similarly to what happens for the nucleon elastic form factors \cite{mds,ff_07}. As we shall see later, the nonrelativistic calculations are improved by taking into account relativistic effects. Since the $\Delta$ resonance  and the nucleon are in the ground state $SU(6)$-configuration, we expect that their internal structures  have strong similarities and that a good description of the $N-\Delta$ transition from factors is possible only with a relativistic approach. Such feature is further supported by the fact that the transitions to the higher resonances are only slightly affected by relativistic 
effects \cite{mds}.

The Roper excitation is reported in Fig.~\ref{p11}. Because of the $\frac{1}{x}$ term in the hypercentral potential of Eq.~(\ref{H_hCQM}), the Roper resonance can be included  in the first resonance region, at variance with h.o.\ models, which predict it to be a 2 $\hbar \omega$ state.  There are problems in the low $Q^2$ region, but for the rest the agreement is interesting, specially if one remembers that the curves are predictions and the Roper has been often been considered a crucial state, non easily included into a constituent quark model description. In particular, the longitudinal excitation is quite different from zero \cite{ab_rop}, in agreement with the hCQM and at variance with the hybrid qqq-gluon model \cite{lb}. In the present model, the Roper is a hyperradial excitation of the nucleon.

\begin{figure}[h]
\includegraphics[width=2.3in]{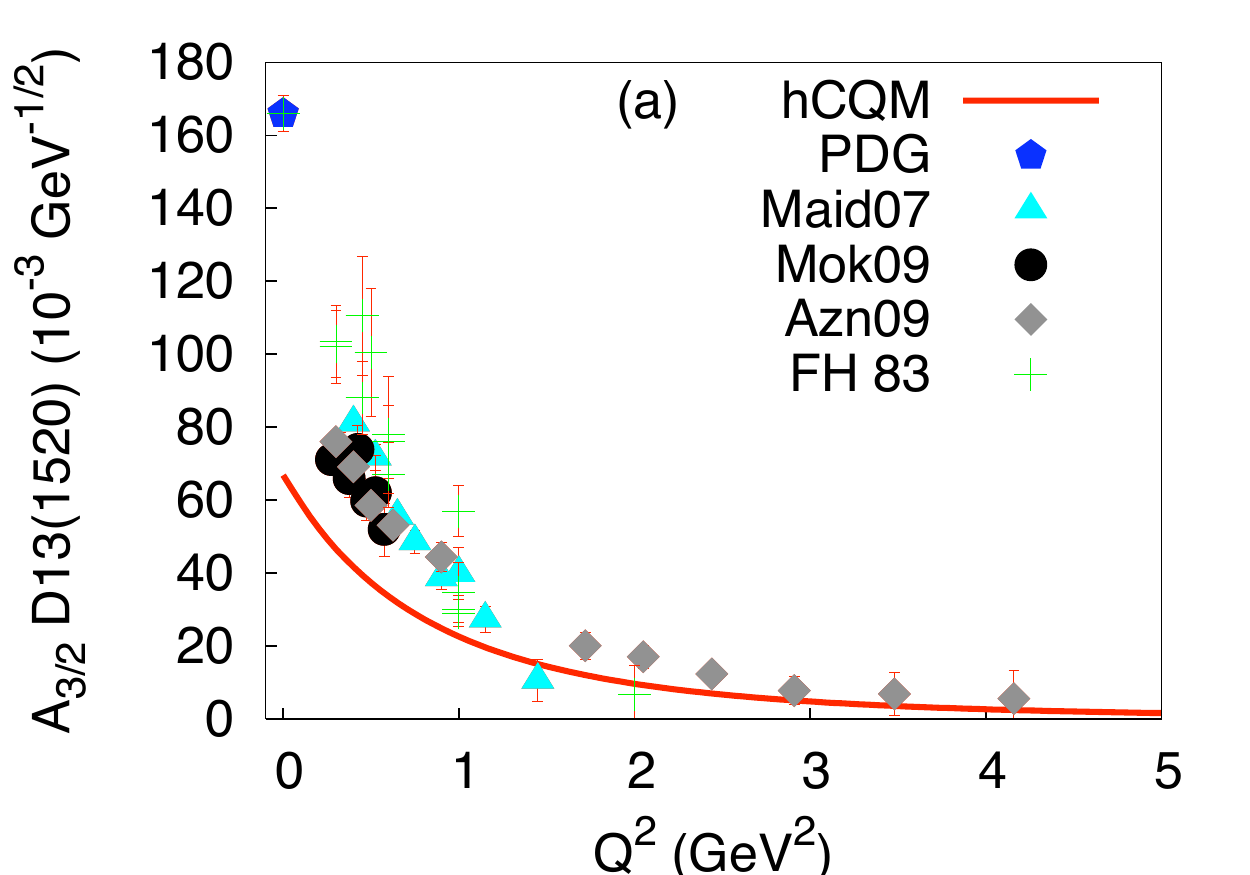} 
\includegraphics[width=2.3in]{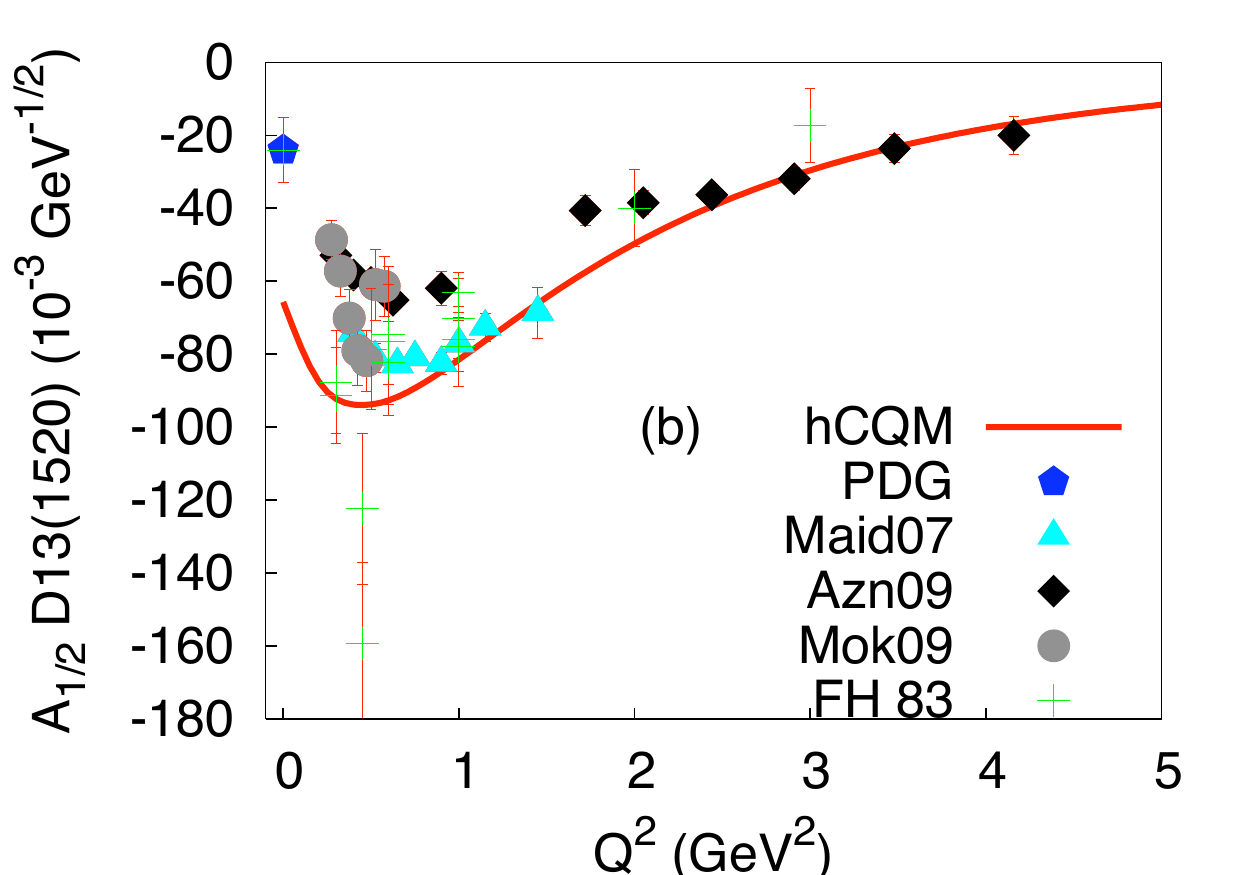}
\includegraphics[width=2.3in]{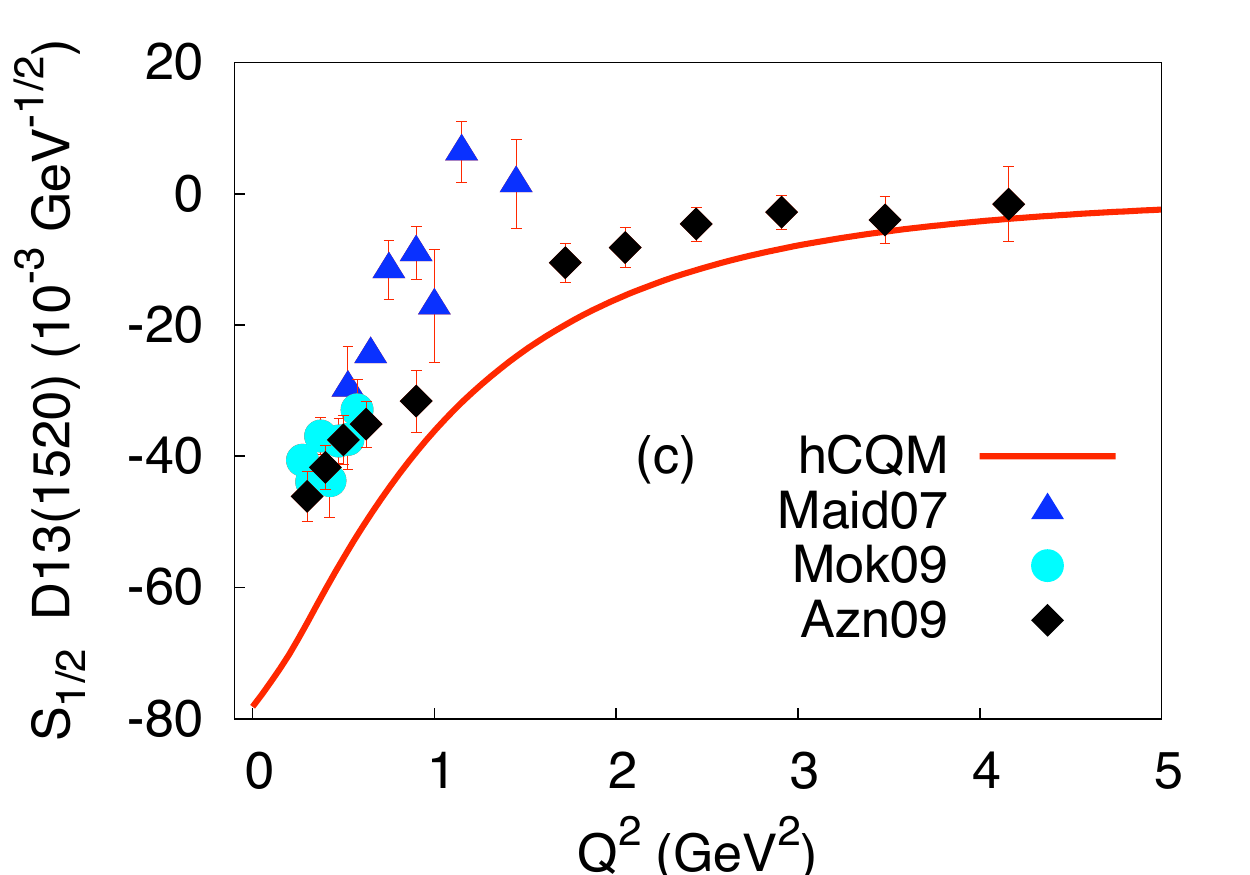}

\caption{(Color on line) The D13(1520)  proton helicity amplitudes predicted by the hCQM (full curves)  $A_{3/2}$ (a), $A_{1/2}$ (b) and $S_{1/2}$ (c), in comparison with the data of refs.~\cite{vm09}, \cite{azn09}, with the compilation reported in refs.~\cite{fh,ger} and the Maid2007 analysis \cite{maid07} of the data by refs.~\cite{joo02} and \cite{lav04}. The PDG points \cite{pdg12}are also shown. The figure is taken from ref.~\cite{sg} (Copyright (2012) by the American Physical Society)}
\label{d13}
\end{figure}

\begin{figure}[t]

\includegraphics[width=2.3in]{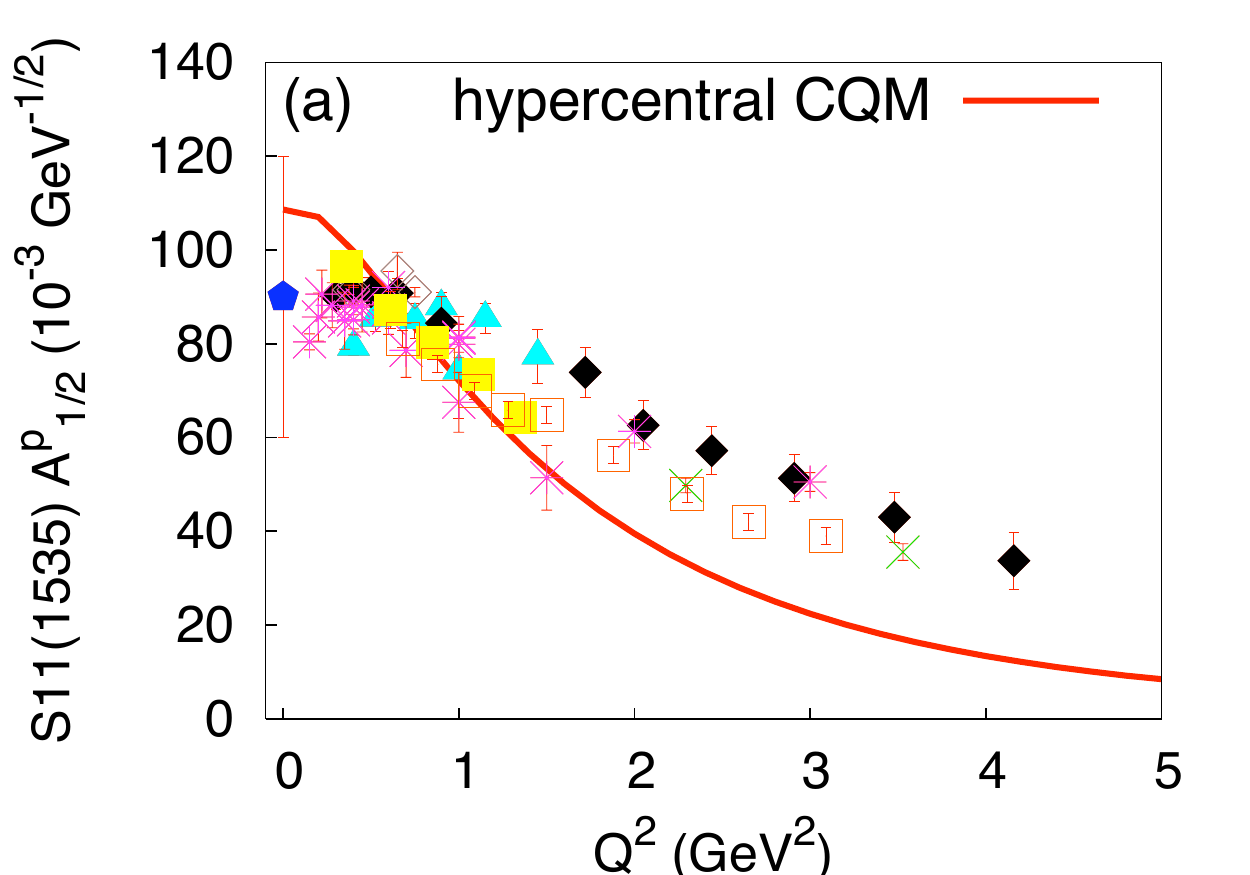}
\includegraphics[width=2.3in]{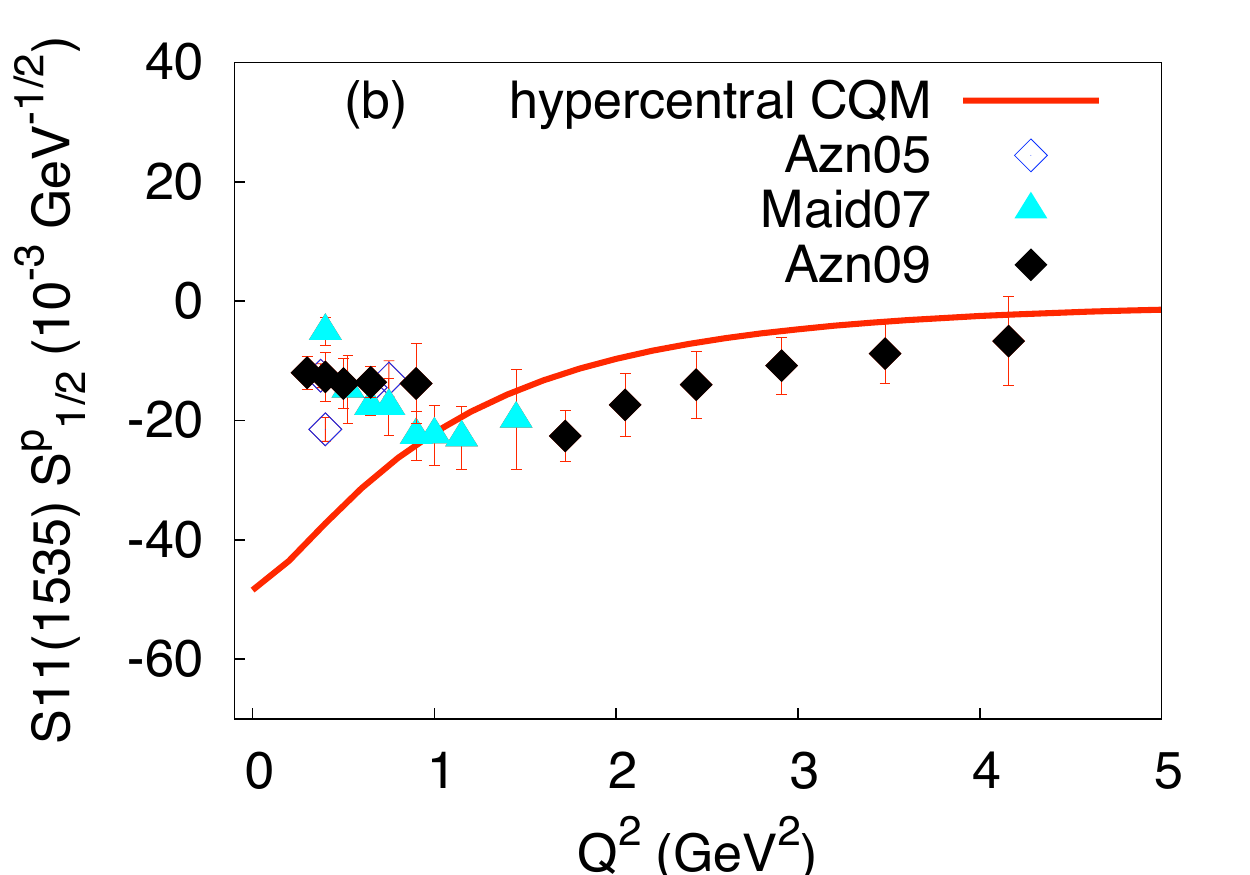}
\caption{(Color on line) The S11(1525) proton transverse (a) and longitudinal (b) helicity amplitudes predicted by the hCQM (full curve), in comparison with the data of refs.~\cite{azn05_1} (open diamonds),  \cite{azn09} (full diamonds), \cite{arm99} (crosses), \cite{den07} (open squares), \cite{thom01} (full squares), the Maid2007 analysis \cite{maid07} (full triangles) of the data by refs.~\cite{joo02} and the compilation of the Bonn-Mainz-DESY data of refs.~\cite{kru,bra,beck,breu} (stars), presented in \cite{thom01}. The PDG point \cite{pdg12} (pentagon) is also shown. The figure is taken from ref.~\cite{sg} (Copyright (2012) by the American Physical Society)}
\label{s11}

\end{figure}

We consider now  the excitations to some negative resonances \cite{aie2,sg}, namely the D13(1520) and the S11(1525) ones, reported in Figs.~\ref{d13} and \ref{s11}, respectively.

\begin{figure}[h]
\centering
\includegraphics[width=2.3in]{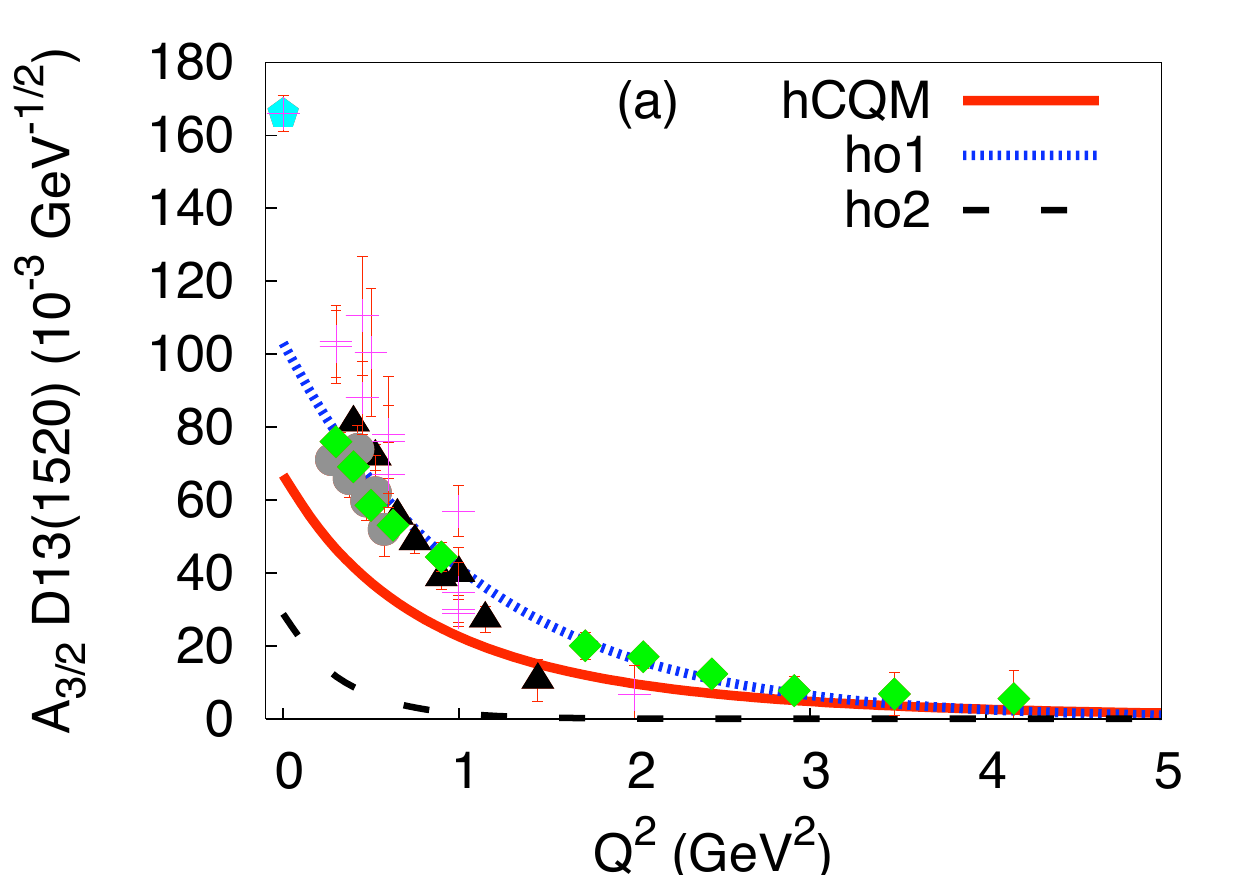}
\includegraphics[width=2.3in]{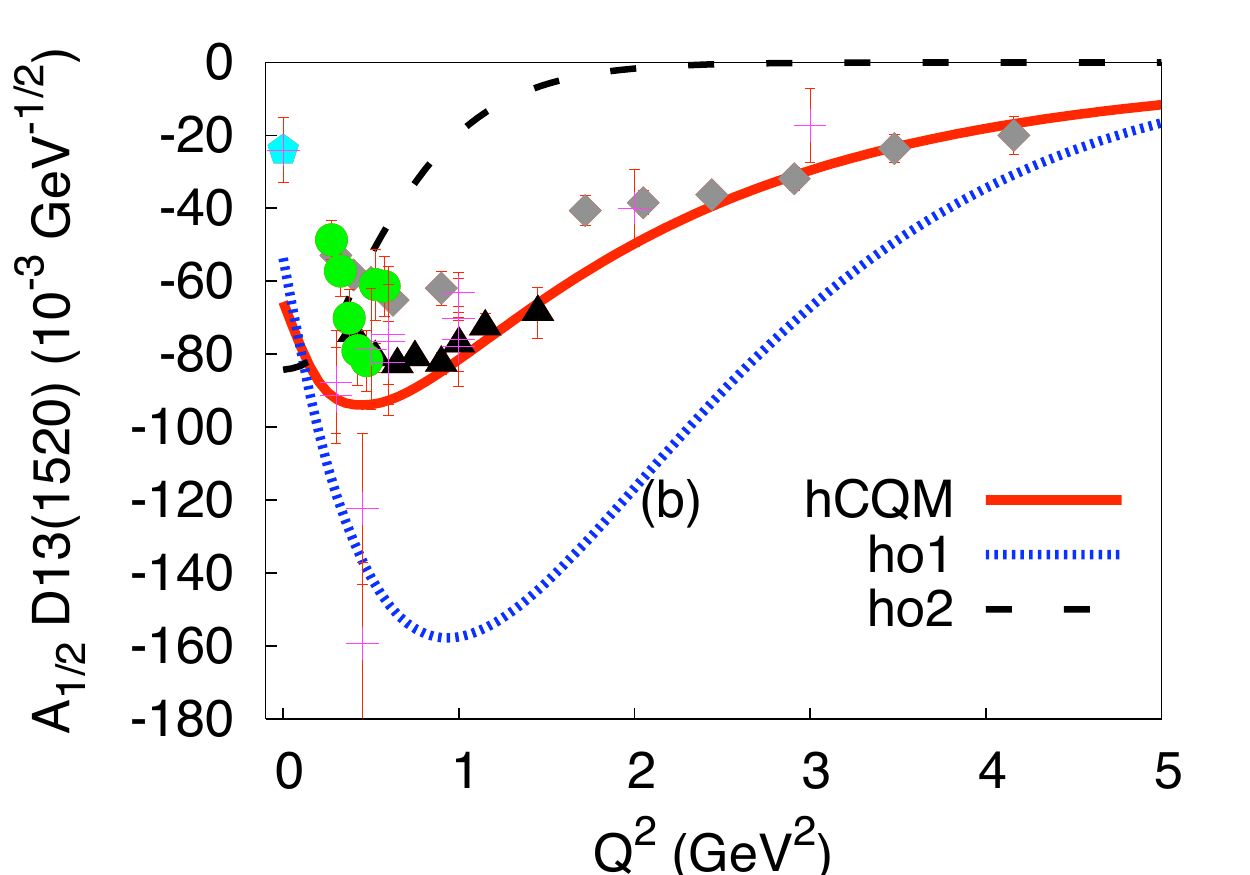}

\caption{(Color on line). The hCQM predictions for the transverse D13(1520) helicity amplitudes (full curve) in comparison with the h.o.\ results corresponding to two values of the proton radius: 0.5 fm (ho1) and 0.86 (ho2). The data are the same as in Fig.~\ref{d13}.}
\label{ho_hel}
\end{figure}

The agreement in the case of the
S11 is remarkable, specially if one considers that  the hCQM curve for the transverse transition has been published three years in
advance \cite{aie2} with respect to the recent TJNAF data \cite{azn09}, \cite{azn05_1}, \cite{den07},  \cite{thom01}. 

It is interesting to discuss the influence of the hyperfine mixing on the excitation of the resonances. Usually there is only a small difference between the values calculated with or without  hyperfine interaction. In some cases, however the excitation strength vanishes in the $SU(6)$ limit, as already mentioned in Table \ref{photo},  the non vanishing result is then entirely due to the hyperfine mixing of states. In the case of  the S11(1650) resonance, the resulting   transverse and longitudinal excitations have a relevant strength.

The three helicity amplitudes of  the D13(1700) resonance are again non zero because of the hyperfine mixing but the  excitation strength is very low. Also in the case of the transverse excitation of the D15(1675), the strength is given by the hyperfine mixing, while the longitudinal amplitude $S_{1/2}$ vanishes also in presence of a $SU(6)$ violation.

The results of the hCQM, as reported in Figs.~\ref{p33}, \ref{p11}, \ref{d13}, \ref{s11} and in ref.~\cite{sg} show some general features. First of all, there is a nice agreement with data at medium-high values of $Q^2$. This is mainly due to the presence of the $1/x$  term in the hCQM hamiltonian. In fact, the helicity amplitudes calculated with the analytical model of refs.~\cite{sig,sig2} have a $Q^2$ dependence  very similar to the complete hCQM and numerical values only slightly different. The wave functions of the analytical model are exactly those determined by the $1/x$ potential, which gives then the main contribution to the transition strength. 

Another important fact which helps in obtaining a  good behaviour of the hCQM helicity amplitudes is the smallness of the resulting proton radius. As already mentioned, the r.m.s.\ radius of the proton calculated with the hCQM wave functions corresponding  to the
parameters of Eq.~(\ref{par}) is $0.48~$fm, which is very near to the value necessary in order to fit
 the $D13$ photocoupling \cite{cko}.

Both these features, the presence of the hyperCoulomb term in the quark potential and the smallness of the proton radius, concur in order to obtain the results shown in the figures. This can be seen also looking at Fig.~\ref{ho_hel}, where the results of Fig.~(\ref{d13}) for the D13(1520) excitation are given in comparison with the h.o.\ curves  corresponding to two values of the proton radius, namely  the experimental one (about 0.86 fm) and the one fitted to the $A^p_{3/2}$ amplitude (about 0.5 fm). The curves with the correct proton radius are completely out of the experimental data because of the gaussian factor  $e^{-\frac{Q^2}{6 \alpha^2}}$, typical of the h.o.;  
$\alpha$ is the h.o.\ constant (see Sec.~2.2) and its value corresponding to the experimental proton radius is 0.229 GeV. In the case of the smaller radius 0.5 fm, with $\alpha$ = 0.41 GeV, the $A^p_{3/2}$ amplitude is of course well reproduced  but the h.o.\ curve for the $A^p_{1/2}$ amplitude is again far from the data. For this reason, we expect that the hCQM should be a good starting point for the description of the non perturbative components of the parton distribution.

There are in general discrepancies at low $Q^2$, displaying a  lack of strength which is typical of  all CQMs. Nevertheless, in many cases the $A^p_{1/2}$ amplitudes are better reproduced than the $A^p_{3/2}$ ones.

These shortcomings of the hCQM results could be ascribed to the non-relativistic character  of
the model. In fact, the electromagnetic excitation leads to a recoil of both the nucleon and the resonance and the effect is expected to increase with the momentum transfer $Q^{2}$, while the wave functions are calculated in the rest frame of each three-quark system. As it will be discussed in Sec.~6.2, it is possible to apply Lorentz boosts in order to bring the nucleon and the resonance to a common Breit frame, but this relativistic corrections produce only a slight modification of the hCQM results \cite{mds2}.

There is a consensus on the fact that the  missing strength at low $Q^2$ is  due to the lack of
quark-antiquark effects \cite{aie2}, probably important in the outer region of the nucleon. This statement receives a strong support by the explicit calculations of the meson cloud contributions to the helicity amplitudes performed in the framework of dynamical models (see ref.~\cite{dmt,sato} and references therein). 

\begin{figure}[h]

\includegraphics[width=5in]{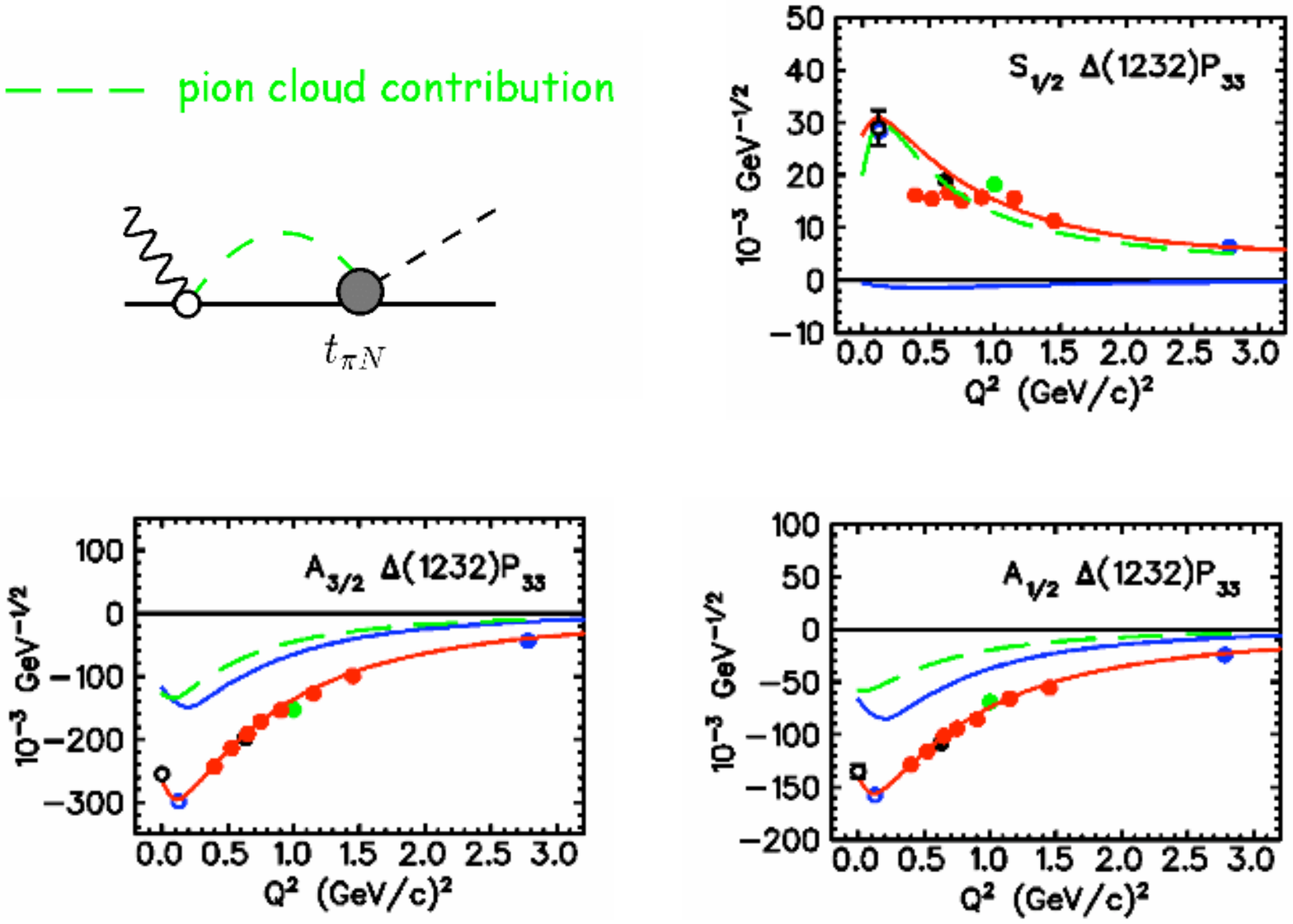}
\caption{(Color on line). The pion cloud mechanism considered in the DMT model \cite{dmt} (upper left). The remaining figures report the hCQM predictions for the $N-\Delta$ excitation (full curves) in comparison with the meson cloud contribution (dashed curves) and the results of the MAID fit \cite{maid99} (full points and the fitted curves passing through them) \cite{ts03}. }
\label{dmt}
\end{figure}

In particular the Dubna-Taipei-Mainz (DMT) model introduces  the pion cloud contribution to the electromagnetic excitation according to the mechanism shown in the upper left part of  Fig.~(\ref{dmt}). In the same figure, the longitudinal $S_{1/2}$ and transverse $A_{3/2} $ and $A_{1/2} $ helicity amplitudes for the $N-\Delta$ are reported \cite{ts03}. The theoretical predictions of the hCQM (full curves) are compared with the results of the MAID fit \cite{maid99} and with the pion cloud contribution calculated by means of the DMT model (dashed curve) \cite{dmt}. The hCQM results are much lower than the experimental data, but the pion cloud contribution gives relevant contributions just where the hCQM is lacking. This is particularly evident in the case of the longitudinal $S_{1/2}$ amplitude: the hCQM predicts an almost vanishing value while the pion alone   seems to be able to account for the data. Of course one cannot simply add the hCQM and pion contributions, since they are calculated in two different and inconsistent frameworks, but it is nevertheless interesting that the pion cloud seems to contribute systematically where the hCQM is lacking, as it can be seen also for the helicity amplitudes of many other resonances \cite{ts03}.

In this way the emerging picture in connection with the electromagnetic  excitation of the nucleon resonances is that of a small confinement zone of about $0.5$ fm surrounded by a  quark-antiquark (or meson) cloud.  The calculations of the meson cloud performed with the DMT model shows that this picture seems to be reasonable, but the problem is how to include in a consistent way the quark-antiquark pair creation mechanisms in the framework of the CQM. This goal can be achieved by unquenching the quark model and, as already mentioned earlier, an important improvement has been achieved by a recent work \cite{sb1,bs,sb2}. We shall come back on this point in the discussion.

\section{The elastic form factors of the nucleon}

\subsection{Introductory remarks}

An important aspect of the quark model predictions concerning the elastic form factors is their $Q^2$ dependence, which is strictly related to the form of the quark wave functions and then of the quark potential. 

We can start studying the nucleon charge form factor in absence of the hyperfine mixing. The nucleon state (see App. B) can be written as
\begin{equation}
|N \rangle  = \psi_{00}(x) \Omega_{[0]00}\frac{1}{\sqrt{2}} (\chi_{MS} \phi_{MS} + \chi_{MA} \psi_{MA}),
\end{equation}
where $\Omega_{[0]00} = 1/(4 \pi)^2 4/\sqrt{\pi}$. 

The nucleon charge form factor is given by the matrix element of the charge density operator of Eq.~(\ref{q_curr})
\begin{equation}
G^N_E(Q^2)~ = ~\langle N| 3 \frac{1+3\tau_0 (3)}{6}e^{-i a k\lambda_z} |N \rangle ,
\end{equation}
where $\tau_0(3)$ is the third component of the isospin operator of the third quark, $a=\sqrt{\frac{2}{3}}$ and $Q^2=k^2$ in the Breit system, according to Eq.~(\ref{breit}). Introducing the hyperspherical coordinates and performing the integrals over the angle and hyperangle variables, one gets
\begin{equation}
G^N_E(Q^2)~ = ~\frac{1+\tau_0}{2} F(k),
\end{equation}
where $\tau_0$ is the third component of the nucleon isospin and 
\begin{equation}
F(k)=\frac{8}{a^2 k^2} \int_0^\infty dx~ x^3 ~\psi_{00}(x)^2 J_2(a k x),
\label{hff}
\end{equation}
$J_2(z)$ being a Bessel function of integer order 2.
Eq.~(\ref{hff}) can be inverted, obtaining an expression of the wave function in terms of the form factor
\begin{equation}
  \psi_{00}(x)^2=\frac{a^4}{8 x^2} \int_0^\infty dk~ k^3 ~F(k)~ J_2(a k x).
\label{hwf}
\end{equation}
In this way, starting from any given form factor it is possible to obtain the appropriate wave function $ \psi_{00}(x)^2$  and then also the potential for which $ \psi_{00}(x)$ is the ground state. Assuming the dipole form factor $1/(1+b^2 k^2)$, the resulting potential is given by
\begin{equation}
V_{dip}(x)= \frac{a^2}{2 b^2}[1- \frac{1}{2}(\frac{K_0(y)}{K_1(y)})^2- 4 \frac{K_0(y)}{y K_1(y)}-\frac{6}{y^2}],
\label{vdip}
\end{equation}
where $y=ax/b$ and $K_0(y),K_1(y)$  are the modified Bessel function of the second kind. It is interesting to note that for large values of y, the potential assumes the form
\begin{equation}
V_{dip}(x) \rightarrow \frac{a^2}{4 b^2}[1- \frac{7}{y}-  \frac{9}{y^2}],
\label{vdip_2}
\end{equation}
that is there is no confinement.

The nucleon charge form factor turns out to be proportional to the charge and therefore it is zero for the neutron. This happens as long as the space part of the state is completely symmetric. Because of the hyperfine interaction, also the state state $|N ^2S_{1/2} \rangle _M$, having mixed space symmetry, contributes to the nucleon (see Eq.~(\ref{nucl})), thereby generating a non-zero neutron form factor \cite{iks}.

In h.o.\ models the ground state wave function is given by a gaussian  $e^{-\frac{\alpha^2 x^2}{2 }}$ which leads to the form factor $e^{-\frac{k^2}{6 \alpha^2}}$, where $\alpha$ is the h.o.\ constant (see Sec.~2.2). Because of the hyperfine mixing, the nucleon state is given by a superposition of $SU(6)$ configurations of Eq.~(\ref{nucl}), but the dominant behaviour of the form factor is in any case  a gaussian, which is too strongly damped with respect to the experimental data.

In the case of the purely hypercoulomb potential, the ground state wave function is given by an exponential function $e^{-gx}$ and the corresponding form factor is
\begin{equation}
G^p_E(Q^2)|_{hC} ~ = ~\frac{1}{(1+\frac{k^2}{6 g^2})^{7/2}}.
\end{equation}
 
 The use of the hCQM with the confinement potential modifies this power-law behaviour, however, even taking into account  the hyperfine mixing, the resulting form factor has a more realistic $Q^2$ behaviour with respect to the h.o.\ model. 
 
 As for the other nucleon form factors,  in the $SU(6)$ limit, there is perfect scaling, similarly to the dipole fit, in the sense that
 \begin{equation}
 G^p_E(Q^2)|_{hCQM} ~ = ~\frac{1}{\mu_p}  G^p_M(Q^2)|_{hCQM} ~ = ~\frac{1}{\mu_n}  G^n_M(Q^2)|_{hCQM} 
 \label{scal}
 \end{equation}
 and
 \begin{equation}
 G^n_E(Q^2)|_{hCQM} ~ = ~0.
  \end{equation}
 
\begin{figure}[h]
\includegraphics[width=6in]{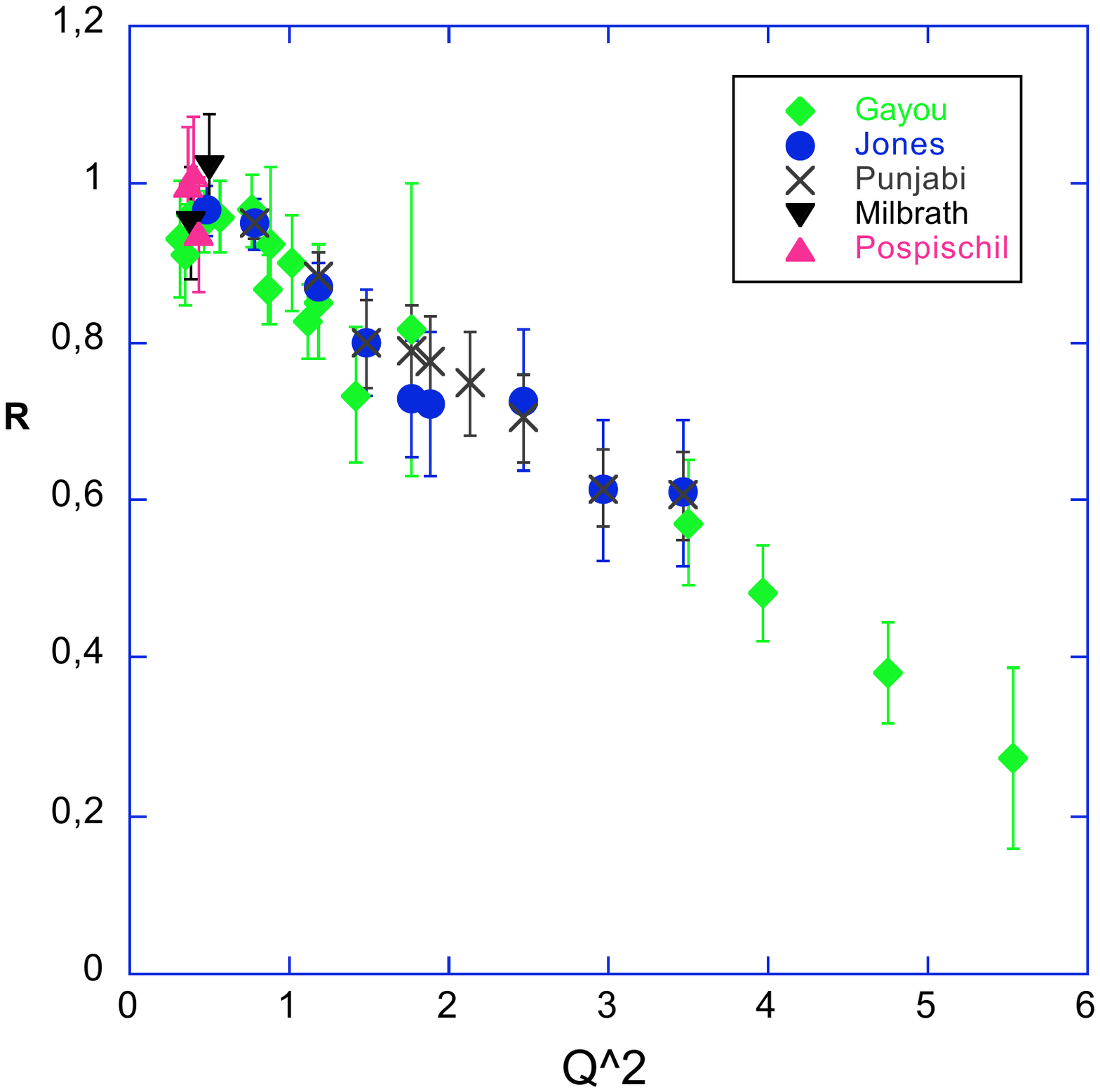}
\caption{(Color on line) The ratio $R=\mu_p \frac{G^p_E(Q^2)}{G^p_M(Q^2)}$ as a function of the mometum transfer $Q^2$. The data are taken from \cite{milb,jones00,gayou01,posp,gayou02,pun}.}
\label{rap07}
\end{figure}

 The hyperfine mixing modifies scarcely the above results, with the already mentioned difference of producing a non zero neutron charge form factor,
 
Comparing the predictions of the hCQM with the experimental data of the nucleon elastic form factor, one is faced with serious discrepancies, which may be due to  two important issues, namely the non-relativistic character of the model and the smallness of the proton radius. 

In the last decade, new and important data on the proton elastic form factors have been obtained by polarization experiments at Jefferson Lab. In fact, if one performs elastic scattering of longitudinally polarized electrons
from unpolarized protons, the recoiling proton has both   longitudinal ($P_l$) and   transverse ($P_t$) polarization  with respect
to the momentum transfer in the scattering plane \cite{pol}. From the ratio $P_t/P_l$, one obtains directly the ratio of the proton electric and magnetic form factors
\begin{equation}
R~ = ~\mu_p \frac{G^p_E(Q^2)}{G^p_M(Q^2)}~ = ~-\mu_p \frac{P_t}{P_l} \frac{E_e+E_e'}{2M_p}
 tan \frac{\theta_e}{2},
 \label{ratio}
\end{equation}
where $\mu_p$ is the proton magnetic moment, $E_e$ is the energy of the incident electron, $E'_e$ is the energy of the scattered electron, $\theta_e$
 is the electron scattering angle and $M_p$ is the proton mass.

At variance with the dipole fit, this ratio is deviating strongly from unity and   it is continuously decreasing towards zero, as shown in Fig.~\ref{rap07}. The measurements have been quite recently extended  up to $Q^2 = 8.5$ GeV$^2$ \cite{puck10}, but they will be discussed later, in comparison with a reanalysis \cite{puck12} of the last three points of ref.~\cite{gayou02}.

 The interesting issue is the existence of a dip in the form factor, as it is suggested by the almost linear decrease of the ratio R.
 Of course, this unexpected result has triggered  many theoretical calculations based on quark models attempting to explain the strong deviation of R from unity and discussing the possibility of a zero in the electric form factor of the proton.

It is interesting to remind, at least for historical purposes, that the idea of a dip in the electric form factor of the proton $G_E^p(Q^2)$ has been considered already at the early stage of the quark model \cite{mo_er}. The problem was that of reconciling the Pauli principle with the completely symmetric character of the three-quark wave function in the $\Delta$ resonance: in absence of the colour degrees of freedom, the only way out was a completely antisymmetric space function, which can be obtained introducing a factor $(r_{12}^2-r_{13}^2)(r_{23}^2-r_{21}^2)(r_{31}^2-r_{32}^2)$ in the wave function.  In general, in presence of a completely antisymmetric space wave function, it is easy to show \cite{mo_er} that the charge density in the origin $\rho(0)$ is zero and therefore $\int dq q^2 G_E^p(q) =0$,  which means that the form factor $G_E^p$ has a dip with a position strongly dependent on the model form factor. This idea has not been further considered since a ground state wave function with nodes is barely acceptable and the introduction of colour made it possible to have completely symmetric space functions.

Actually, there is a calculation, prior to the recent R data, which has predicted a dip in $G_E^p(Q^2)$. It has been shown \cite{holz96} that the simple Skyrme soliton model, with vector meson corrections, leads to a form factor with a dip at $Q^2~ \sim 1$ GeV$^2$ and that, after boosting the initial and final proton state to the Breit system, the zero is shifted up to $10$ GeV$^2$, a value highly compatible with the data of Fig.~\ref{rap07}. In later versions of the calculation the zero was pushed to $\sim 16$ GeV$^2$ \cite{holz00}.

Furthermore, in ref.~\cite{fjm}, a calculation of the electromagnetic proton form factors using a relativistic light-cone constituent quark model has been performed. The plot of R derived from their results exhibits a strong
decrease with $Q^2$ that is due to a zero in the electric form factor at  $6$ GeV$^2$ \cite{mill}.

As a final remark, we mention the semiphenomenological fit of the nucleon form factors performed in ref.~\cite{ijl}. The parametrized formula contains an intrinsic contribution, that now can be ascribed to the quark core, and a vector meson cloud. For the charge form factor of the proton, the data were available only up to $3$ GeV$^2$, but if one extrapolates the fitted formula to higher momentum transfer, one is faced with the remarkable fact of a dip  at about $9$ GeV$^2$. However, in ref.~\cite{ijl}, the intrinsic (quark core) contribution  was not considered for the isovector Pauli form factor $F^V_2$. Its  inclusion  in $F^V_2$ has been performed in a subsequent paper \cite{bi}, but in this case the zero, if present, is shifted beyond $10$ GeV$^2$.

Coming back to the hCQM, the ratio R, according to Eq.~(\ref{scal}), is identically 1 in the $SU(6)$ limit and it remains practically 1 also in presence of the hyperfine interaction. The hCQM described up to know is nonrelativistic and it is  worthwhile to analyse the effect of introducing relativity, which, as we shall see in the next subsection, is quite beneficial for the description of the decrease of the ratio R.

\subsection{Introducing relativity}

In order to describe the electromagnetic form factors both in the elastic and inelastic (helicity amplitudes) cases, one has to calculate matrix elements of the type
\begin{equation}
\langle  \Psi_F | J_\mu | \Psi_I  \rangle ,
\label{matr}
\end{equation}
where $ \Psi_I,  \Psi_F$ are the initial and final baryon states, respectively and $J_\mu$ is the quark current of Eq.~(\ref{qcurr}).
The wave functions of the three quark systems  are determined in the respective rest frames, but the calculation of the matrix elements of Eq.~(\ref{matr})  requires that the quark states be boosted to the Breit system. As long as the momentum transfer increases, one expects that  the effects of the relativistic transformations  are more important and therefore relativistic corrections must be introduced. In the case of the excitation to the baryon resonances, the state B has a mass which can be much greater than the nucleon one and in this case  the recoil is expected to be less relevant. However, for the ground states, nucleon and $\Delta$ in the $SU(6)$ limit, the relativistic corrections are unavoidable.

Relativistic corrections to the matrix elements can be introduced starting from the thee-quark wave function in momentum space in the rest frame $\psi(\vec{p}_\rho,\vec{p}_\lambda)$ and applying  the Lorentz boosts \cite{mds}
\begin{equation}
\Psi_I~ = ~\prod_{i=1}^3 B_i~u(p_i) ~\psi_I(\vec{p}_\rho,\vec{p}_\lambda),
\end{equation}
where $B_i (i=1,2,3)$ are the usual Dirac boost operators
that transform the quark spinors $u(p_i)$ from the nucleon rest frame to the Breit one; the quark momenta $p_i$ are in the baryon rest frame and the variables $\vec{p}_\rho,\vec{p}_\lambda$ are conjugate to the standard coordinates $\vec{\rho},\vec{\lambda}$, always in the rest frame. The final state is written in a similar way. 

In order to get the first order corrections  to the nonrelativistic approach, the matrix elements are expanded keeping the first order in the quark momenta, but any order in the momentum transfer \cite{mds}.

\begin{figure}[t]

\includegraphics[width=2.9in]{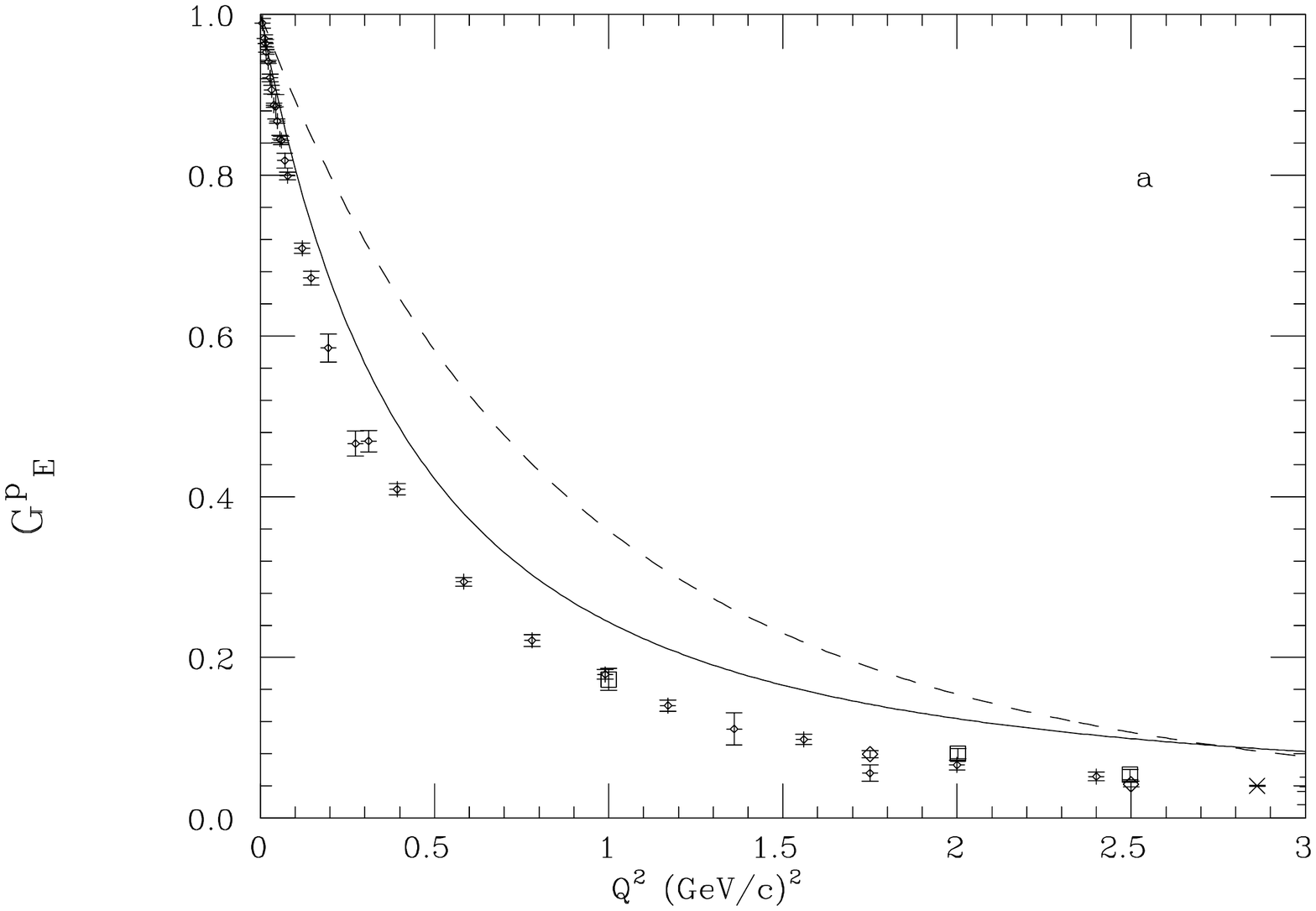}
\includegraphics[width=2.9in]{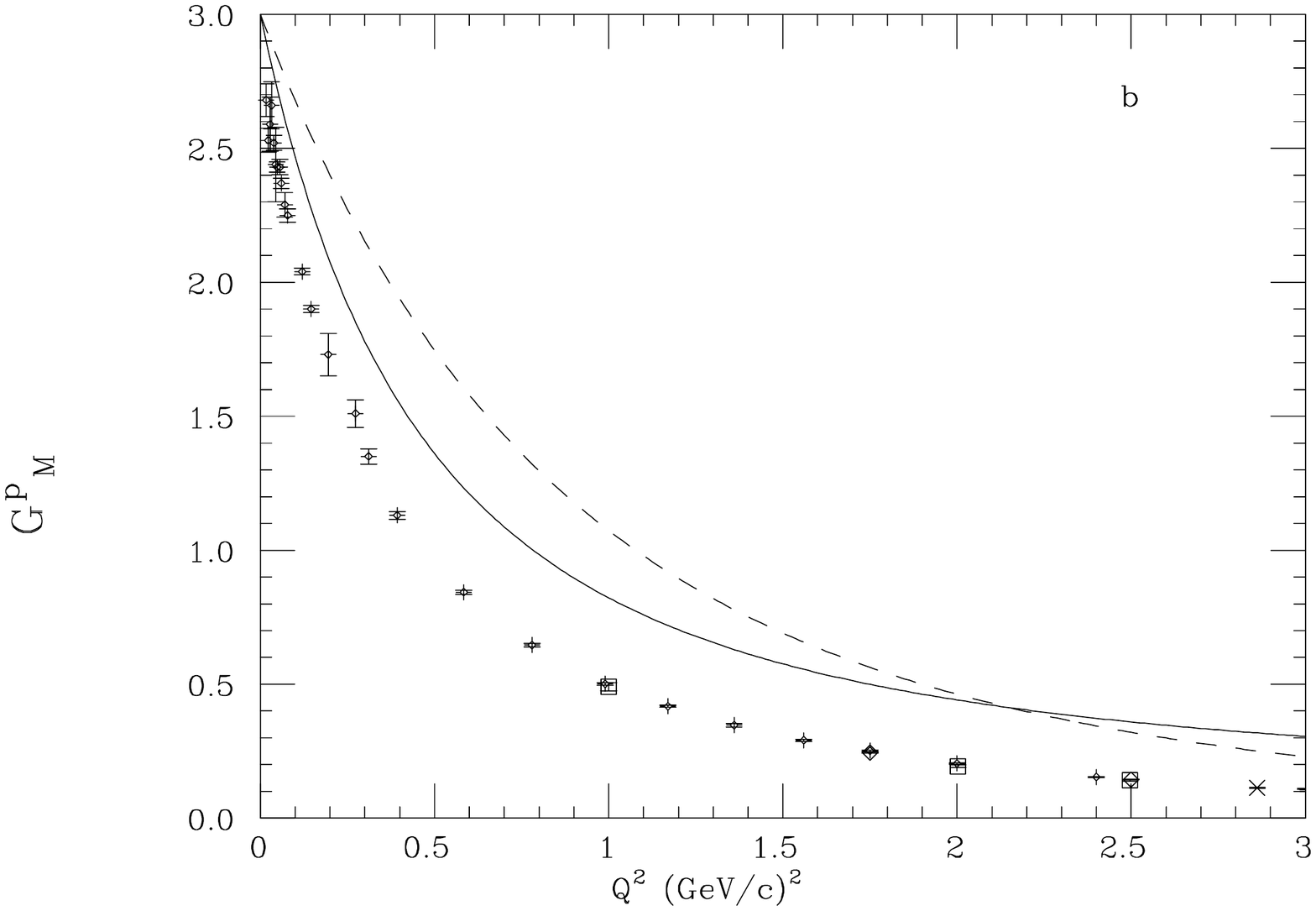}
\caption{The charge (a) and magnetic (b) form factors of the proton. The dashed curves are the predictions  of the nonrelativistic hCQM, the full curves are obtained taking into account the corrections given in Eqs.~(\ref{GE_b}) and  (\ref{GM_b}). The data are from a compilation of ref.~\cite{card_95}.  The figure is taken from ref.~\cite{rap} (Copyright (2000) by the American Physical Society).}
\label{ff_b}

\end{figure}

\begin{figure}[t]
\includegraphics[width=4in]{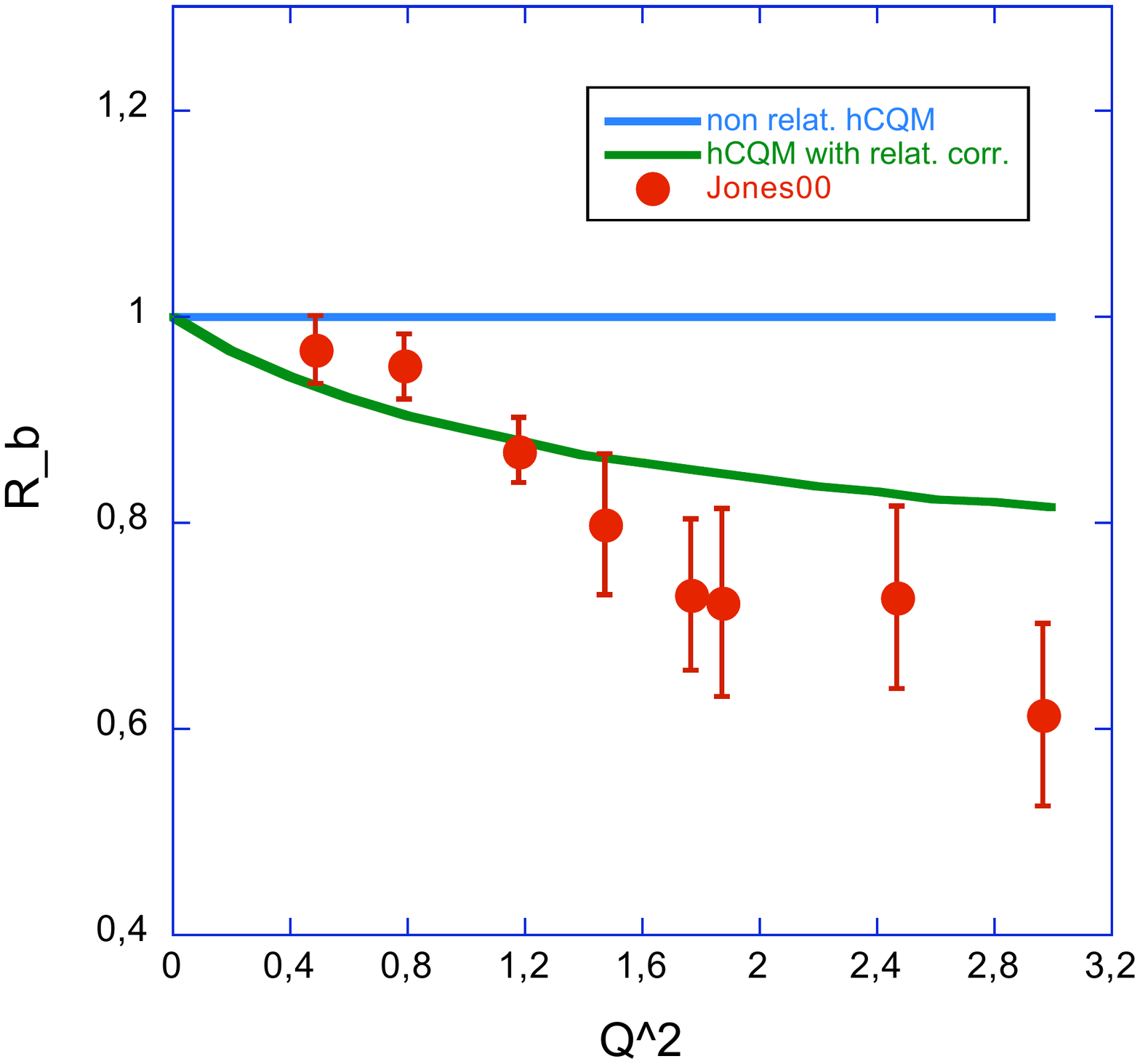}
\caption{The ratio R predicted by the hCQM  with the relativistic corrections \cite{mds2,rap}; the data are from ref.~\cite{jones00}.}
\label{rap_b}
\end{figure}

This approach can be applied to the calculation of the elastic form factors of the nucleon, obtaining  simple analytical forms \cite{mds}
\begin{equation}
G_E(Q^2)~ = ~ f_E G_E^{nr}(Q^2 \frac{M^2}{E^2}),
\label{GE_b}
\end{equation}
\begin{equation}
G_M(Q^2)~ = ~f_M G_E^{nr}(Q^2 \frac{M^2}{E^2}),
\label{GM_b}
\end{equation}
where $G_E^{nr}, G_M^{nr}$ are  the electric and magnetic form factors  predicted by the nonrelativistic hCQM, M is the nucleon mass, E its energy in the Breit frame and $f_E, f_M$ are known kinematic factors \cite{mds}:
\begin{equation}
f_E~ = ~\frac{E}{M} (t_S)^2 t_I, ~~~~f_M~ = ~\frac{E}{M} (t_S)^2 t_I \frac{g_\sigma}{2m},
\end{equation}
where
\begin{equation}
t_S = \frac{1}{Mm} (E \eta_S -\frac{M}{E} \frac{Q^2}{12}), ~~~~t_I=\frac{Mm}{E \eta_I + \frac{MQ^2}{6E}}, ~~~~ g_\sigma=\frac{2}{3} \eta_I+ \frac{4}{9}M,
\end{equation}
with
\begin{equation}
 \eta_S=(m^2+ \frac{M^2 Q^2}{36E^2})^{1/2}, ~~~~\eta_I=(m^2+ \frac{M^2Q^2}{9E^2})^{1/2}.
\end{equation}
One sees from Eqs.~(\ref{GE_b}) and (\ref{GM_b}) that, besides the multiplicative factors, the effect of relativity is to rescale the argument of the nonrelativistic form factors. It is just thanks to this mechanisms that in ref.~\cite{holz96}, the zero of the soliton form factor has been shifted to higher values.

In Fig.~\ref{ff_b} (a) and (b) the results for the electric and magnetic form factor of the proton are shown. The predictions of the hCQM are higher than the data, mainly because the proton radius is about 0.5 fm, however the relativistic corrections are beneficial, although there is still a discrepancy with respect to the data.

\begin{figure}[h]

\includegraphics[width=4in]{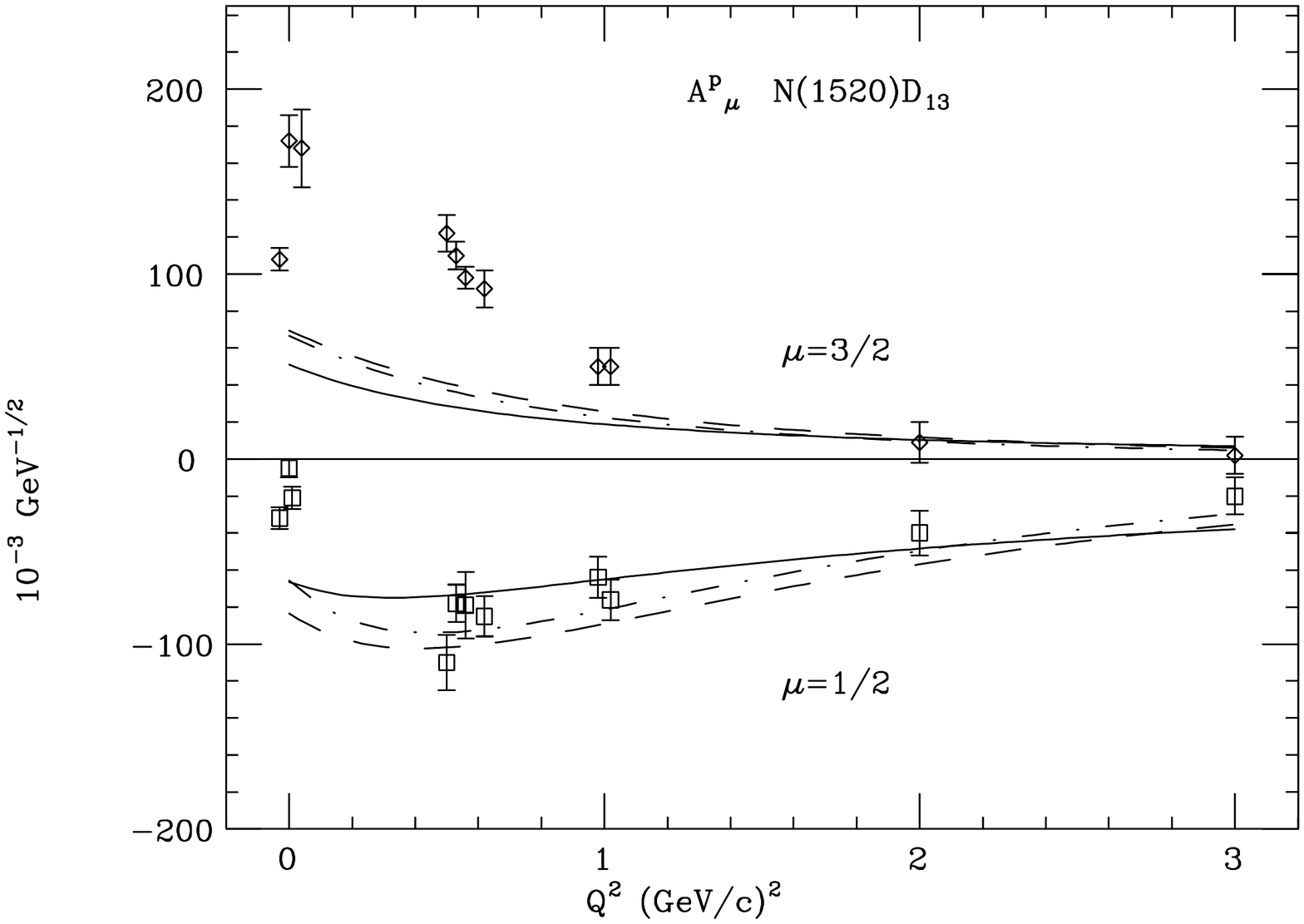}
\caption{The transverse  helicity amplitudes for the D13(1520) resonance. The full curves are the predictions of the hCQM, reported in \cite{aie2,sg}. The dot-dashed (dashed) curves are the results with the relativistic corrections in the Breit (equal velocity) frame (From ref.~\cite{mds2}).}
\label{helamp_b}
\end{figure}

The interesting point in connection with Eqs.~(\ref{GE_b}) and (\ref{GM_b})  is that the ratio R of Eq.~(\ref{ratio}) acquires a $Q^2$ dependence because of the factor $g_\sigma$ \cite{rap}. In Fig.~\ref{rap_b}, the ratio R obtained applying the boosts to the nonrelativistic hCQM is given by the full curve, which deviates from the horizontal line, that is the prediction of the nonrelativistic hCQM.

One can conclude from Fig.~\ref{ff_b} that the decrease of  R  can be ascribed to relativistic effects \cite{mg00,rap}. This can be understood if one considers that the nonrelativistic results for the form factors are obtained in the nucleon rest frame and that the charge and magnetization densities, which are equal in the rest frame, behave in a different manner when a Lorentz boost   to the Breit frame is applied.

A similar approach can be applied also to the helicity amplitudes predicted by the hCQM. In this case one obtains again simple analytical forms \cite{mds2}, which are omitted here for simplicity. The corrections have been calculated for the transverse excitation of various resonances, both in the Breit and in the equal velocity  frames, showing that in any case the deviation from the nonrelativistic predictions are very small. As an example, we show in Fig.~\ref{helamp_b} the results for the transverse helicity amplitudes of the D13(1520) resonance \cite{mds2}.
A similar situation is valid also for the S11(1535), S11(1650), S31(1620) and D33(1700) states \cite{mds2}.

\section{The relativistic hCQM and the elastic form factors}

\subsection{The relativistic formulation}

The inclusion of relativistic effects applying Lorentz boosts to the nonrelativistic wave functions is an approximation which is beneficial, as shown in the preceding Section, but certainly not sufficient. What is needed is a relativistically covariant theory, which can be achieved by means of the Bethe-Salpeter approach or considering one of the forms of relativistic dynamics introduced by Dirac \cite{dirac}.

A relativistic covariant theory is characterized by its behaviour under the transformations of the Poincar\'{e} group. To this end it is sufficient to consider the generators of the infinitesimal transformations $P_\mu (\mu=0,1,2,3)$ for the space-time translations and $J_i$ and $K_i  (i=1,2,3)$  for the space rotations and boosts, respectively. These generators  obey to the following commutation relations ($P_0=H$)

\begin{equation}
[P_\mu, P_\nu] = 0, ~~~[J_i, P_j]  = \epsilon_{ijk} P_k, ~~~[J_i, H]  =0,
\label{poi_1}
\end{equation}
\begin{equation}
[J_i, J_j]  = \epsilon_{ijk} J_k, ~~~[K_i, K_j]  = ~- \epsilon_{ijk} J_k,
\label{poi_2}
\end{equation}
\begin{equation}
[K_i, P_j]  = \delta_{ij} H, ~~~[K_i, H]=iP_i, ~~~ [K_i, J_k]  =  \epsilon_{ikj} K_j.
\label{poi_3}
\end{equation}

The problem of building a relativistic theory is equivalent to finding a solution to the Eqs.~(\ref{poi_1}),  (\ref{poi_2}) and (\ref{poi_3}). Some of the ten quantities $P_\mu, J_i, K_i$ are complicated by their dependence on the hamiltonian, while the remaining ones are  interaction free. As shown in ref.~\cite{dirac}, there are three different types of solutions, called instant, point and front form, differing in the type and numbers of interaction free quantities. In practical calculations all of them can be used, but here we shall concentrate ourselves on the Point Form (PF), for which the angular momentum operators $J_i$ and the boosts $K_i$ are interaction free \cite{klink}, while the four-momentum operator $P_\mu$ contain the interaction. 

The general three quark states are defined on the product space $H_1 \otimes H_2 \otimes H_3$ of the
one-particle spin-1/2, positive energy representations  $\it{H_i}=\it{L^2}
(\it{R}^3) \times \it{S}^{1/2}$ (i=1,2,3) of the Poincar\'{e} group \cite{pv,ff_10} and can be written
as
\begin{equation}
|p_1, p_2, p_3, \lambda_1,  \lambda_2, \lambda_3 \rangle  ~ = ~| p_1,\lambda_1  \rangle  | p_2, \lambda_2 \rangle  |p_3, \lambda_3  \rangle ,
\end{equation}
where $p_i$ is the four-momentum of the i-th quark and $\lambda_i$ the
z-projection of its spin. The states $| p_i,\lambda_i  \rangle $, i=1,2,3 are given by the Dirac spinors $u(p_i)$. We introduce the so called velocity states \cite{klink,pv} as
\begin{equation}
|v, k_1, k_2, k_3, \lambda_1,  \lambda_2, \lambda_3 \rangle  ~ = ~U_{B(v)} | k_1, k_2, k_3, \lambda_1,  \lambda_2, \lambda_3  \rangle _0 ,
\label{vel}
\end{equation}
where the suffix 0 means that  the quark three-momenta $\vec{k}_i$ satisfy the rest frame  condition $\sum_{i=1}^3 ~\vec{k}_i~= ~ 0$. With canonical boosts $U_{B(v)}$, the transformed quark tetramomenta 
 are given by $p_i = U_{B(v)} k_i$ and satisfy the relation
\begin{equation}
\sum_{i=1}^3 ~p_i^{\mu}~=
~\frac{P^{\mu}}{M} \sum_{i=1}^3 ~\epsilon (\vec{k_i}),
\end{equation}
where $\epsilon (\vec{k_i})$ is
the rest frame quark energy, $ P^{\mu}$ is the observed nucleon 4-momentum
and M its mass. In this way the conditions for the PF are satisfied \cite{klink,melde}, so that the rest frame quark momenta $\vec{k}_i$ and the quark spins undergo the same Wigner rotation when the boost is applied. A particularly appealing consequence is that the combination of the angular momentum states can be performed using the Clebsch-Gordan coefficients as in the nonrelativistic theory \cite{klink,klink2}.

The hCQM can now be formulated in a consistent relativistic approach. The model hamiltonian of Eq.~(\ref{H_hCQM}) is substituted with the mass operator M \cite{ff_07} 
\begin{equation}
M~=~ \sum_{i=1}^3 \sqrt{\vec{k}_i^2 + m^2}-\frac{\tau}{x} + \alpha x + H_{hyp},
\label{mass}
\end{equation}
the rest frame quark momenta $\vec{k}_i$ satisfy the condition $\sum_{i=1}^3 ~\vec{k}_i~= ~ 0$. In the calculation, the internal quark momenta are substituted with the Jacobi ones 
$\vec{p}_\rho = \frac{1}{\sqrt{2}}(\vec{k}_1-\vec{k}_2) $, $\vec{p}_\lambda = \frac{1}{\sqrt{6}} (\vec{k}_1+\vec{k}_2 -2 \vec{k}_3) $, which are compatible with the rest frame condition and undergo the same Wigner rotations as the internal momenta $\vec{k}_i$. 

The mass operator M can be considered as the result of  a Bakamjian-Thomas (BT) construction \cite{BT}, in which the interaction is introduced by adding to the free mass operator $M_0 =  \sum_{i=1}^3 \sqrt{\vec{k}_i^2 + m^2}$ an interaction $M_I$ in such a way that
$M_I$ commutes with the non interacting Lorentz generators $J_k, K_i$. The momentum operator $P_\mu$ can be written as $MV_\mu$, where the velocity operator $V_\mu$ commutes
with the total mass operator M, which  is independent of the non interacting four velocity \cite{kp}.

The mass operator M can be diagonalized and the resulting  baryon spectrum turns out to be not much different from the nonrelativistic one,  with only a slight modification of the fitted parameters $\alpha$ and $\tau$. One obtains in this way the three-quark wave functions $\psi(\vec{k}_\rho, \vec{k}_\lambda)$. The bound states in the rest frame are
\begin{equation}
 \psi(\vec{k}_\rho, \vec{k}_\lambda) | k_1, k_2, k_3, \lambda_1,  \lambda_2, \lambda_3 \rangle _0 ,
\end{equation}
which are then boosted to any reference frame according to the Eq.~(\ref{vel})
\begin{equation}
U_{B(v)} \psi(\vec{k}_\rho, \vec{k}_\lambda) | k_1, k_2, k_3, \lambda_1,  \lambda_2, \lambda_3 \rangle _0  ~ = ~      \psi(\vec{k}_\rho, \vec{k}_\lambda) |v,  k_1, k_2, k_3, \lambda_1,  \lambda_2, \lambda_3 \rangle ,
\label{bar}
\end{equation}
the baryon states in any frame are then given by superpositions of velocity states.

\subsection{The elastic form factors in the relativistic theory}

In order to describe the electromagnetic properties of the baryons, it is necessary to introduce the quark current. For the i-th quark the current is given by
\begin{equation}
\overline{u}(p_i) j_\mu(i) {u}(p_i') ~ = ~     \overline{u}(p_i) e \gamma _\mu(i) {u}(p_i'),
\end{equation}
($u(p_i)$ are the quark spinors), with which one can calculate the matrix elements of the total  quark current  $J_\mu$ in the space of the single quark free spinor states:
\begin{eqnarray}
\begin{array}{rcl}
\langle p_1, p_2, p_3, \lambda_1,  \lambda_2, \lambda_3| J_\mu |p_1', p_2', p_3', \lambda_1',  \lambda_2', \lambda_3'  \rangle  & ~ = ~& \\
& & \\
\sum_i \overline{u}(p_i) j_\mu(i) {u}(p_i') \overline{u}(p_j)  {u}(p_j') \delta(p_j-p_j') \overline{u}(p_k)  {u}
(p_k')  \delta(p_k-p_k'), & & 
\end{array}
\end{eqnarray}
the current conservation is ensured by applying the simple substitution  $J_\mu'= J_\mu - q_\mu (q \cdot J)/q^2$, where $q_\mu$ is the photon virtual momentum.

\begin{figure}[t]
\includegraphics[width=2.5in]{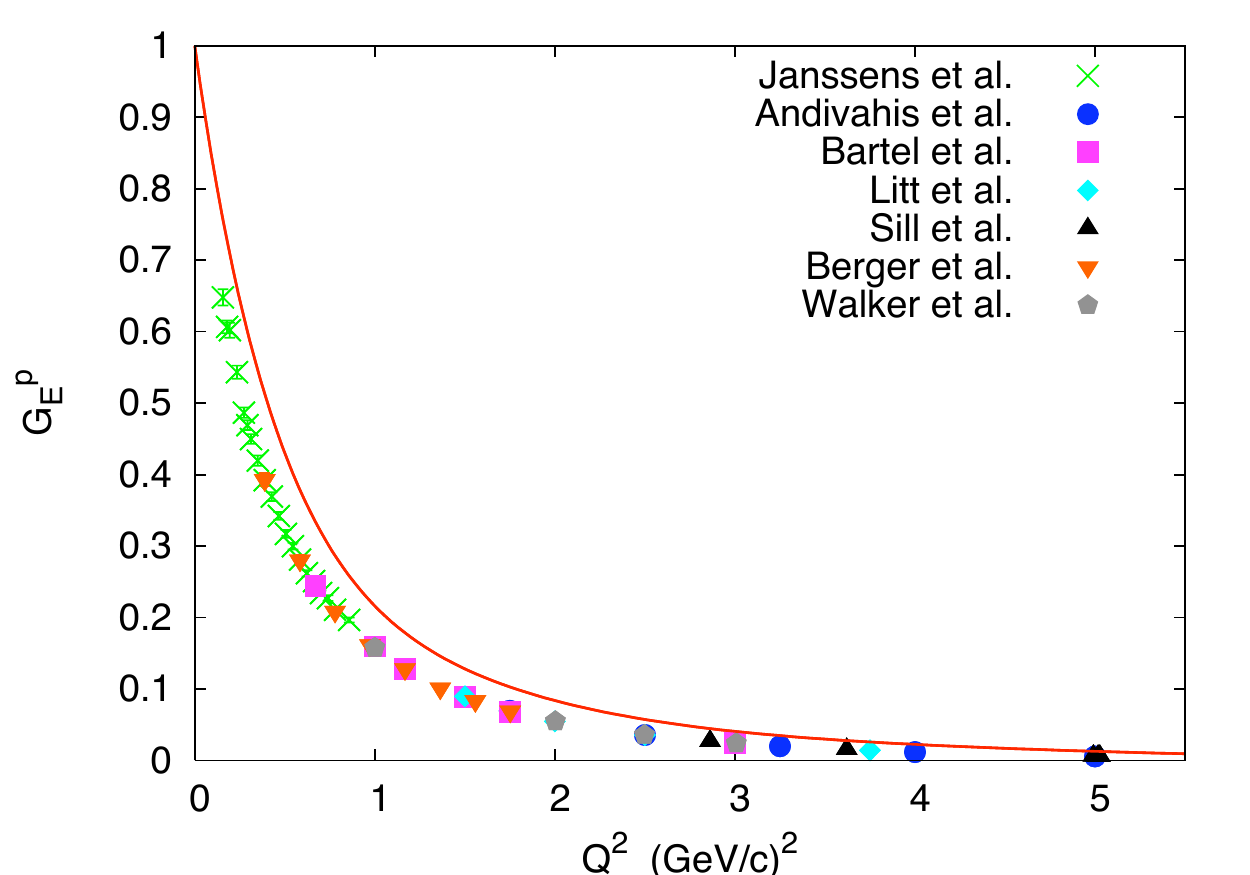}
\includegraphics[width=2.5in]{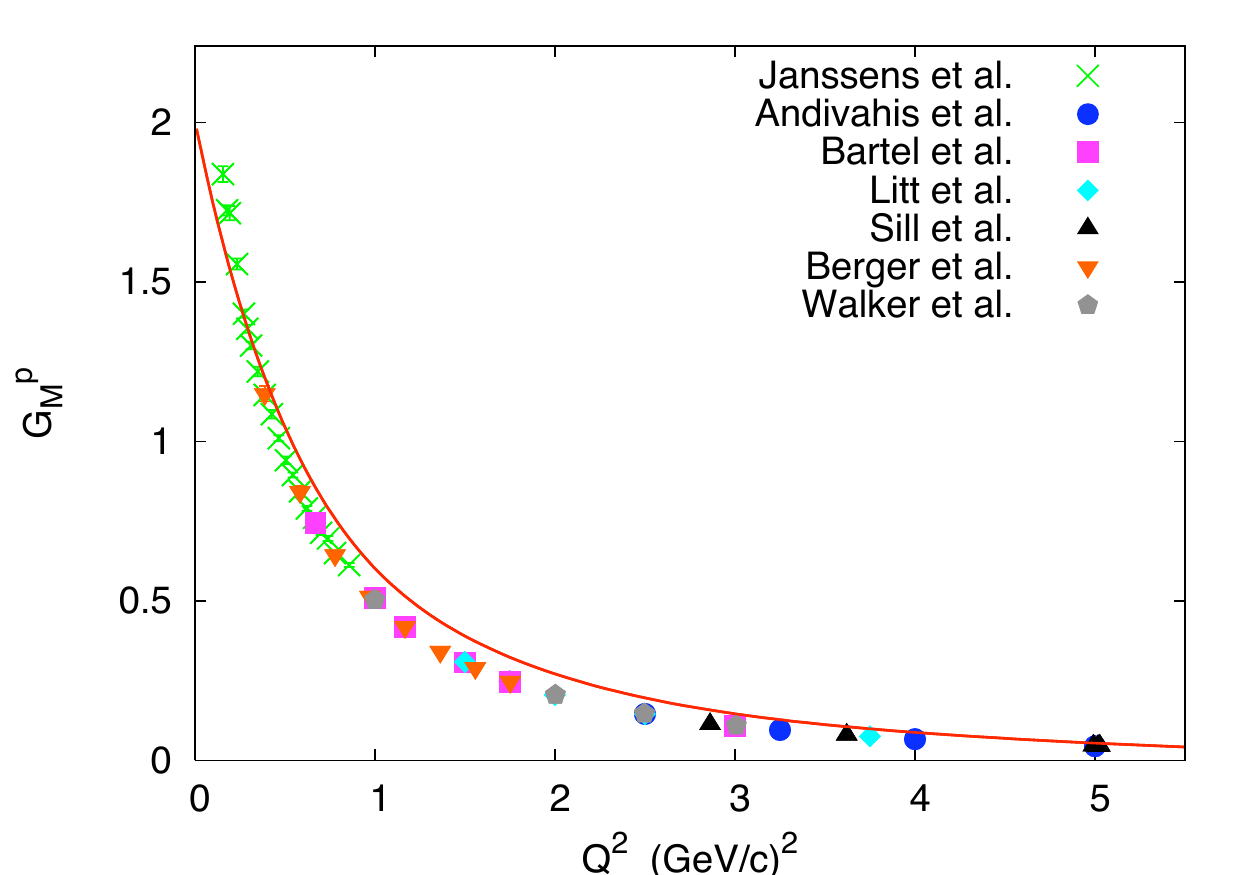}
\includegraphics[width=2.5in]{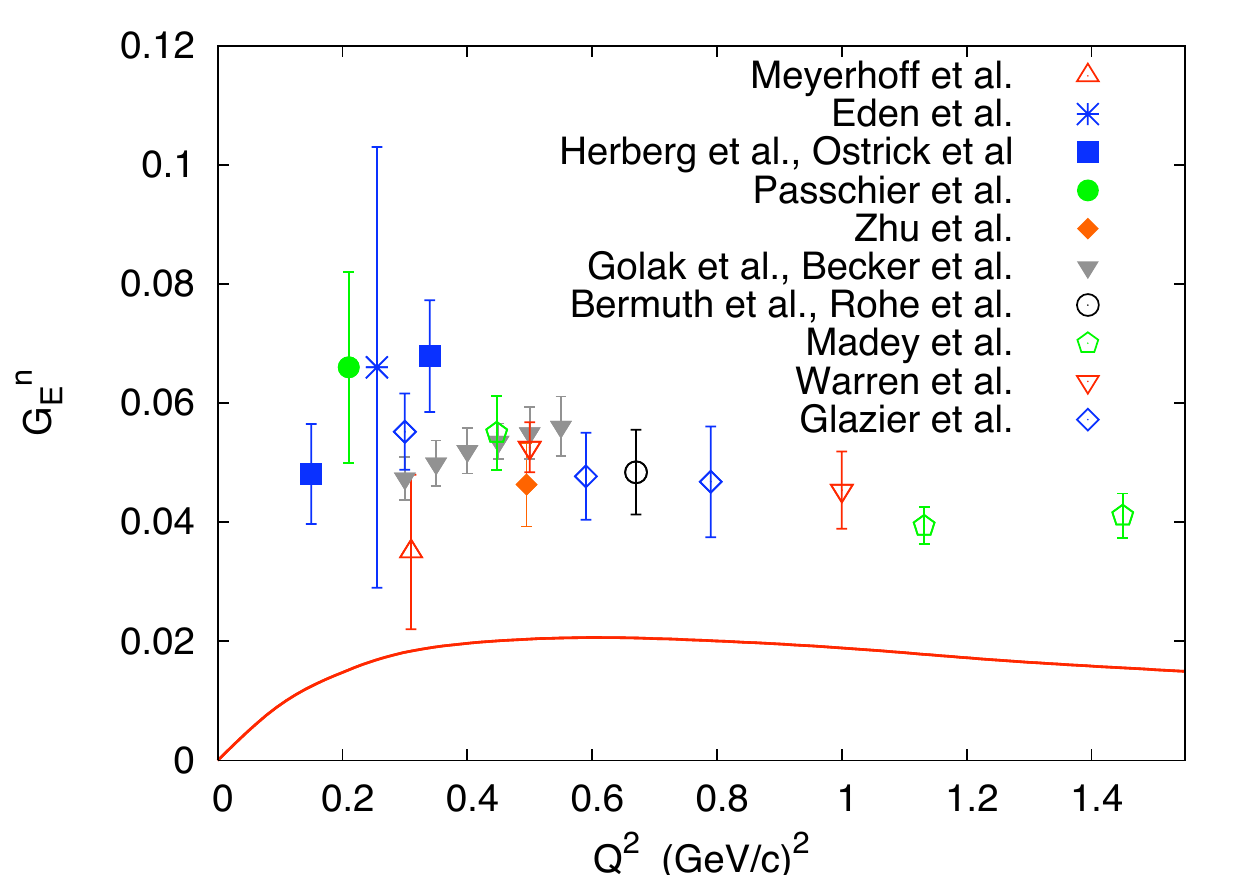}
\includegraphics[width=2.5in]{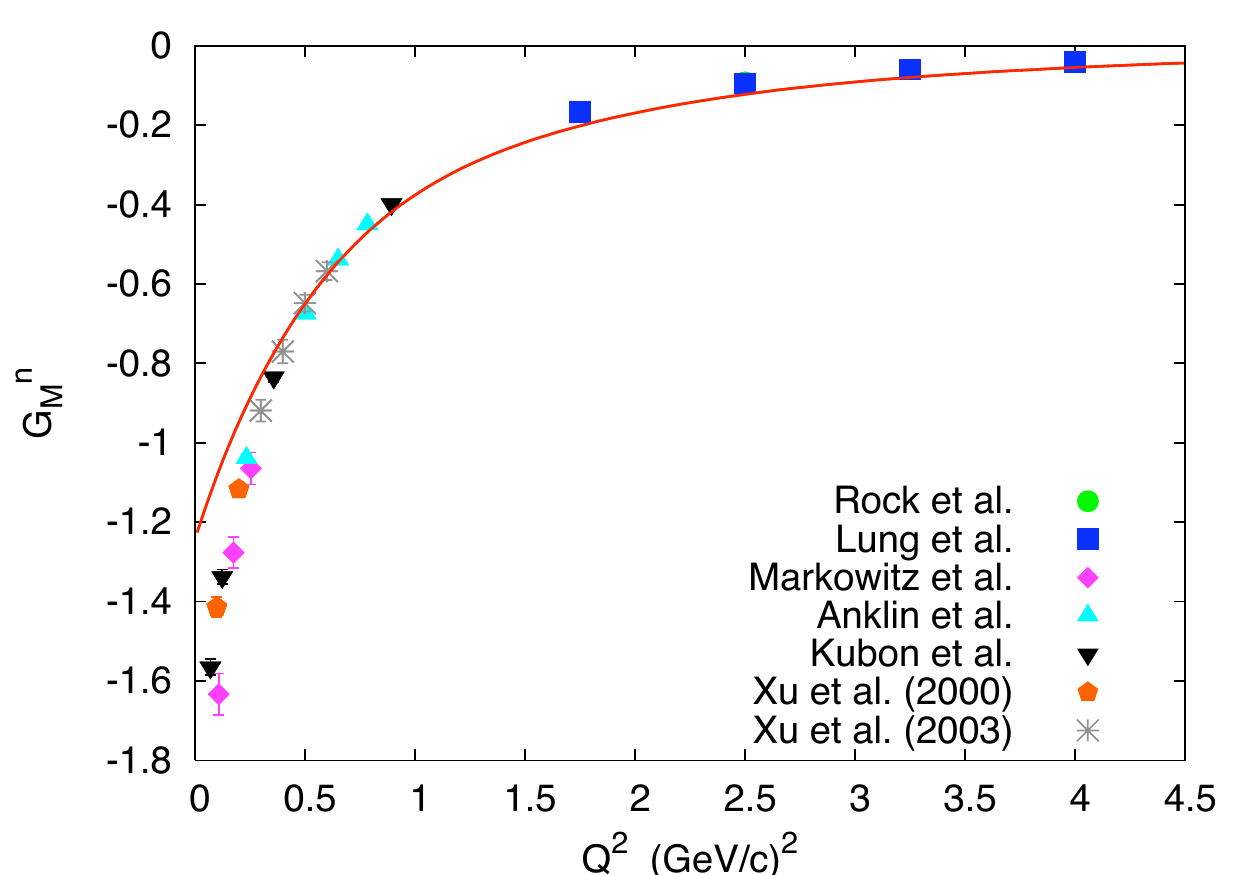}
\caption{The nucleon elastic form factors predicted by the relativistic hCQM (full line) \cite{ff_07}.  The $G_M^p$ data are taken from the reanalysis made in \cite{brash} of the data from \cite{gmp}. The $G_E^p$ data are obtained from the  $G_M^p$ ones and the fit \cite{brash} of the Jlab data on the ratio R; the data for $G_E^n$ and $G_M^n$ are from refs.~\cite{gen} and \cite{gmn}, respectively. The figure is taken from ref.~\cite{ff_07} (Copyright (2007) by the American Physical Society).}
\label{ff_nqf}
\end{figure}

The elastic form factors are extracted from the matrix elements of the current $J_\mu'$ between two nucleon states of the type of Eq.~(\ref{bar}), provided that the boosts are chosen in such a way to transform the initial and final nucleon states to the Breit system.
In this way one can calculate the predicted values of the nucleon form factors in the hCQM in the relativistic version \cite{ff_07}. The theoretical results are compared with the experimental data in Fig.~\ref{ff_nqf}. 

The ratio R turns out to deviate from unity more strongly than in the semirelativistic analysis shown in Sec.~6.2, in which it becomes flat at a value of about 0.6.

The relativistic calculations are not very far from  the experimental data and this is a nice achievement if one thinks that the curves in Fig.~\ref{ff_nqf} are predictions, however  the residual discrepancy indicates that something is missing in the theoretical description. 

We should remind that the CQM accounts only for the contribution of a quark core and that the proton radius corresponding to the hCQM wave functions is much lower that the experimental one. In fact, according to the analysis of the helicity amplitudes for the electroexcitation of the nucleon resonances, the comparison with the experimental data and the evaluation of the pion cloud terms has shown that there is the need of a meson or quark-antiquark  cloud surrounding the  three valence quarks. In this respect it is interesting to note that the fit performed in ref.~\cite{ijl} is based on expressions of the form factors including both an intrinsic and meson contribution; moreover the intrinsic contribution corresponds to a radius ranging from 0.34 fm to 0.55, according to the type of form factor used, dipole or monopole, respectively. There are various attempts in the literature to add meson contributions to the quark core ones, but the problem of a consistent treatment of the quark-antiquark pair creation mechanism, that is of unquenching the CQM, is very complicated and only recently there has been an important improvement for the case of baryons \cite{sb1,bs,sb2}.

There is another relevant feature which is missing. The valence quarks can be viewed as effective degrees of freedom which take into account implicitly complicated interactions involving also gluons and quark-antiquark pairs. As a consequence of such interactions, the quarks can acquire a mass and also a size. This statement is supported by the  analysis of the deep inelastic electron-proton scattering performed in ref.~\cite{ricco} within the Bloom-Gilman duality: the inelastic scattering is due an elastic scattering on constituent quarks with an approximate scaling rule given by the quark form factor. Therefore it seems reasonable to take into account the possibility of quark form factors also in the description of the elastic electron-nucleon scattering. 

\begin{figure}[h]
\includegraphics[width=2.5in]{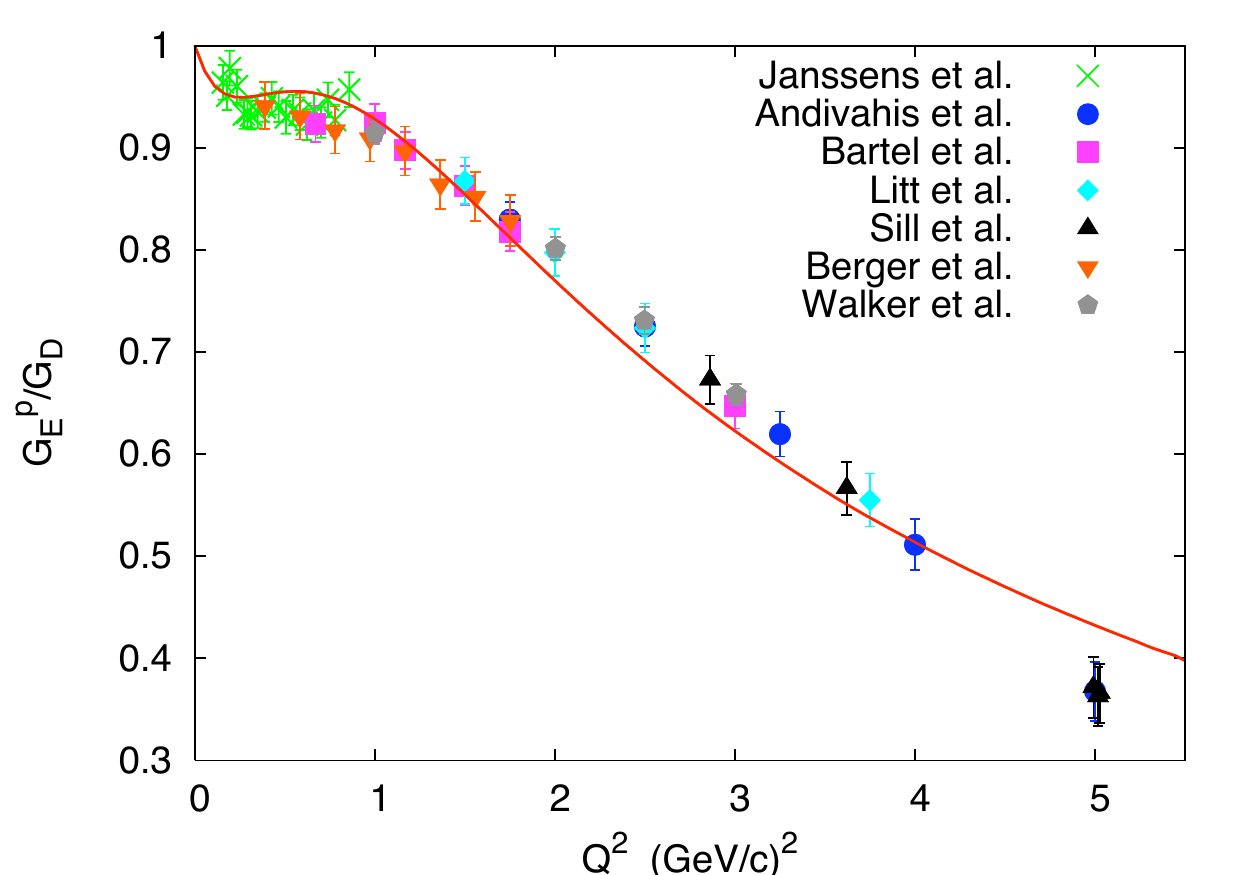}
\includegraphics[width=2.5in]{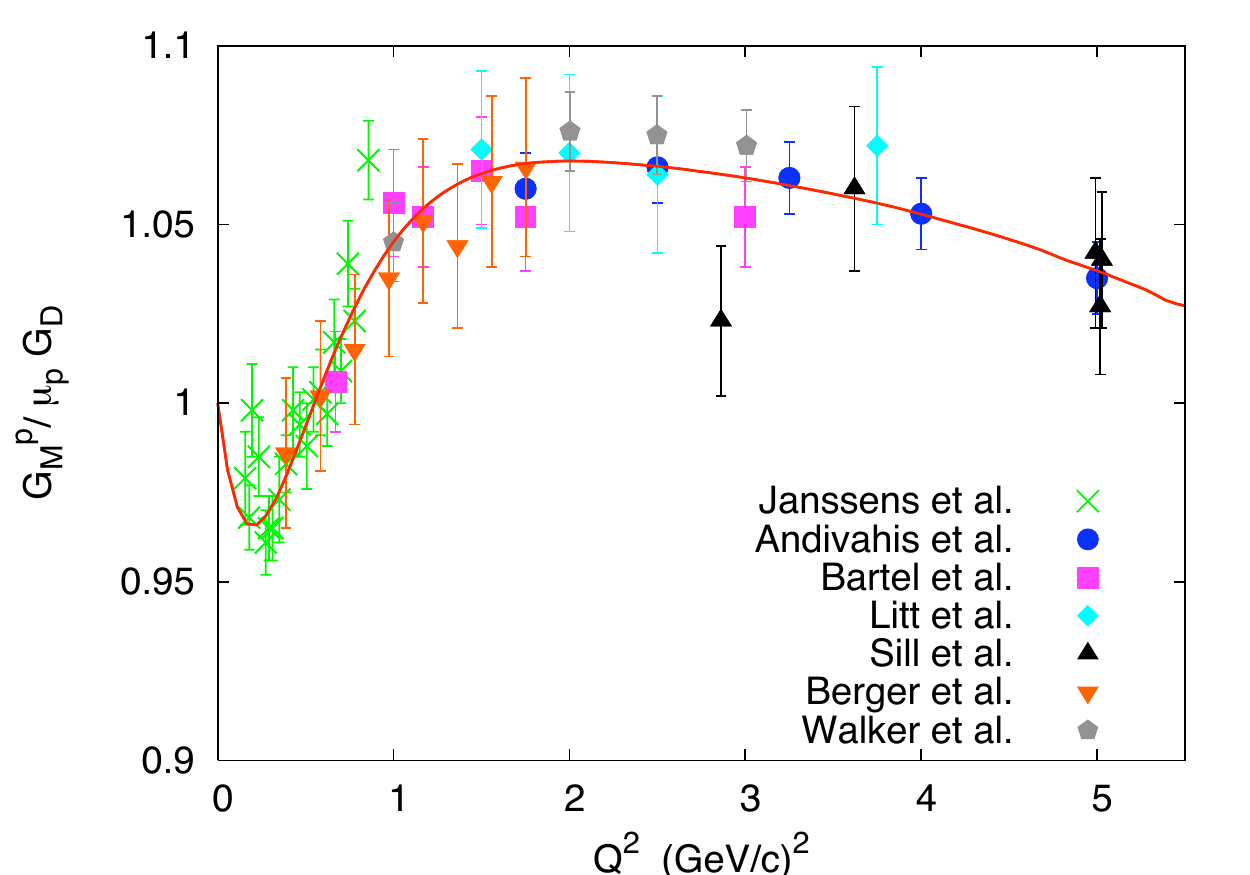}
\includegraphics[width=2.5in]{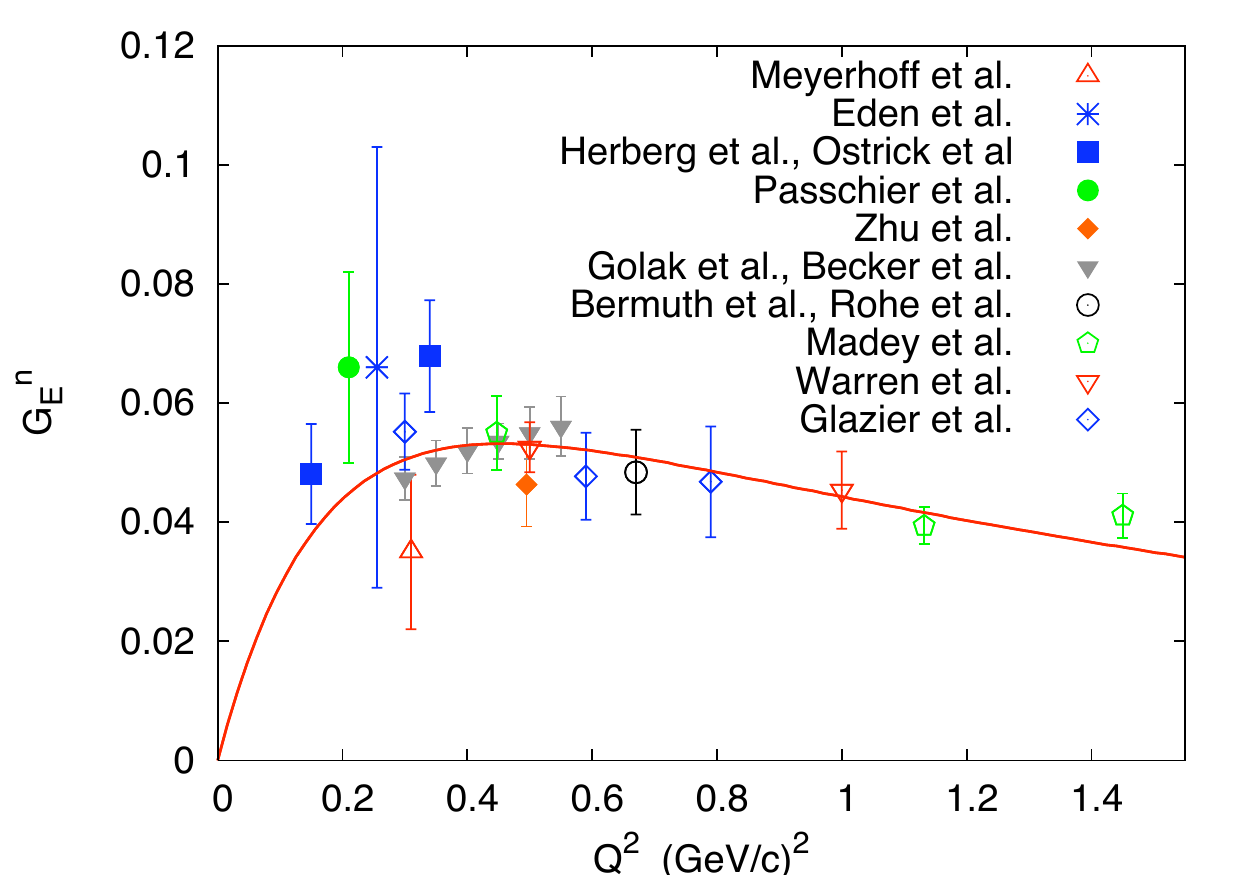}
\includegraphics[width=2.5in]{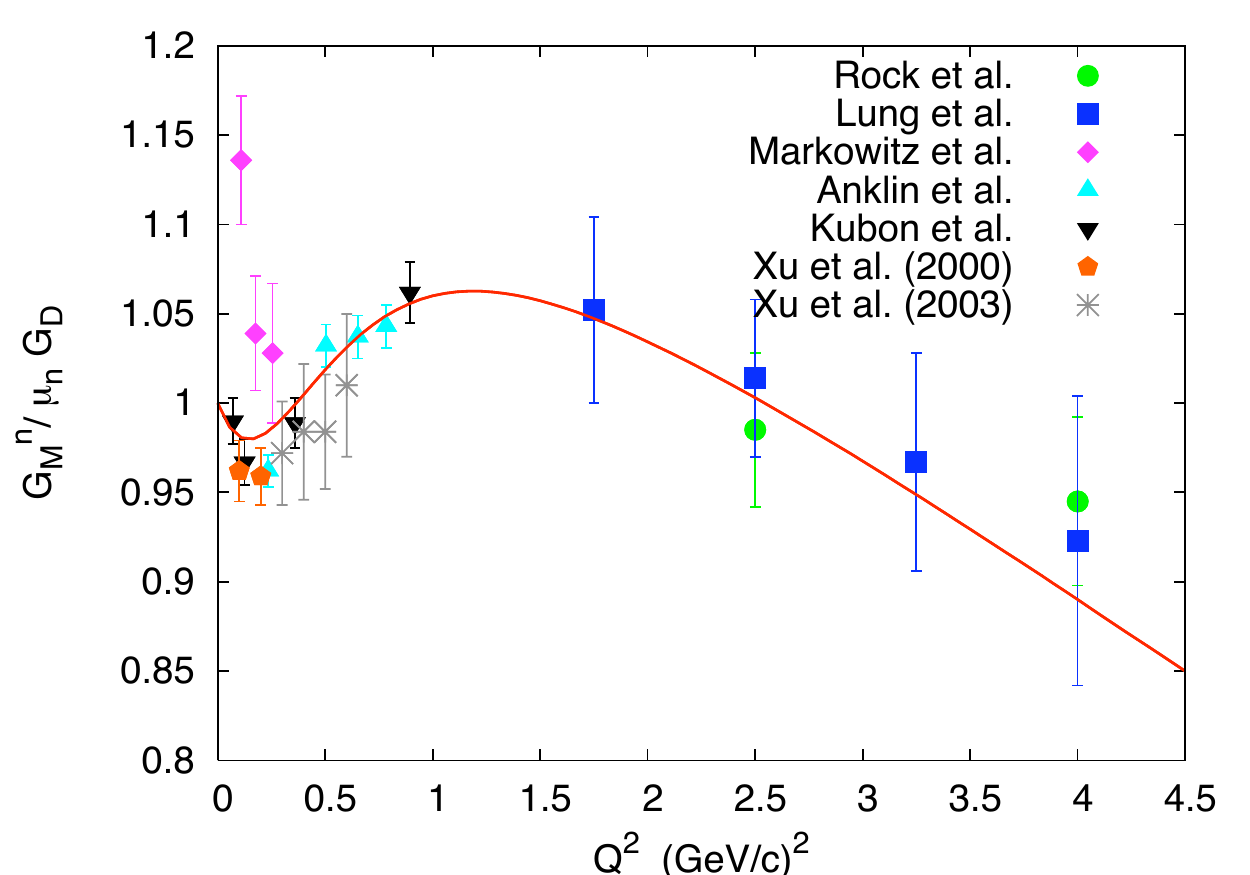}
\caption{The nucleon elastic form factors obtained by the relativistic hCQM (full line) with quark form factors \cite{ff_07}.  The data are as in Fig.~\ref{ff_qff}. The figure is taken from ref.~\cite{ff_07} (Copyright (2007) by the American Physical Society).}
\label{ff_qff}
\end{figure}

A detailed expression of these quark form factors is not known, then one can assume some phenomenological form  and use it in order to get a better fit of the elastic nucleon form factors. Presently the role of these form factors is actually  to parametrize the intrinsic structure of the constituent quarks, but also any 
other  effect which is not included in the theory, as for instance the quark-antiquark pair or meson cloud 
contributions.

\begin{figure}[h]
\includegraphics[width=3in]{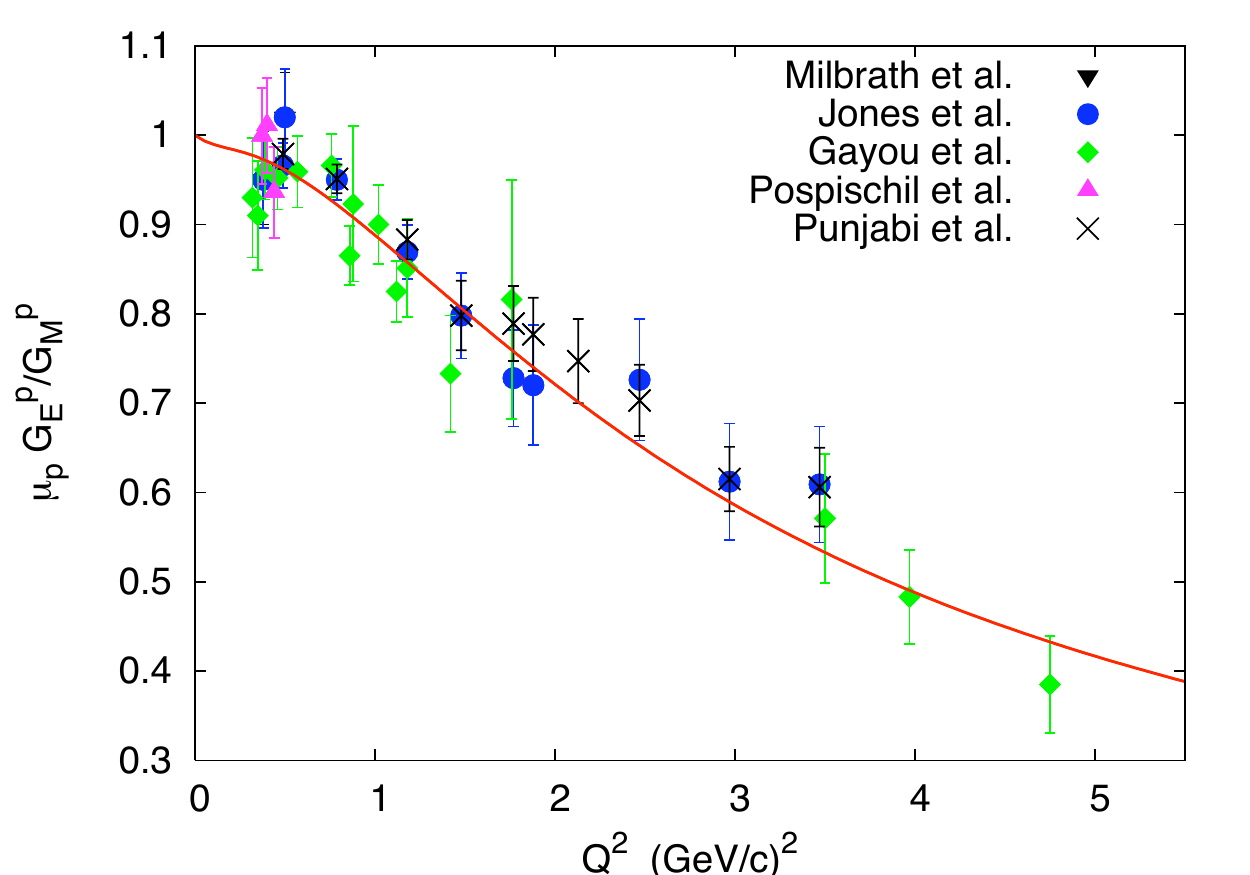}
\caption{The ratio $\mu_p G_E^P(Q^2)/G_M^P(Q^2)$ obtained by the relativistic hCQM (full line) with quark form factors \cite{ff_07}.  The data are as in Fig.~\ref{rap07}. The figure is taken from ref.~\cite{ff_07} (Copyright (2007) by the American Physical Society).}
\label{rap_qff}
\end{figure}

Quark form factors as superpositions of monopole and dipole functions has been used within the relativistic hCQM \cite{ff_07}, leading to a very nice agreement with data. The results of the hCQM taking into account  the quark form factors are shown in Fig.~\ref{ff_qff}, where, for the $G_E^p, G_M^P$ and $G_M^N$ form factors, the ratio to the dipole fit is reported. The fit is performed using the data for the form factors $G_E^N, G_M^P$ and $G_M^N$ and for the ratio $R= \mu_p G_E^p/ G_M^P$, while the curve for $G_E^p$ is derived from the ratio R and $G_M^P$. The resulting ratio R is given in Fig.~\ref{rap_qff}. 

The theoretical description of the experimental data is quite accurate, specially considering the fact that any discrepancy is certainly enhanced by the fact that  the ratio to the dipole form factor is presented.

\subsection{The new data and the problem of a dip in the electric form factor of the proton}

\begin{figure}[h]
\includegraphics[width=2.5in]{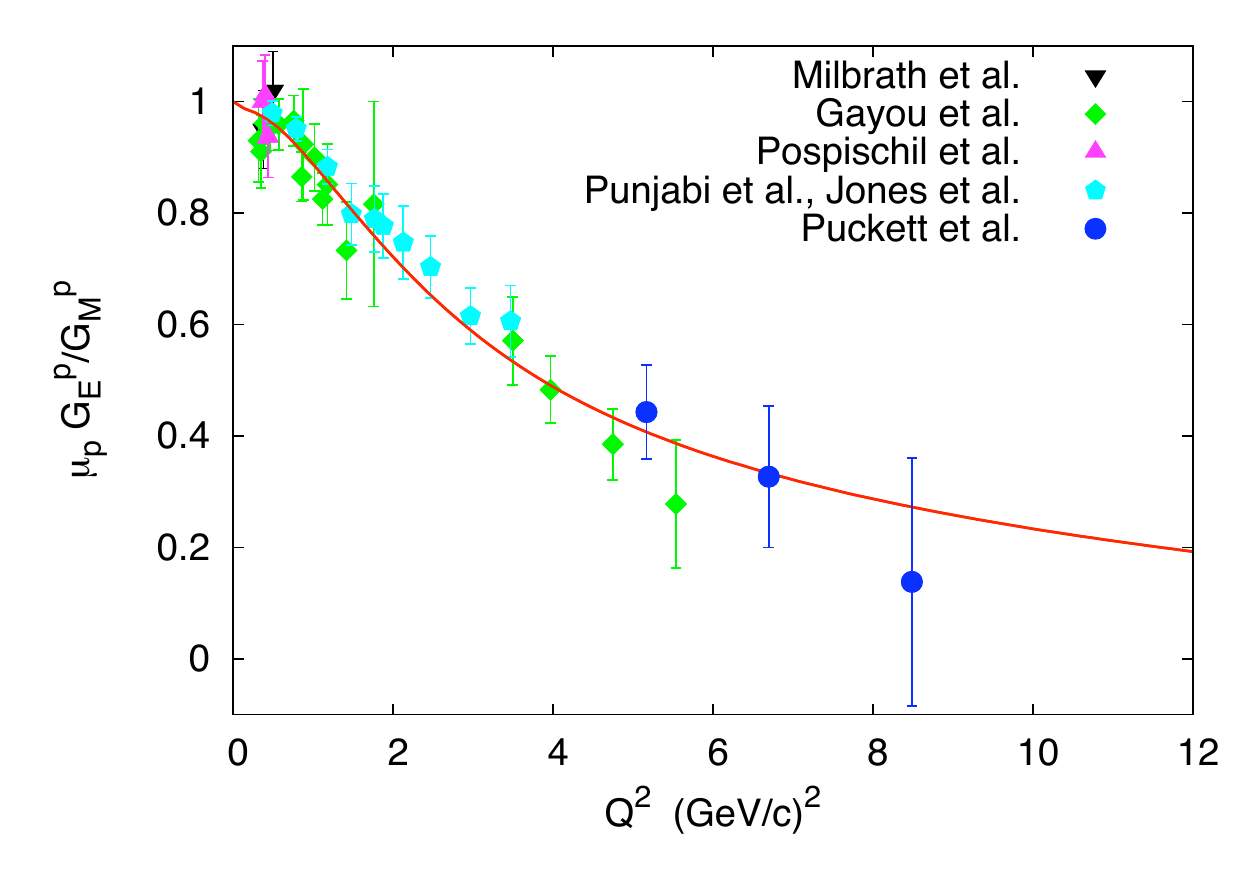}
\includegraphics[width=2.5in]{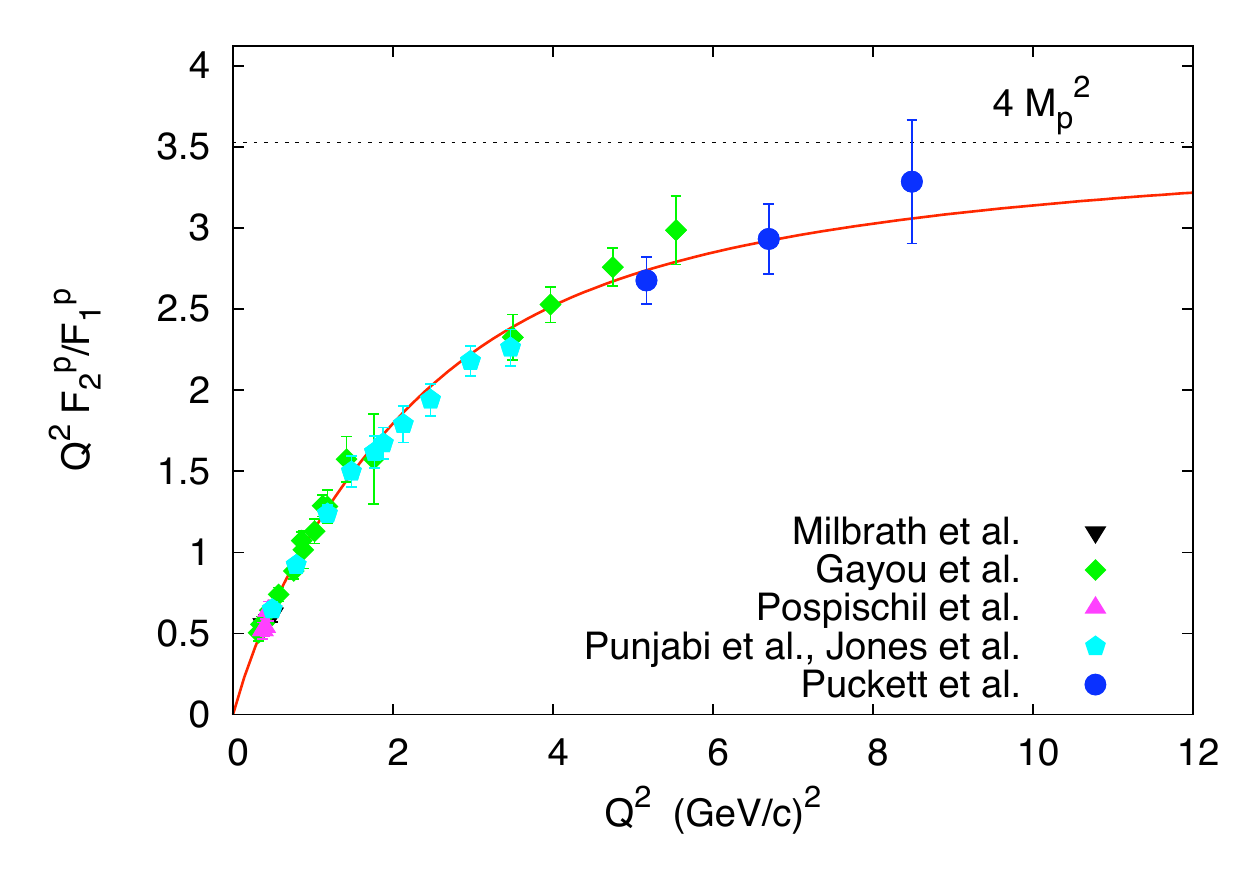}
\caption{Left: The ratio $\mu_p G_E^P(Q^2)/G_M^P(Q^2)$ obtained by the relativistic hCQM (full line) with quark form factors \cite{ff_10}.  The data are as in Fig.~\ref{rap07} and from \cite{puck10}. Right: the ratio $F_p= Q^2 F^p_2(Q^2)/F^p_1(Q^2)$ obtained by the relativistic hCQM (full line) with quark form factors \cite{ff_10}; the data are as in the left side. The figure is taken from ref.~\cite{ff_10} (Copyright (2010) by the American Physical Society) 
 }
\label{rap_10}
\end{figure}

Quite recently, the measurement of the ratio R has ben extended up to $Q^2= 8.5$ GeV$^2$ \cite{puck10}. The curves of ref.~\cite{ff_07} have been extrapolated, without any new parameter fit,  to $Q^2= 12$ GeV$^2$ and compared with the new data \cite{ff_10}. As shown in the left side of  Fig.~\ref{rap_10}, the agreement of the extrapolation with the new data is quite good, although some doubt is cast because of the large error in the last point. The latter is compatible with a dip, while the theoretical curve actually continues to decrease, but  there seems to be no  indication of a zero at finite values of $Q^2$.

The high $Q^2$ behaviour of the proton charge form factor can be studied in an alternative way by introducing the ratio
\begin{equation}
F_p~=~ Q^2\frac {F^p_2(Q^2)}{F^p_1(Q^2)},
\end{equation}
where $F^p_1(Q^2)$ and $F^p_2(Q^2)$ are, respectively, the Pauli and Dirac proton form factors. According to the analysis of Ref.~\cite{brodsky-farrar,brodsky-lepage}, such ratio should reach an asymptotic constant value. It is interesting to note that in correspondence of a zero of the proton form factor $G^p_E(Q^2)$, $F_p$ must cross the value $4 M_p^2$ \cite{ff_10}, where $M_p$ is the proton mass. Using the curves for the ratio R and for  the proton magnetic form factor  $G^p_M(Q^2)$ evaluated in \cite{ff_10}, it is easy to determine $F^p_1(Q^2)$ and $F^p_2(Q^2)$. The results are shown in the right side of Fig.~\ref{rap_10}.

Of course, the two plots in Fig.~\ref{rap_10} contain the same information and no further comment is needed.

The new data by ref.~\cite{puck10} have some problem of compatibility with the previous ones. In fact, the trend of the ratio was, mainly thanks to the last points of ref.~\cite{gayou02}, a linear decrease to a zero given by the fit \cite{ppv}
\begin{equation}
R ~=~ 1.0587 - 0.14265 Q^2.
\end{equation}
As for the data in ref.~\cite{puck10}, the last point, with a very large error, is compatible with such trend, but the other two definitely no. This question has been studied in ref.~\cite{puck12}, where the last points of ref.~\cite{gayou02} have been reanalyzed, with the result  that the measured values of R for $3.5$ GeV$^2 \leq Q^2 \leq  5.6$ GeV$^2$ should be increased, improving the consistency with the higher $Q^2$ data. The presence of a dip in the charge form factor of the proton is still an open problem, which will hopefully solved by the planned experiments at the 12 GeV upgrade of the Jefferson Lab electron accelerator.

Coming back to the hCQM, the theoretical curve reported in Fig.~\ref{rap_10} shows no evidence of dip, in agreement with the new data and the modified \cite{gayou02} points. It should be noted that in ref.~\cite{puck10} the present data   have been fitted by a curve for which the dip, if any, is pushed well beyond $10$ GeV$^2$.

The problem of the R ratio has been dealt with also by other groups involved in the building of CQMs. 

The Rome group \cite{card_00} have calculated the R ratio using the model of ref.~\cite{gi} in the front from dynamics, obtaining a decrease with $Q^2$ given by the Melosh rotations. Using the GBE model with point form dynamics, the Pavia-Graz group \cite{wagen,boffi} reproduced the form factor up to $4$ GeV$^2$ and a decrease of the ratio. A similar behaviour \cite{mert} is obtained by the Bonn group within the Bethe-Salpeter approach with instanton interaction. For further details, the reader is referred to refs.~\cite{gao,kees}.

\subsection{Relativistic transition form factors}

As quoted in Sec.~6.2, the helicity amplitudes seem to be practically unaffected by the application of Lorentz boosts, with the possible ecception of the $N-\Delta$ excitation. These statements can now be checked by calculating the helicity amplitudes with the relativistic hCQM described in the previous sections. These work is presently in progress and for the moment there are only preliminary results on the $N-\Delta$ transition, which is expected to be more sensitive to relativistic corrections, as it happens with the elastic form factor. One should not forget that in the $SU(6)$ limit, the nucleon and the $\Delta$ states are degenerate and belong to the lowest shell.

\begin{figure}[h]
\includegraphics[width=2.5in]{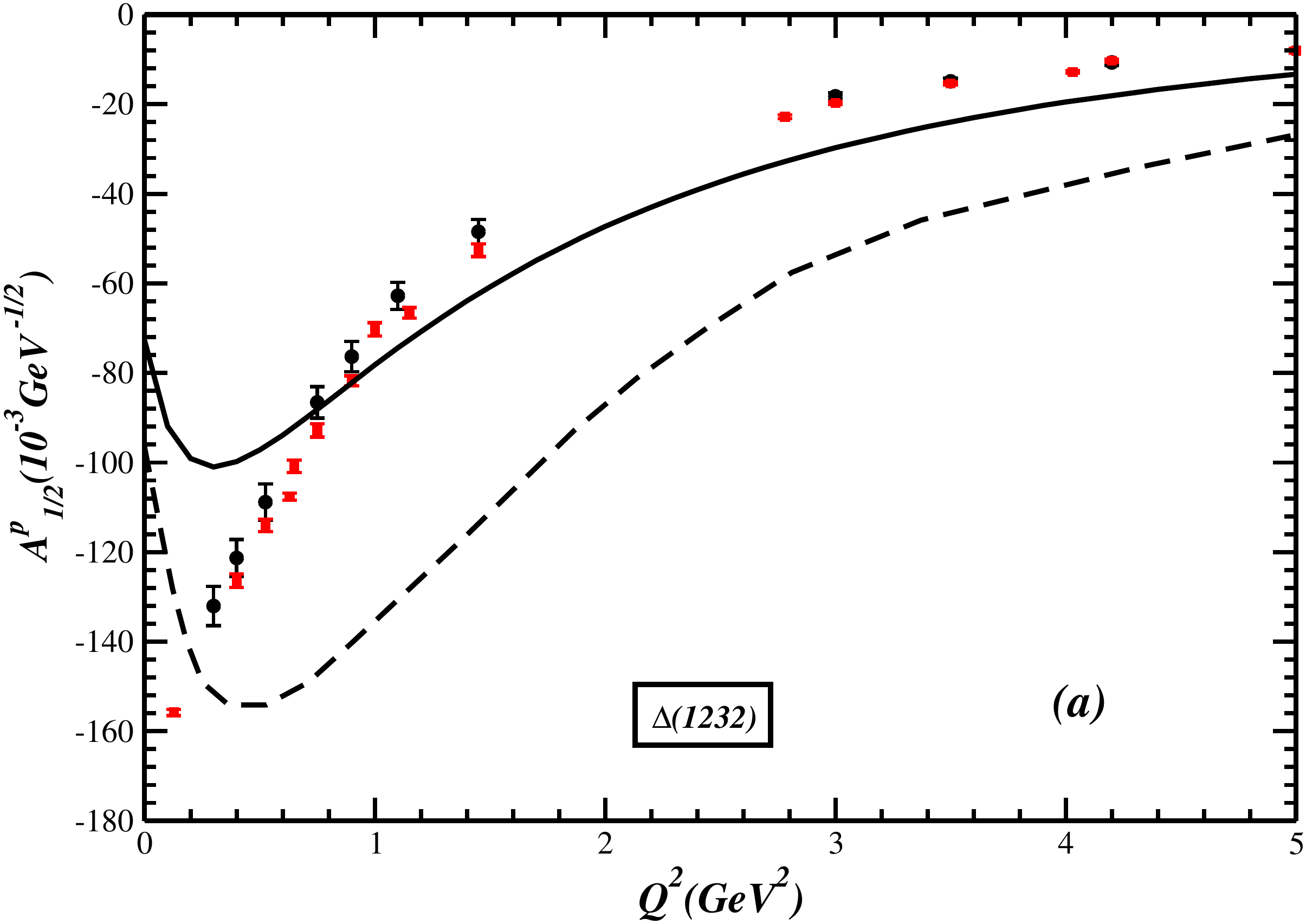}
\includegraphics[width=2.5in]{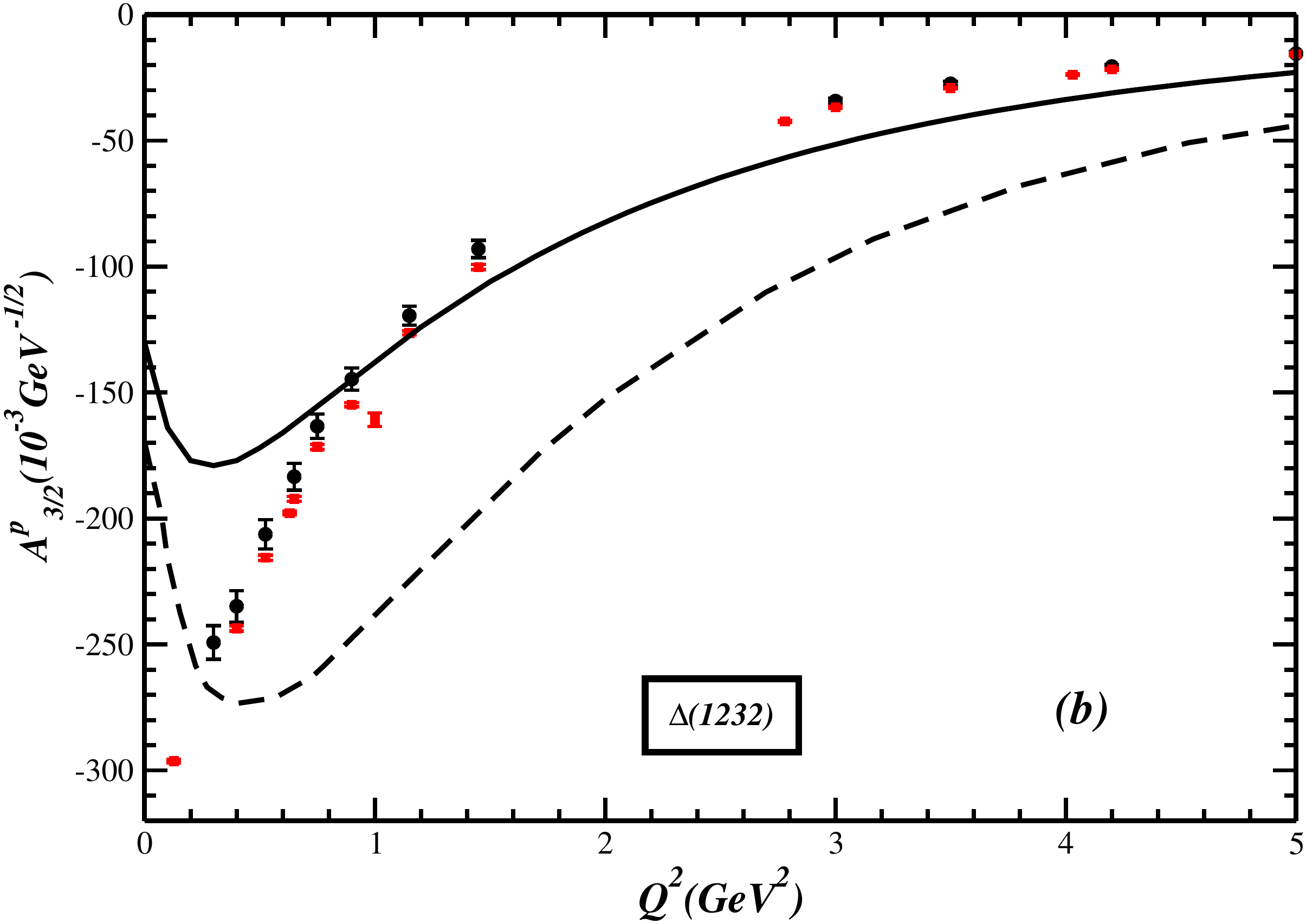}
\caption{ The transverse helicity amplitudes for the $N-\Delta$ excitation. The results of the relativistic hCQM  (full curves) are compared with the nonrelativistic ones (dashed curves) and with the experimental data. a) $A^P_{1/2}$ amplitude; b) $A^P_{3/2}$ amplitude. The data are from \cite{azn09,maid07}(From ref.~\cite{fb22}).}
\label{fb22}
\end{figure}

The procedure for the calculation of the $N-\Delta$ helicity amplitude is the same as in the case of the elastic form factor, apart from the fact  that in this case the final baryon state is given by a $\Delta$. The results \cite{fb22} are given in Fig.~\ref{fb22}.

As expected, the calculations in the fully consistent relativistic hCQM produce a sensible improvement of the results. There is still a lack of strength at low $Q^2$, which, as mentioned in Sec.~5.3, is due to the missing quark-antiquark pair creation mechanisms.

 For the remaining resonances,  the approximate inclusion of relativistic effects performed in Sec.~6.2 is certainly preliminary and a program of a systematic relativistic calculation of both the longitudinal and transverse helicity amplitudes is desirable. In fact there is an important experimental program at the 12 GeV upgrade of Jlab and the availability of a consistent quark model will be useful for the analysis of the high $Q^2$ behaviour of inelastic processes, which are expected to provide valuable information on the short distance properties of the nucleon \cite{wp,wp2}.

\section{Summary and conclusions}

The hCQM provides a good basis for the description of many baryon properties, such as the spectrum, the transition amplitudes for the photo- and electro-excitation of the resonances and the elastic nucleon form factors.

The model  potential (see Eq.~(\ref{H_hCQM})) has, as it happens in all CQMs, a dominant $SU(6)$-invariant part 
and a perturbative term producing the $SU(6)$ violation, a structure which is present already in the early Lattice QCD calculations \cite{wil,deru}. The spin-flavour independent potential with its hC  and linear confinement terms  has strong similarities with the Cornell one \cite{corn} and the static $q \overline{q}$ interaction derived from LQCD in the $SU(3)$ limit \cite{LQCD}, while the $SU(6)$ violating term  is based on the One-Gluon-Exchange mechanism \cite{deru}.

The model has only three free parameters, which are fixed in order to describe the experimental non strange baryon spectrum, obtaining a good agreement in particular for the negative parity resonances \cite{aie2}. Once the parameters have been determined, all the baryon states can be consistently constructed and used  for parameter-free calculations.

Among these there are the predictions of the hCQM for the helicity amplitudes, which are fairly described. The medium-high $Q^2$ behaviour is generally well reproduced, while there is a lack of strength al low $Q^2$, a feature which is in common with all CQMs. The results for the 1/2 amplitudes are usually better than for the 3/2 ones; this is  particularly true for the S11(1535) transverse excitation, whose behaviour has been predicted \cite{aie2} three years in advance with respect to data  \cite{azn09}, \cite{azn05_1}, \cite{den07},  \cite{thom01}. It should be reminded that the calculated proton radius turns out to be $0.48~$fm; such a value was obtained by the fit of the $D13$ photocoupling \cite{cko}, while in the hCQM case is a consequence of the fit to the spectrum. 

 It should be observed that the hC term 1/x \cite{pl,sig,sig2} plays a crucial role in obtaining a good agreement with the experimental data. First of all,  there is a perfect degeneracy between  the first negative parity level $1^-_M$ and the first hyperradial excitation $0^+_S$, a feature which is only slightly modified by the confinement term and is at variance with any two-body potential. This fact provides a better starting point for the description  of the Roper resonance,  although its precise position can be given only if flavour dependent interactions are considered. Moreover, the particular structure of the 1/x spectrum allows to include within the first three shells all the observed 4- and 3- star negative parity resonances, while in the h.o.\ case one should involve  the $3 \hbar \omega$ leveles, the more so if one considers also the 2-star states \cite{gs15}.
 Finally, the presence of the hC potential leads to more realistic quark wave functions, which allow a better description of the helicity amplitudes \cite{aie2,sg}. This statement is supported by the fact that the analytical model \cite{sig,sig2}, where the confinement is treated as a perturbation, makes use only of the hC wave functions and the resulting helicity amplitudes are very similar to those predicted by the complete model of Eq.~(\ref{H_hCQM}).
 
 The nonrelativistic hCQM has been extended in order to include also isospin dependent interactions \cite{iso} and to describe strange baryon resonances \cite{gr}.
 
 The above results have been obtained in the nonrelativistic version of the model and are quite satisfactory, apart from the elastic nucleon form factor. In fact, as long as the transfer momentum $Q^2$ is increased, the theoretical predictions deviate from the experimental data, both because of the small theoretical proton radius and because of an incorrect account of the recoil in passing from the rest frame, where the calculations are performed, to the Breit frame which is more appropriate for the interpretation of the form factors as related to the charge and magnetic distributions. In principle, these considerations are valid also for the inelastic transitions to the resonances, but in this case, because of  the large masses of the resonances, the recoil is less important, apart from the lightest resonance, that is the $\Delta(1232)$. On the other hand, one should not forget that the nucleon and the $\Delta(1232)$ are, in the $SU(6)$ limit, both in the ground state configuration. Actually, the relativistic effects turn out to be important for both the elastic nucleon form factors and for the electromagnetic excitation of the $\Delta(1232)$ resonance.
  
Relativity can be introduced in the hCQM calculations  of the form factors, both elastic and inelastic, by simply applying Lorentz boost to the nonrelativistic three quark states \cite{mds,rap,mds2}. As expected, the helicity amplitudes for the higher resonances are not much affected by this relativistic corrections. On the contrary, the boosted elastic nucleon form factors are improved but are still  not in agreement with the  experimental data. 

A consistent way of constructing a relativistic quark model is provided by the forms of relativistic dynamics introduced by Dirac \cite{dirac}. We have chosen the Point Form (PF), which allows to combine the angular momentum states in the same way as in the nonrelativistic case and considered a mass operator according to a Bakamjian-Thomas (BT) construction \cite{BT} and containing the quark interaction of the hCQM, that is a  hC potential and  a linear confinement. Refitting the parameters, which turn out to be very similar to the nonrelativistic ones, it is possible to predict the elastic form factors \cite{ff_07,ff_10}, which are now very near to the data. Also the predictions for the $\Delta$ excitation are improved if the relativistic PF is used. A very good description of the elastic form factors is achieved if quark form factors are introduced \cite{ff_07}. 

An important issue in connection with the elastic nucleon form factors is the observed behaviour of the ratio $R=\mu_p \frac{G^p_E(Q^2)}{G^p_M(Q^2)}$ with increasing values of the momentum transfer. In the nonrelativistic model, this ratio is practically one, while already the simple application of Lorentz boosts leads to a ratio R which deviates from one, agreement with the idea that  the depletion is a relativistic effect \cite{mg00,rap}. In fact the relativistic hCQM, with the inclusion of quark form factors, is able to describe the behaviour of R as well \cite{ff_07}. As for the possible presence of a dip in the electric form factor, the situation is not yet clear and hopefully it will be settled by the future Jlab experiments. Extending the relativistic hCQM calculations up to $12$ GeV$^2$, there seems to be no indication of a zero in the proton electric form factor.

The hCQM shares with all other models the drawback that the predicted  resonance states have zero width, that is there is no coupling to the continuum. To this end  one can introduce some mechanism for the production of mesons and in this way it is possible to calculate  the baryon decays as for instance in refs.~\cite{ki,cr,cr3,bil3,melde}. However the coupling to the continuum is important also for its consequences on the spectrum, in particular for the possible presence of meson cloud or quark-antiquark components in the states. The latter feature can be introduced by simply considering higher Fock components in the baryon states, either in the form of a meson-baryon or $4q \overline{q}$ configurations. A more consistent procedure is preferable and this  can be achieved by generalizing the quark interaction in order to take into account the coupling to the continuum and then perform ab initio calculations of the various quantities of interest. An attempt in this direction is the work of ref.~\cite{blask}, where however quark and meson-baryon degrees of freedom are treated on the same footing. 
An important progress in this direction is provided by the recent  unquenching of the CQM formulated in refs.~\cite{sb1,bs,sb2}, in which in particular the quark-antiquark pair contributions to the spin and the flavor asymmetry in the proton have been calculated consistently. As already mentioned, the quark-antiquark pair creation mechanism is expected to be the main contribution to the electromagnetic excitation strength at low $Q^2$. In any case, these works open the way to a consistent quark model calculation of the baryon spectrum, the strong decays,  the meson electroproduction and the elastic form factors.

\section*{Appendix A. The hyperradial wave functions}
\setcounter{equation}{0}
\setcounter{table}{0}

The hyperradial wave functions  $\psi_{\gamma \nu} (x)$ are solutions of the hyperradial Sch\"{o}dinger equation (\ref{hyrad}) of the text
\begin{equation}
[~\frac{{d}^2}{dx^2}+\frac{5}{x}~\frac{d}{dx}-\frac{\gamma(\gamma+4)}{x^2}]
~~\psi_{\gamma \nu}(x)
~~=~~-2m~[E-V_{3q}(x)]~~\psi_{\gamma \nu}(x).
\end{equation}
they are normalized according to the condition
\begin{equation}
\int_0^\infty dx ~x^5 |\psi_{\gamma \nu} (x)|^2~=~1.
\end{equation}
The reduced wave function $u_{\gamma \nu}(x)$ is defined as
\begin{equation}
\psi_{\gamma \nu} (x)~=~\frac{u_{\gamma \nu}(x)}{x^{5/2}}
\end{equation}
and obeys to the reduced equation
\begin{equation}
\frac{{d}^2u_{\gamma \nu}(x)}{dx^2}-[\gamma (\gamma+4)-\frac{15}{4}]\frac{u_{\gamma \nu}(x)}{x^2}
+\frac{2m}{ \hbar^2}~[E-V_{3q}(x)]~~u_{\gamma \nu}(x)~~=~~0;
\label{red}
\end{equation}
the behaviour for small values of x is given by 
\begin{equation}
u_{\gamma \nu}(x) \sim x^{5/2+\gamma}.
\end{equation}

\begin{table}[t]
\caption{The hyperradial wave functions $\psi_{\gamma \nu} (x)$ for the harmonic (h.o.) and hyperCoulomb (hC) potentials, up to the second energy shell. In the h.o.\ case $\alpha=\sqrt{3 K m}/\hbar$, $N_{h.o.} = 2 \nu + \gamma$ and an overall common factor   is omitted. For the hC case $N_{hC}= \nu + \gamma$ and the constant g is given by $\tau m/(\nu+\gamma+5/2)$.  
\label{wf}}
\vspace{0.4cm}
\begin{center}
\begin{tabular}{|c|c|c|c||c|c|}
\hline
& & & & &   \\
$\gamma$ & $\nu$ & $N_{h.o.}$ & $\psi_{\gamma \nu} (x)$ (h.o.) &$ N_{hC}$ &   $\psi_{\gamma \nu} (x)$ (hC) \\
& & & & &  \\
\hline
& & & & &  \\
 0 & 0 & 0 & $\alpha^3 $ & 0 &  $\frac{(2g)^3}{\sqrt{5!}} e^{-gx}$  \\
& & & & &  \\
1 & 0 & 1 & $\frac{\alpha^4}{\sqrt{3}} x  $ & 1 &   $\frac{(2g)^4}{\sqrt{7!}} x e^{-gx}$  \\
& & & & &  \\
0 & 1 & 2 & $\frac{\alpha^3}{\sqrt{3}} (\alpha^2 x^2 - 3) $ & 1 &  $\frac{(2g)^3}{\sqrt{7!}} \sqrt{6} (5-2 g x) e^{-gx}$ \\
& & & & &  \\
2 & 0 & 2 &  $\alpha^5 x^2$ & 2 & $\frac{(2g)^5}{3\sqrt{8!}} x^2 e^{-gx}$ \\
 & & & & &  \\
1 & 1 & 3 & - & 2 &  $\frac{(2g)^4}{3 \sqrt{7!}} x (7-2 g x) e^{-gx}$ \\
& & & & &  \\
0 & 2 & 4 & - & 2  & $\frac{(2g)^3}{3 \sqrt{2} \sqrt{6!}} (30-24 g x + 4 g^2 x^2) e^{-gx}$  \\
& & & & &  \\
\hline
\end{tabular}
\end{center}
\end{table}

In the case of the h.o.\ interaction $V_{3q}(x)=3/2 k x^2$, the asymptotic behaviour is determined by the gaussian factor 
\begin{equation}
u_{\gamma \nu}(x) \sim e^{-\alpha^2 x^2/2},
\end{equation}
for the reduced wave function one can then assume the form, apart from a normalization factor,
\begin{equation}
u_{\gamma \nu}(x) ~=~  x^{5/2+\gamma} e^{-\alpha^2 x^2/2} P(x).
\end{equation}
In order that Eq.~(\ref{red}) be satisfied, the function P must obey to
\begin{equation}
\frac{d^2 P}{dy^2} + (\frac{a}{y}-1) \frac{dP}{dy}+\nu \frac{P}{y} ~=~0,
\label{hyconf}
\end{equation}
where the variable $y=\alpha^2 x^2$ has been introduced and
\begin{equation}
a~=~3+\gamma, ~~~~~~~-\nu ~=~\frac{3+\gamma}{2} - \frac{\epsilon}{4 \alpha^2},
\end{equation}
with $\epsilon=\frac{2m}{ \hbar^2} E$. Eq.~(\ref{hyconf}) is the hypergeometric confluent equation and P is then given by the hypergeometric confluent function $F(-\nu, 3+\gamma,y)$. In order to preserve the correct behaviour for small and large x, $\nu$ has to be a non negative integer and P(y) is a polynomial  of order $\nu$. The energy values are then given by
\begin{equation}
E~=~(3 + 2\nu + \gamma) \hbar \omega,
\end{equation}
with $\omega=\sqrt{\frac{3K}{m}}$.

The h.o.\ wave functions for the first three shells are reported in Table \ref{wf}.

For the hC potential $-\frac{\tau}{x}$, one can proceed in an analogous way. The reduced equation (\ref{red}) is written as
\begin{equation}
\frac{{d}^2u_{\gamma \nu}(x)}{dx^2}-[\gamma (\gamma+4)-\frac{15}{4}]\frac{u_{\gamma \nu}(x)}{x^2}
+\frac{2m}{ \hbar^2}~[E-V_{3q}(x)]~~u_{\gamma \nu}(x)~=~0.
\label{redhC}
\end{equation}
In this case the asymptotic behaviour is determined by the factor $e^{-gx}$, where
\begin{equation}
g^2~=~-\sqrt{2 m E},
\end{equation}
($\hbar$ is taken equal to 1). One can then assume
\begin{equation}
u_{\gamma \nu}(x) ~=~  x^{5/2+\gamma} e^{-g x} P(x).
\end{equation}
Eq.~(\ref{red}) is satisfied if the function P obeys to
\begin{equation}
\frac{d^2 P}{dy^2} + (\frac{a}{y}-1) \frac{dP}{dy}+\nu \frac{P}{y} ~=~0,
\label{conf}
\end{equation}
where the variable y is now defined as y=2gx and the parameters a and $\nu$ are given by
\begin{equation}
a~=~5+2 \gamma, ~~~~~~~-\nu ~=~\frac{5}{2} + \gamma - \frac{m \tau}{g }.
\end{equation} 
From the last formula one obtains finally \cite{sig2}
\begin{equation}
g~=~\frac{m \tau}{\frac{5}{2}+ \gamma + \nu}
\end{equation} 
and 
\begin{equation}
E~=~- \frac{g^2}{2m}~=~-\frac{m \tau^2}{2[\frac{25}{4}+(\gamma+\nu)(\gamma+\nu+5)]}.
\end{equation}

The quantities P(y) are associated Laguerre polynomials (see Eq.~(\ref{eigom}) in the text). Also for the hC wave functions for the first three shells are reported in Table \ref{wf}.

\section*{Appendix B. The baryon states}
\setcounter{equation}{0}
\setcounter{table}{0}

The baryon states are superpositions of $SU(6)-$configurations, which can be factorized as follows (see Eq.~(\ref{3q}) of the text) :
\begin{equation}\label{psi_tot1} 
\Psi_{3q} = \theta_{colour} \cdot \chi_{spin} \cdot \Phi_{isospin}
   \cdot \psi_{3q}(\vec{\rho},\vec{\lambda}).
\end{equation}

As already mentioned in the text, the various parts must be combined in order to have a
completely antisymmetric three-quark wave function. To this end it is necessary to study
the behaviour of the different factors with respect to the permutations of three objects
(that is with respect to the group $S_3$). In general, any three particle wave function
 belongs to one of the following symmetry types: antisymmetry (A), symmetry (S), mixed
symmetry with symmetric pair (MS) and mixed symmetry with antisymmetric pair (MA).

For the colour part $\theta_{colour}$ one must choose the antisymmetric colour singlet
combination.

The three-quark spin states are defined as:

\begin{equation}
\chi_{MS}~=~|((\frac{1}{2},\frac{1}{2})1,\frac{1}{2})\frac{1}{2} \rangle ,\nonumber
\end{equation}
\begin{equation}
\chi_{MA}~=~|((\frac{1}{2},\frac{1}{2})0,\frac{1}{2})\frac{1}{2} \rangle ,\nonumber
\end{equation} 
\begin{equation}
\chi_{S}~=~|((\frac{1}{2},\frac{1}{2})1,\frac{1}{2})\frac{3}{2} \rangle .\nonumber
\end{equation} 

The antisymmetric combination is absent because there are only two states at disposal
for three particles.

Similarly one can define the isospin states $\phi_{MS}, \phi_{MA}, \phi_{S}$.

If the interaction is spin and isospin (flavour) independent, one has to introduce
products of $\chi-$ and $\phi-$ states with definite $S_3-$ symmetry. Here we give the
explicit forms only for the case that both factors have mixed symmetry, the remaining
ones being trivial:

\begin{equation}
\Omega_S~=~\frac{1}{\sqrt{2}}~[\chi_{MA} \phi_{MA} + \chi_{MS} \phi_{MS}],
\nonumber
\end{equation}
\begin{equation}
\Omega_{MS}~=~\frac{1}{\sqrt{2}}~[\chi_{MA} \phi_{MA} - \chi_{MS} \phi_{MS}],
\nonumber
\end{equation}
\begin{equation}
\Omega_{MA}~=~\frac{1}{\sqrt{2}}~[\chi_{MA} \phi_{MS} + \chi_{MS} \phi_{MA}],
\nonumber
\end{equation}
\begin{equation}
\Omega_A~=~\frac{1}{\sqrt{2}}~[\chi_{MA} \phi_{MS} - \chi_{MS} \phi_{MA}].
\nonumber
\label{om}
\end{equation}

The  space wave function is given by
\begin{equation}
\psi_{3q}(\vec{\rho},\vec{\lambda})~=
\psi_{\nu\gamma}(x)~~
{Y}_{[{\gamma}]l_{\rho}l_{\lambda}}({\Omega}_{\rho},{\Omega}_{\lambda},\xi),
\label{psi1}
\end{equation}
where $\gamma= 2n +l_\rho + l_\lambda$; the hyperspherical functions are given by \cite{baf}
\begin{equation}
{Y}_{[{\gamma}]l_{\rho}l_{\lambda}}({\Omega}_{\rho},{\Omega}_{\lambda},\xi)~ = ~Y_{l_\rho m_\rho} (\Omega_\rho)~Y_{l_\lambda m_\lambda} (\Omega_\lambda)~^{(2)}P^{l_{\lambda} l_{\rho}}_N(\xi),
\end{equation}
where
\begin{equation}
^{(2)}P^{l_{\lambda} l_{\rho}}_N(\xi)~ = ~C_{n l_\rho l_\lambda} (cos \xi)^{l_\lambda} (sin \xi)^l_{\rho} P_n^{l_\rho+1/2, l_\lambda+1/2}(cos 2\xi),
\end{equation}
with
\begin{equation}
C_{n l_\rho l_\lambda} ~ = ~ \sqrt{\frac{2(2 \gamma +2) \Gamma(\gamma+2-n) \Gamma(n+1)}{\Gamma(n+l_\rho+3/2) \Gamma(n+l_\lambda+3/2}}
\end{equation}
 and $P^{\alpha, \beta}_n(z)$ is a Jacobi polynomial.

The symmetry property of the wave function Eq.~(\ref{psi1}) is determined by the hyperspherical part only, since the hyperradius $x$ is completely
symmetric. In Table \ref{config} we report the combinations of the hyperspherical
harmonics having definite $S_3-$symmetry.

\begin{table}[h]
\caption{Combinations $(Y_{[\gamma]l_{\rho}l_{\lambda}})_{S_3}$  \cite{sig2} of 
the hyperspherical
harmonics $Y_{[\gamma]l_{\rho}l_{\lambda}}$ that have definite $S_3-$symmetry. For
simplicity of notation, in the third column we have omitted the coupling of $l_{\rho}$
and
$l_{\lambda}$  to the total orbital angular momentum $L$. Each combination is labelled
as $L^P_{t}$, specifying the total orbital angular momentum $L$, the parity $P$ and the
$S_3-$symmetry type $t=A, MA, MS, S$.
\label{config}}
\vspace{0.4cm}
\begin{center}
\begin{tabular}{|cc|cc|cc|c|}
\hline
& & & & & & \\
$\gamma$ & & $L^P_t$ & & $(Y_{[\gamma]l_{\rho}l_{\lambda}})_{S_3}$ & &
$S_3$ \\
& & & & & & \\
\hline
\hline
& & & & & & \\
$0$ & & $0^+_S$ & & $Y_{[0]00}$ & & $S$ \\
& & & & & & \\
$1$ & & $1^-_M$ & & $Y_{[1]10}$ & & $MA$ \\
& & & & & & \\
& & $ $ & & $Y_{[1]01}$ & & $MS$ \\
& & & & & & \\
$2$ & & $2^+_S$ & & $\frac{1}{\sqrt{2}}[Y_{[2]20}+Y_{[2]02}]$
& & $S$ \\
& & & & & & \\
& & $2^+_M$ & & $Y_{[2]11}$ & & $MA$ \\
& & & & & & \\
& & & & $\frac{1}{\sqrt{2}}[Y_{[2]20}-Y_{[2]02}]$ & &
$MA$ \\
& & & & & & \\
& & $1^+_A$ & & $Y_{[2]11}$ & & $A$ \\
& & & & & & \\
& & $0^+_M$ & & $Y_{[2]11}$ & & $MA$ \\
& & & & & & \\
& & & & $Y_{[2]00}$ & & $MA$ \\
& & & & & & \\
\hline
\end{tabular}
\end{center}
\end{table}

\begin{table}[t]
\caption{Three-quark states with positive parity \cite{sig2}. The second,
third and fourth  columns show the angular momentum, parity and $S_3$-symmetry,
$L^P_{S_3}$, the spin, $S$, and isospin, $T$. States are shown in the last column  and
are written in terms of  the hyperradial wave functions, $\psi_{\nu \gamma}$,
 of the hyperspherical harmonics, $(Y_{[\gamma]})_{S_3}$ of Table 
\ref{config} and of the spin and isospin states.
\label{statpos}}
\vspace{0.4cm}
\begin{center}
\begin{tabular}{cccccc}
\hline
State & $L^P_{S_3}$ & S & T & & SU(6) configurations \\ 
\hline
$P11$  &  $0^+_S$ & $\frac{1}{2}$ & $\frac{1}{2}$ & & ${\psi}_{00}~Y_{[0]00}
~\Omega_S $
\\
&  $0^+_S$ & $\frac{1}{2}$ & $\frac{1}{2}$ & & ${\psi}_{10}~  Y_{[0]00}
~ \Omega_S$ \\
&  $0^+_S$ & $\frac{1}{2}$ & $\frac{1}{2}$ & & ${\psi}_{20}~  Y_{[0]00}
~ \Omega_S$ \\
&  $0^+_M$ & $\frac{1}{2}$ & $\frac{1}{2}$ & & ${\psi}_{22}~  \frac{1}
{\sqrt{2}}~ [ Y_{[2]00} ~\Omega_{MS} +  Y_{[2]11} ~\Omega_{MA}] $ \\
&  $2^+_M$ & $\frac{3}{2}$ & $\frac{1}{2}$ & & ${\psi}_{22}  ~\frac{1}
{\sqrt{2}}~ [\frac{1}{\sqrt{2}}~ (Y_{[2]20} - Y_{[2]02 })~\phi_{MS} +  
Y_{[2]11} ~\phi_{MA} ]~\chi_S$ \\

$P13$ &  $2^+_M$ & $\frac{1}{2}$ & $\frac{1}{2}$ & & ${\psi}_{22}
~\frac{1}{\sqrt{2}}~[\frac{1}{\sqrt{2}} ~(Y_{[2]20} - Y_{[2]02 })
~\Omega_{MS} +  Y_{[2]11} ~\Omega_{MA} ] $
\\
&  $2^+_M$ & $\frac{3}{2}$ & $\frac{1}{2}$ & & ${\psi}_{22}
~  \frac{1}{\sqrt{2}}~[\frac{1}{\sqrt{2}}~ (Y_{[2]20} - Y_{[2]02 })
~\phi_{MS} +  Y_{[2]11}~ \phi_{MA} ]~\chi_S$ \\
&  $0^+_M$ & $\frac{3}{2}$ & $\frac{1}{2}$ & & ${\psi}_{22}
~\frac{1}{\sqrt{2}}~ [Y_{[2]00} ~\phi_{MS} +  Y_{[2]11} ~\phi_{MA}]~\chi_S$\\
&  $2^+_S$ & $\frac{1}{2}$ & $\frac{1}{2}$ & & ${\psi}_{22}
~\frac{1}{\sqrt{2}}~ [ Y_{[2]20} + Y_{[2]02 }] ~\Omega_S$\\

$F15$ &  $2^+_M$ & $\frac{1}{2}$ & $\frac{1}{2}$ & & ${\psi}_{22}
~ \frac{1}{\sqrt{2}}~[\frac{1}{\sqrt{2}}~ (Y_{[2]20} - Y_{[2]02 })
~\Omega_{MS} +  Y_{[2]11} ~\Omega_{MA} ] $
\\
&  $2^+_M$ & $\frac{3}{2}$ & $\frac{1}{2}$ & & ${\psi}_{22}
~\frac{1}{\sqrt{2}}~[\frac{1}{\sqrt{2}}~ (Y_{[2]20} - Y_{[2]02 })~\phi_{MS} 
+  Y_{[2]11} ~\phi_{MA} ] ~\chi_S$\\
&  $2^+_S$ & $\frac{1}{2}$ & $\frac{1}{2}$ & & ${\psi}_{22}
~\frac{1}{\sqrt{2}}~ [Y_{[2]20} + Y_{[2]02 }]~ \Omega_S$\\

$F17$  &  $2^+_M$ & $\frac{3}{2}$ & $\frac{1}{2}$ & & ${\psi}_{22}
~\frac{1}{\sqrt{2}} ~[\frac{1}{\sqrt{2}} ~(Y_{[2]20} - Y_{[2]02 })~\phi_{MS} 
+  Y_{[2]11} ~\phi_{MA} ] ~\chi_S$\\

$P31$  &  $2^+_S$ & $\frac{3}{2}$ & $\frac{3}{2}$ & & ${\psi}_{22}
~\frac{1}{\sqrt{2}}~ [ (Y_{[2]20} + Y_{[2]02 }] ~\chi_S ~\phi_S$\\
&  $0^+_M$ & $\frac{1}{2}$ & $\frac{3}{2}$ & & ${\psi}_{22}
~\frac{1}{\sqrt{2}}~ [ Y_{[2]00} ~\chi_{MS} + Y_{[2]11} ~\chi_{MA}] ~\phi_S$
\\

$P33$  &  $0^+_S$ & $\frac{3}{2}$ & $\frac{3}{2}$ & & ${\psi}_{00}~Y_{[0]00} 
~\chi_S ~\phi_S$\\
&  $0^+_S$ & $\frac{3}{2}$ & $\frac{3}{2}$ & & ${\psi}_{10}~Y_{[0]00} ~\chi_S 
~\phi_S$\\
&  $0^+_S$ & $\frac{3}{2}$ & $\frac{3}{2}$ & & ${\psi}_{20}~Y_{[0]00} ~\chi_S~ 
\phi_S$\\
 &  $2^+_S$ & $\frac{3}{2}$ & $\frac{3}{2}$ & & ${\psi}_{22}~\frac{1}{\sqrt{2}}
~ [Y_{[2]20} + Y_{[2]02 }] ~\chi_S ~\phi_S$\\
&  $2^+_M$ & $\frac{1}{2}$ & $\frac{3}{2}$ & & ${\psi}_{22}~\frac{1}{\sqrt{2}}
 ~ [\frac{1}{\sqrt{2}} ~(Y_{[2]20} - Y_{[2]02 })~\chi_{MS} +  Y_{[2]11} 
~\chi_{MA} ]~\phi_S$\\

$F35$  &  $2^+_M$ & $\frac{1}{2}$ & $\frac{3}{2}$ & & ${\psi}_{22}
~\frac{1}{\sqrt{2}} ~ [\frac{1}{\sqrt{2}}~ (Y_{[2]20} - Y_{[2]02 })~\chi_{MS} 
+  Y_{[2]11}~\chi_{MA} ] ~\phi_S$\\
 &  $2^+_S$ & $\frac{3}{2}$ & $\frac{3}{2}$ & & ${\psi}_{22}~\frac{1}{\sqrt{2}}
~ [Y_{[2]20} + Y_{[2]02 }] ~\chi_S ~\phi_S$\\

$F37$   &  $2^+_S$ & $\frac{3}{2}$ & $\frac{3}{2}$ & & ${\psi}_{22}
~\frac{1}{\sqrt{2}}~ [ Y_{[2]20} + Y_{[2]02 }] ~\chi_S ~\phi_S$\\
\hline
\end{tabular}
\end{center}
\end{table}

\begin{table}[t]
\caption{Three quark states with negative parity \cite{sig2}. Notation as in Table \ref{statpos}
\label{statneg}}
\vspace{0.4cm}
\begin{center}
\begin{tabular}{ccccccc}
\hline
Resonances & & $L^P_{S_3}$ & S & T & & States \\ 
\hline
$S11$  & & $1^-_M$ & $\frac{1}{2}$ & $\frac{1}{2}$ & & ${\psi}_{11}
~\frac{1}{\sqrt{2}} ~[Y_{[1]10} ~\Omega_{MA} +  Y_{[1]01} ~\Omega_{MS}]$\\
& & $1^-_M$ & $\frac{1}{2}$ & $\frac{1}{2}$ & & ${\psi}_{21}~\frac{1}{\sqrt{2}} 
~[Y_{[1]10} ~\Omega_{MA} +  Y_{[1]01}] ~\Omega_{MS}$\\
& & $1^-_M$ & $\frac{3}{2}$ & $\frac{1}{2}$ & & ${\psi}_{11}~\frac{1}{\sqrt{2}} 
~[Y_{[1]10} ~\phi_{MA} +  Y_{[1]01} ~\phi_{MS}] ~\chi_S$\\
& & $1^-_M$ & $\frac{3}{2}$ & $\frac{1}{2}$ & & ${\psi}_{21}~\frac{1}{\sqrt{2}}
~[Y_{[1]10} ~\phi_{MA} +  Y_{[1]01} ~\phi_{MS}] ~\chi_S$\\

$D13$  & & $1^-_M$ & $\frac{1}{2}$ & $\frac{1}{2}$ & & ${\psi}_{11} 
~\frac{1}{\sqrt{2}} 
~[Y_{[1]10} ~\Omega_{MA} +  Y_{[1]01} ~\Omega_{MS}]$\\
& & $1^-_M$ & $\frac{1}{2}$ & $\frac{1}{2}$ & & ${\psi}_{21}~ \frac{1}{\sqrt{2}}
~ [Y_{[1]10}~ \Omega_{MA} +  Y_{[1]01}] ~\Omega_{MS}$\\
& & $1^-_M$ & $\frac{3}{2}$ & $\frac{1}{2}$ & & ${\psi}_{11}
~\frac{1}{\sqrt{2}}~ [Y_{[1]10} ~\phi_{MA} +  Y_{[1]01}~\phi_{MS}] ~\chi_S$\\
& & $1^-_M$ & $\frac{3}{2}$ & $\frac{1}{2}$ & & ${\psi}_{21}~\frac{1}{\sqrt{2}} 
~[Y_{[1]10} ~\phi_{MA} +  Y_{[1]01} ~\phi_{MS}] ~\chi_S$\\

$D15$  & & $1^-_M$ & $\frac{3}{2}$ & $\frac{1}{2}$ & & ${\psi}_{11}
~\frac{1}{\sqrt{2}} ~[Y_{[1]10} ~\phi_{MA} +  Y_{[1]01} ~\phi_{MS}] ~\chi_S
$\\
& & $1^-_M$ & $\frac{3}{2}$ & $\frac{1}{2}$ & & ${\psi}_{21}~\frac{1}{\sqrt{2}} 
~[Y_{[1]10} ~\phi_{MA} +  Y_{[1]01} ~\phi_{MS}] ~\chi_S$\\

$S31$  & & $1^-_M$ & $\frac{1}{2}$ & $\frac{3}{2}$ & & ${\psi}_{11}~
\frac{1}{\sqrt{2}}~[Y_{[1]10} ~\chi_{MA} +  Y_{[1]01} ~\chi_{MS}] ~\phi_S$\\
& & $1^-_M$ & $\frac{1}{2}$ & $\frac{3}{2}$ & & ${\psi}_{21}~\frac{1}{\sqrt{2}} 
~[Y_{[1]10} ~\chi_{MA} +  Y_{[1]01} ~\chi_{MS}] ~\phi_S$\\

$S33$  & & $1^-_M$ & $\frac{1}{2}$ & $\frac{3}{2}$ & & ${\psi}_{11}
~\frac{1}{\sqrt{2}} ~[Y_{[1]10} ~\chi_{MA} +  Y_{[1]01} ~\chi_{MS}] ~\phi_S
$\\
& & $1^-_M$ & $\frac{1}{2}$ & $\frac{3}{2}$ & & ${\psi}_{21}
~\frac{1}{\sqrt{2}}~ [Y_{[1]10} ~\chi_{MA} +  Y_{[1]01} ~\chi_{MS}] ~\phi_S$
\\
\hline
\end{tabular}
\end{center}
\end{table}

  In Tables \ref{statpos} and
\ref{statneg}, we give the explicit  form of the three-quark states with 
positive and negative parity, respectively. In these Tables the hyperradial wave
functions $\psi_{\nu\gamma}$ are solutions of the hyperradial equation Eq.~(\ref{hyrad}) of the text;
their form depends of course on the hypercentral potential.

\end{document}